\documentclass[prd,aps,twocolumn,nofootinbib,preprintnumbers,superscriptaddress,preprintnumbers,balancelastpage,longbibliography]{revtex4-2}
\usepackage[utf8]{inputenc}
\usepackage[colorlinks=true,citecolor=blue,linkcolor=blue]{hyperref}
\usepackage[normalem]{ulem}
\usepackage{amsmath,amssymb, mathrsfs,xfrac}
\usepackage{epsfig}
\usepackage{graphicx}               
\usepackage{color}
\usepackage{slashed}
\usepackage{multirow}
\usepackage{placeins}
\usepackage[dvipsnames]{xcolor}
\usepackage{epstopdf}
\usepackage{soul}
\usepackage{tikz}
\usepackage[capitalise, english]{cleveref}
\usepackage{siunitx}
\usepackage{xspace}
\usepackage{booktabs}
\usetikzlibrary{trees}
\usepackage{graphicx}
\usepackage[export]{adjustbox}
\usetikzlibrary{decorations.pathmorphing}
\usetikzlibrary{decorations.markings}
\usepackage{aas_macros}
\usepackage{lettrine}
\usepackage{subfigure}
\usepackage{float}

\LettrineTextFont{\itshape}
\usepackage{comment}

\newcommand\myshade{80}
\colorlet{mylinkcolor}{ForestGreen}
\colorlet{mycitecolor}{Aquamarine}
\colorlet{myurlcolor}{violet}

\hypersetup{
  linkcolor  = mylinkcolor!\myshade!black,
  citecolor  = mycitecolor!\myshade!black,
  urlcolor   = myurlcolor!\myshade!black,
  colorlinks = true
}

\definecolor{jblue}{RGB}{20,50,100}
\definecolor{npurple}{RGB} {153, 51, 204}
\definecolor{wred}{RGB}{217,0,56}
\definecolor{white}{RGB}{255,255,255}

\definecolor{korange}{RGB}{235, 80,  43}
\definecolor{korange2}{RGB}{245, 100,  63}
\definecolor{kyelloworange}{RGB}{255, 210,  110}
\definecolor{kyelloworange2}{RGB}{240, 170,  90}
\definecolor{kred}{RGB}{204,  102, 153}
\definecolor{kpurple}{RGB}{153,  61, 190}
\definecolor{kpurplelight}{RGB}{213,  161, 230}

 \definecolor{tobycolour}{rgb}{.5,.0,.5}

\DeclareSIUnit\year{yr}
\DeclareSIUnit\pc{pc}
\DeclareSIUnit\ergs{ergs}
\DeclareSIUnit\msun{\ensuremath{M_\odot}}
\allowdisplaybreaks

\makeatletter
\providecommand*{\diff}%
  {\command{\lmultau}{\ensuremath{L_\mu-L_\tau}\xspace}
\new@ifnextchar^{\DIfF}{\DIfF^{}}}
\def\DIfF^#1{%
  \mathop{\mathrm{\mathstrut d}}%
    \nolimits^{#1}\gobblespace}
\def\gobblespace{%
  \futurelet\diffarg\opspace}
\def\opspace{%
  \let\DiffSpace\!%
  \ifx\diffarg(%
    \let\DiffSpace\relax
  \else
    \ifx\diffarg[%
      \let\DiffSpace\relax
    \else
        \ifx\diffarg\{%
        \let\DiffSpace\relax
      \fi\fi\fi\DiffSpace}

\usepackage{tikz,xcolor,hyperref}

\definecolor{lime}{HTML}{A6CE39}
\DeclareRobustCommand{\orcidicon}{\hspace{-1mm}
	\begin{tikzpicture}
	\draw[lime, fill=lime] (0,0) 
	circle [radius=0.16] 
	node[white] {{\fontfamily{qag}\selectfont \tiny \,ID}};
	\draw[white, fill=white] (-0.0525,0.095) 
	circle [radius=0.007];
	\end{tikzpicture}
	\hspace{-3mm}
}

\foreach \x in {A, ..., Z}{\expandafter\xdef\csname orcid\x\endcsname{\noexpand\href{https://orcid.org/\csname orcidauthor\x\endcsname}
			{\noexpand\orcidicon}}
}


\newcommand{\mytitle}{Road through Dark$\nu$ess: Probing dark matter-neutrino interactions using KM3-230213A }

\begin{document}
	
	
	\title{\mytitle}

\author{Ranjini Mondol\orcidA{}}
\email{ranjinim@iisc.ac.in}
\affiliation{Centre for High Energy Physics, Indian Institute of Science, C.\,V.\,Raman Avenue, Bengaluru 560012, India}

\author{Subhadip Bouri\orcidB{}}
\email{subhadipb@iisc.ac.in}
\affiliation{Department of Physics, Indian Institute of Science, C. V. Raman Avenue, Bengaluru 560012, India}
\affiliation{Centre for High Energy Physics, Indian Institute of Science, C.\,V.\,Raman Avenue, Bengaluru 560012, India}

\author{Akash Kumar Saha\orcidC{}}
\email{akashks@iisc.ac.in}
\affiliation{Centre for High Energy Physics, Indian Institute of Science, C.\,V.\,Raman Avenue, Bengaluru 560012, India}

\author{Ranjan Laha\orcidD{}}
\email{ranjanlaha@iisc.ac.in}
\affiliation{Centre for High Energy Physics, Indian Institute of Science, C.\,V.\,Raman Avenue, Bengaluru 560012, India}

	\date{\today}
	
	
	\begin{abstract}
		KM3NeT has recently reported an event where a muon of energy $120^{+110}_{-60}$ PeV was observed at its ARCA detector, which can stem from a very high-energy neutrino interaction in the vicinity of the detector. Besides revolutionizing our understanding of high-energy neutrino sources, this event can serve as a valuable probe for studying Beyond the Standard Model (BSM) interactions of neutrinos. In this work, we study the dark matter (DM)-neutrino interaction by assuming the neutrino for the event KM3-230213A is originated from a blazar. The flux of such neutrinos, traveling through DM distributed across astrophysical and cosmological scales, can get attenuated due to DM interactions. The detection of such event by KM3NeT allows us to place constraints on the interaction cross section at highest-ever neutrino energy. We derive both conservative constraints—neglecting flux attenuation from the host halo—and optimistic ones by including host halo contributions. Our results show that the energy-independent constraints are weaker than previous bounds. For energy-dependent case, the extreme energy of the event allows us to set some of the strongest limits on scattering cross sections. In future, more such neutrino events with well-understood origin will be essential in constraining or potentially discovering DM-neutrino interactions. 
	\end{abstract}
	
\maketitle
	
\section{Introduction}
\label{Introduction}
Neutrinos act as excellent messengers, since their weak interaction with other Standard Model (SM) particles enables them to travel long distances with minimal deflection and attenuation. This allows them to carry information about both their sources and the intervening medium. Several terrestrial detectors have detected neutrinos over a wide energy range over the last few decades, opening up a new window to probe our Universe\,\cite{Super-Kamiokande:1998kpq,Super-Kamiokande:2001ljr, Super-Kamiokande:2005mbp, Borexino:2008dzn,SNO:2002tuh,SNO:2001kpb, IceCube:2013low} . In the high energy frontier, the IceCube neutrino telescope, located in Antarctica, has detected several Very-High-Energy (VHE) neutrino events in the energy range of $\mathcal{O}(10\,\, \mathrm{TeV})$ to $\mathcal{O}(\mathrm{few \, PeV})$\,\cite{IceCube:2013low, IceCube:2013cdw, IceCube:2014stg, IceCube:2018cha, IceCube:2018dnn,LIGOScientific:2017ync, IceCube:2021rpz, IceCube:2023ame,IceCube:2015gsk}. In addition, telescopes like ANTARES and Baikal-GVD have also detected high-energy neutrinos in the last few years\,\cite{ANTARES:2012xie, Baikal-GVD:2022fmn,Baikal-GVD:2022fis}, opening up a new chapter in high-energy neutrino astroparticle physics.
\begin{figure*}
    \centering
    \includegraphics[width=1.0\linewidth]{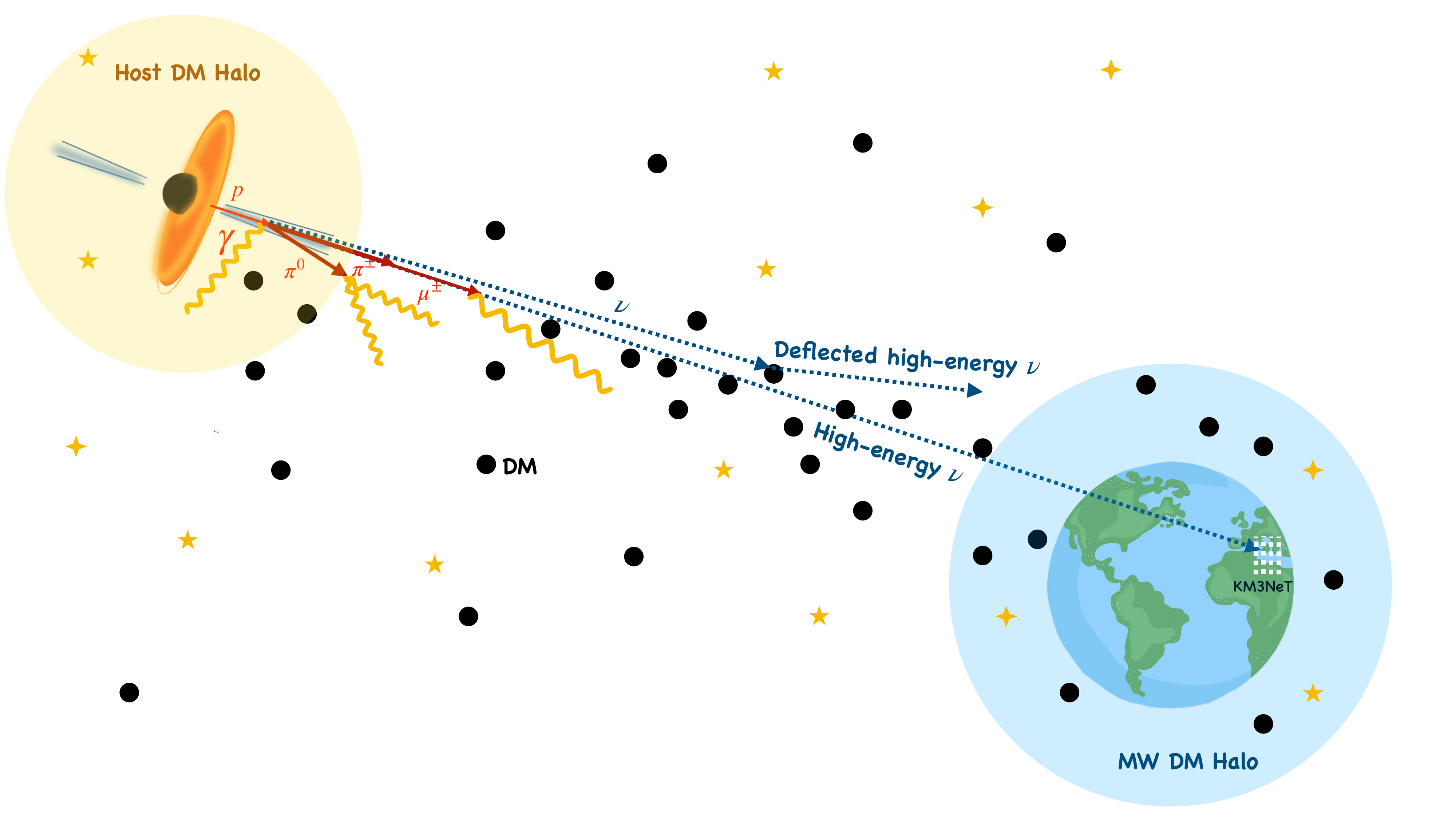}
    \caption{Schematic of the central idea of our work. If KM3-230213A is of a blazar origin\,\cite{KM3-Blazar}, then the neutrinos have to traverse through the host DM halo, extragalactic DM, and Milky Way DM halo. Existence of a DM-neutrino scattering can attenuate/\,deflect the high energy neutrino flux and thus the detection of KM3-230213A can be used to probe DM-neutrino interaction. }
    \label{fig:enter-label}
\end{figure*}

Recently, the KM3NeT collaboration has reported the detection of a through-going muon of energy $120^{+110}_{-60}$ PeV at their ARCA detector on 13th February 2023\,\cite{KM3NeT-Nature,KM3NeT:2025vut,KM3NeT:2025bxl,KM3NeT:2025aps}. This event has been named KM3-230213A. The simulated median energy of the parent high-energy neutrino\footnote{Given that KM3NeT cannot distinguish between neutrinos and antineutrinos at these energy ranges, we will refer to them both as `neutrino'} that can produce such a muon event is $220$ PeV, with an uncertainty spanning $110-790$ PeV (1$\sigma$) and $72\,\mathrm{PeV}-2.6\,\mathrm{EeV}$ ($2\sigma$)\,\cite{KM3NeT-Nature}.  Although IceCube has previously reported multiple high-energy neutrino events, KM3-230213A stands out as the most energetic detection to date, exceeding IceCube’s highest-energy event by approximately a factor of $\mathcal{O}(10)$\,\cite{KM3NeT-Oscillation}. However, the origin of the neutrino for the event KM3-230213A remains uncertain. In Refs.~\cite{KM3NeT-Nature, KM3-Blazar, KM3NeT:2025aps, KM3NeT:2025vut}, the collaboration has identified different possible origins of this event. We note that even though IceCube has a larger exposure, it has not yet detected such a high energy neutrino event. This has resulted in a 2\,-\,3.5$\sigma$ tension between KM3NeT and IceCube depending on the origin of KM3-230213A\,\cite{Li:2025tqf}. Various astrophysical\,\cite{Li:2025tqf,Muzio:2025gbr,Fang:2025nzg,Filipovic:2025ulm,Neronov:2025jfj,Wang:2025lgn} and beyond the Standard Model (BSM) scenarios\,\cite{Boccia:2025hpm,Borah:2025igh,Brdar:2025azm,Airoldi:2025opo,Dev:2025czz,Farzan:2025ydi} have been proposed as the source of KM3-230213A.

The non-gravitational nature of DM remains one of the most intriguing questions in science\,\cite{Bertone:2016nfn}. 
Various space-based as well as terrestrial experiments have been trying to probe the non-gravitational nature of DM, in turn offering valuable insights into the fundamental nature of DM\,\cite{ Cirelli:2024ssz, Strigari:2012acq, Slatyer:2017sev, Lin:2019uvt} . One such possibility is that neutrinos interact with DM particles. Previously, the effects of DM-neutrino interaction have been discussed in the context of laboratory-based experiments \cite{Berryman:2022hds, Dev:2024twk}, cosmological\,\cite{Boehm:2000gq, Boehm:2003xr, Boehm:2000gq, Mangano:2006mp, Wilkinson:2014ksa, Escudero:2018thh, Boehm:2014vja, Brax:2023tvn}, and astrophysical observations\,\cite{Farzan:2014gza,Arguelles:2017atb,Kelly:2018tyg,ChoiIceCube, Murase-Neutrino-AGN, Murase:2019xqi, Ferrera-TXS,Carpio:2022sml, Cline-TXS,Eskenasy:2022aup, Fujiwara:2023lsv,KA:2023dyz}.

As neutrinos travel toward Earth, they pass through various astrophysical environments. However, because of the extremely low densities of Standard Model (SM) particles in these regions, the resulting flux attenuation due to the well-tested weak interactions is expected to be negligible.  Owing to their high density and unknown properties (e.g., mass, number density, and interactions), DM can significantly attenuate the neutrino flux if the interaction cross section is large enough.
In this work, we probe the DM-neutrino interaction using the KM3-230213A event, under the assumption that the event has a blazar origin. The DM-neutrino interaction en route to Earth can attenuate the neutrino flux. Thus, the observation of such an event enables us to place constraints on both energy-independent and energy-dependent DM–neutrino scattering cross-sections. Previously, similar limits have been derived for neutrinos detected at different energies\,\cite{Mangano:2006mp,Choi:2019ixb, Cline-TXS, Cline:NGC-1068,Heston:2024ljf,Fujiwara:2023lsv,Fujiwara:2024qos,Chauhan:2025hoz,Zapata:2025huq,Trojanowski:2025oro} (see \cite{Ng:2014pca,Mazumdar:2020ibx, Chauhan:2024fas,Das:2021lcr,Esteban:2021tub,Das:2024ghw} for the effects of neutrino self-interactions on astrophysical neutrinos detected by IceCube). KM3-230213A has already been used to probe different BSM scenarios including various DM candidates and their interactions\,\cite{KM3NeT:2025mfl, Satunin:2025uui,Cattaneo:2025uxk,Kohri:2025bsn,Narita:2025udw,He:2025bex}. Our work uses the highest-ever energy neutrino event, KM3-230213A to probe the DM-neutrino interaction.

The paper is organized as follows: In Sec.~\ref{sec:Origin} we discuss the possible origin for the event KM3-230213A. In Sec.~\ref{sec:Distribution} we present distribution of DM along the line of sight ($l.o.s$) between the neutrino source and the observer. In Sec.~\ref{sec:scattering} we study how KM3-230213A can be used to put limits on DM-neutrino cross section in both model-independent and model-dependent approaches. We present our results in Sec.~\ref{sec:Result} and finally, in Sec.~\ref{sec:conclusion}, we conclude.

\section{Blazar: A Plausible source of KM3-230213A}

\label{sec:Origin}
The reconstructed direction of the event KM3-230213A is centered on celestial coordinates $ \mathrm{RA} = 94.3^{\circ} $ and $ \mathrm{Dec} = -7.8^{\circ}$ within a region of radius $ 3^{\circ}$ with 99\% confidence\,\cite{KM3NeT-Nature,KM3-Blazar}. 
Several hypotheses for the origin of this neutrino have been tested, including scenarios involving galactic and local sources, transients, and extragalactic sources\,\cite{KM3NeT-Nature, KM3-Blazar, KM3NeT:2025aps, KM3NeT:2025vut}. No potential galactic or transient source has been found in existing gamma-ray, optical, or transient object catalogs that could be associated with this event\,\cite{KM3NeT-Nature}. This implies that the neutrino probably has an extragalactic origin. 

Among extragalactic sources, one of the most compelling possibilities is that VHE neutrinos originate from active galactic nuclei (AGN)\,\cite{Murase:2019vdl,Smith:2020oac, IceCube:2021pgw,Kheirandish:2021wkm,Murase:2022feu}. Supermassive black holes (SMBHs) can become active galactic nuclei (AGN) when they are fueled by the gravitational infall of surrounding matter.  As this matter accretes onto the SMBH, it can convert into radiation, often resulting in jets and winds\,\cite{Murase-Neutrino-AGN, Oikonomou-Neutrino-Blazar}. Blazars are a particular category of AGNs, with their jets oriented along our $l.o.s$\,\cite{Cavaliere:2001gb,Hovatta:2019ulp}. In this work, we take into account the possibility that the KM3-230213A event is sourced by a blazar in its direction.

\begin{figure}[!h]
    \centering
    \includegraphics[width=1.0\linewidth]{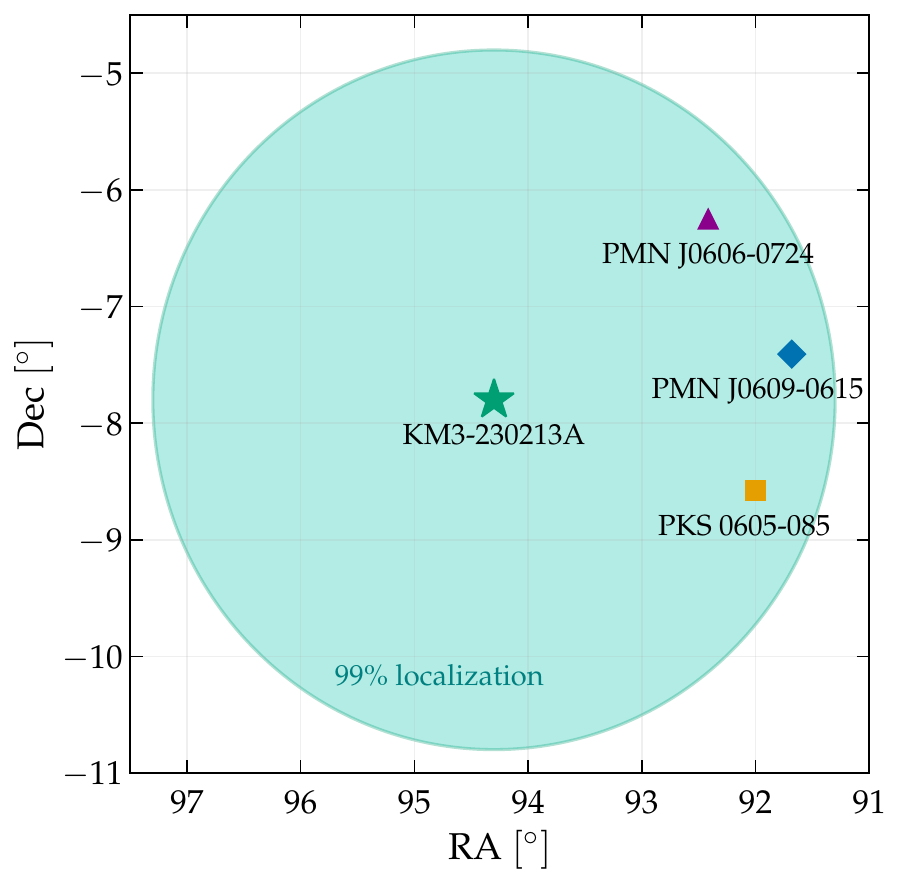}
    \caption{Positions of possible blazar candidates for the event KM3-230213A ($ \mathrm{RA} = 94.3^{\circ} $ and $ \mathrm{Dec} = -7.8^{\circ}$). Details about the event localization can be found in Ref.\,\cite{KM3-Blazar}. We present the details of these blazars in Table\,\ref{tab:Source-Coordinates}. We note that out of the 17 possible blazar candidates presented in Ref.\,\cite{KM3-Blazar}, only these three sources have spectroscopically confirmed redshifts, which is essential for probing DM-neutrino interactions.  }
    \label{fig:blazar_loc}
\end{figure}

Given the directionality of KM3-230213A, a list of seventeen blazars within this angular region has been proposed as potential candidates\,\cite{KM3-Blazar,Dzhatdoev:2025sdi}. These sources have been observed across multiple wavelengths, including radio\,\cite{2025ApJS..276...38P,2018ApJS..234...12L,Aller:2014nwa}, infrared\,\cite{vallenari2023gaia,2010AJ....140.1868W,2011ApJ...731...53M}, X-ray\,\cite{2000yCat.9031....0W,Voges:1999ju,2004ApJ...611.1005G,2005SSRv..120..165B,2021A&A...656A.132S}, and gamma-ray bands\,\cite{2012ApJ...748...49S,2015Ap&SS.357...75M,Ballet:2023qzs}. Among these candidates, we consider only the three sources with known redshifts confirmed by spectroscopic methods, listed in Table\,\ref{tab:Source-Coordinates}. The locations of these three blazars are shown in Fig.\,\ref{fig:blazar_loc}.
\vspace{-0.4 cm}
\begin{table}[h]
\caption{\label{tab:Source-Coordinates}%
List of blazars with known redshifts that might be associated with the event KM3-230213A\,\cite{KM3-Blazar,KM3NeT-Nature}. Here $z$, $\rm M_{\mathrm{BH}}$ and $\rm M_{vir}$ are the redshift, measured mass of the blazar candidates and estimated DM mass of the corresponding host halo of blazar candidates, respectively. The mass of PMN J0606-0724 is yet to be measured. } 
\begin{ruledtabular}
\begin{tabular}{lcll}
\textrm{Source Name} &  $z$ & $ log_{10}(\frac{ M_{\mathrm{BH}}}{ M_{\odot}})$ & $ log_{10}(\frac{M_{\rm vir}}{M_{\odot}})$\\
\colrule
PKS 0605-085 & 0.87 & $8.6^{+0.3}_{-0.3}$\cite{PKSMBH}&$12.9^{+0.2}_{-0.1}$  \\
PMN J0609-0615 & 2.219  & $9.0, 8.89$\,\cite{PMNMBH} &$13.2,13.3$ \\
PMN J0606-0724  & 1.277  & -- \\
\end{tabular}
\end{ruledtabular}
\end{table}

The detection of IceCube-170922A from TXS~0506+056 at $3.5\sigma$ significance has established blazars as promising candidates for VHE neutrinos. Neutrino emission from blazars is not yet very well understood. Following the detection of the event IceCube-170922A, several studies have explored the possibility of neutrinos originating from the broad-line region (BLR) surrounding the blazar TXS~0506+056\,\cite{Padovani:2018acg,Dermer:2014vaa}. In Ref.~\cite{Padovani-BLR-Size-TXS}, the radius of the BLR for TXS~0506+056 was estimated to be $R_{\mathrm{BLR}} \sim 0.021\, \mathrm{pc}$. The authors in Ref.~\cite{Padovani-BLR-Size-TXS} use an empirical relation connecting $R_{\mathrm{BLR}}$ to the monochromatic luminosity at 5100~\AA, $L_{\lambda}(5100\,\text{\AA})$ mentioned in Ref.~\cite{Ghisellini-BLR-Size}. However, determining the BLR size for blazars is challenging because broad emission lines are often overwhelmed by the jet’s continuum emission, rendering them too faint to detect\,\cite{Cavaliere:2001gb}. Nevertheless, the blazars with known masses listed in Table~\ref{tab:Source-Coordinates} are classified as Flat Spectrum Radio Quasars (FSRQs)~\cite{Dzhatdoev-PKS, PMNMBH}, which typically have BLR size of $R_{\mathrm{BLR}} \sim 0.1\, \mathrm{pc}$~\cite{Dermer:2014vaa,Tavecchio:2016dcj,Zheng-FSRQ_BLR,Fan:2023fzj}. Therefore, in this work, we assume a neutrino emission radius in the range $R_{\mathrm{em}} = 0.01-1\, \mathrm{pc}$ considering the possibility of neutrino emission both from the BLR region as well as the inner jet~\cite{Viale-Neutrino-Emission, Xue-Emission-BLR}.

In the BLR regions, the dominant VHE neutrino production channel is photopion production \cite{Oikonomou-Neutrino-Blazar,Dermer:2014vaa}. High-energy protons interact with the UV/X-ray photons to produce neutrinos of energies $\sim$ TeV--PeV through the interaction $p\gamma \to \pi^{+}n$, followed by $\pi^{+} \to \mu^{+} \nu_{\mu}$ and $\mu^{+} \to e^{+}\nu_{e}\bar{\nu_{\mu}}$ \cite{Oikonomou-Neutrino-Blazar, Zhu-BLR-Jet-Production}. On the other hand, jets act like giant cosmic particle colliders, using shock waves or magnetic fields to accelerate protons to PeV--EeV energies. These relativistic jets with protons undergo hadronic interactions with ambient matter and radiation, leading to the production of mesons, primarily pions. The decay of charged pions results in the production of high-energy neutrinos\,\cite{Zhu-BLR-Jet-Production}. In BLR regions neutrinos can still be produced by $pp$ channels. However, it is difficult to produce neutrinos via $p\gamma$ interaction in the outer jets, due to the low abundance of target photons\, \cite{Zhu-BLR-Jet-Production,Murase-Neutrino-AGN,Murase:2018iyl,Xue:2019txw,Zhang:2019dob,Gasparyan-Neutrino-TXS}.

\section{DM Distribution and Neutrino Propagation}
\label{sec:Distribution}
Neutrinos, originated from blazar, travel through DM present in between the source and the observer. Here, we discuss the DM distribution across a wide range of length scales, from galactic to extragalactic (EG), with varying DM densities. Neutrinos emitted near a blazar can interact with DM present in the intervening medium, potentially altering direction, energy, or flavor. As a result, the observed neutrino flux can be suppressed or have its direction altered. We study the effect of DM inside the host DM halo, the EG medium, and the Milky Way (MW) DM halo on flux attenuation. DM surrounding blazars can adiabatically accrete to form a central DM spike in a DM halo. Additionally, if DM particles undergo self-annihilation, this can lead to a flattened density profile. To be exhaustive in our approach, we take into account the host density profile for different possible scenarios and study their effect on constraining the DM-neutrino interaction cross-section. Since the incoming direction of the KM3NeT event does not pass through the MW galactic center, we do not consider any DM density spike profile inside our galaxy.

The amount of suppression in the observed neutrino flux due to neutrino scattering with DM, can be expressed in terms of the optical depth $\tau$,
\begin{equation}
    \tau = \int_{l.o.s}n_{\chi}(l)\,\sigma_{\chi\nu}\,dl,
\end{equation}
 where $l$ is the line-of-sight ($l.o.s$) distance between the source and the observer and $n_{\chi}$ is the DM number density that varies along the $l.o.s$. The initial neutrino flux, $\Phi_{\mathrm{ini}}$ gets attenuated to the observed flux, $\Phi_{\mathrm{obs}}$ as
 \begin{equation}
     \Phi_{\mathrm{obs}} = \Phi_{\mathrm{ini}}\,e^{-\tau}\,\,.
     \label{eq:OpticalDepth}
 \end{equation}
 The optical depth $\tau$ receives contributions from DM present at three different regions, i) inside the host halo, ii) across the EG medium, and iii) inside the MW halo. Thus, the optical depth for neutrinos can be written as, 
\begin{eqnarray}
   \tau = \frac{\sigma_{\chi\nu}}{m_{\chi}}\Bigg(\int\rho_{\mathrm{host}}(r)dr + c\int \rho_{_{\rm EG}}(z)\frac{dt}{dz}dz \nonumber \\ + \int \rho_{\mathrm{gal}}(r(s, l, b))ds\Bigg) \label{eq:DiffCtbnLos},
\end{eqnarray}
where we have made the substitution  $n_{\chi} = \rho/m_{\chi}$ with  $\rho$ being the DM density at different regions and $m_{\chi}$ being the particle DM mass. In the above equation, $\rho_{\mathrm{host}}$, $\rho_{_{\rm EG}}$, and $\rho_{\mathrm{gal}}$ are the DM densities in the host halo, EG medium, and MW halo, respectively.  The integrated DM density along the $l.o.s$ is defined as column density, $\Sigma = \int_{l.o.s} \rho(l)dl$. Below we discuss the possible DM distribution across different scales along the propagation path of the neutrinos and how they contribute to $\Sigma$. 
 
\subsection*{Host halo}

 Neutrinos produced near the blazar, close to the center of its host galaxy, propagate radially outward through DM halo of the host. Therefore, it is reasonable to consider the $l.o.s$ distance within the host halo, as its virial radius, inside which the halo is assumed to be gravitationally bound and dynamically stable. We assume that the DM energy density of the host halo, $\rho_{\mathrm{host}}$, follows a Navarro-Frenk-White (NFW) profile, which is motivated from various N-body simulations\,\cite{Navarro:1996gj, 2010arXiv1010.2539D, Enomoto:2024twc,Angulo:2021kes},
\begin{equation}
    \rho_{\mathrm{host}}(r) = \frac{\rho_{s}}{\left(\frac{r}{r_s}\right)\Big(1+ \frac{r}{r_s}\Big)^{2}} \label{eq: hostProfile},
\end{equation}
where $\rho_{s}$ and $r_{s}$ are the characteristic density and the scale radius of the host halo.\,\,To find $\rho_{s}$ and $r_{s}$, we first estimate the virial mass, $M_{\mathrm{vir}}$, from the empirical $M_{\rm BH}-M_{\mathrm{vir}}$ scaling relation given in Refs.\,\cite{PowellMBHMH:, MarascoMBHMH},
\begin{equation}
\log_{10} \left( \frac{M_{\mathrm{BH}}}{M_{\odot}} \right) = 1.62 \log_{10} \left( \frac{M_{\mathrm{vir}}}{M_{\odot}} \right) - 12.38\,\,, \label{eq:MBH-Mhalo}
\end{equation}
where $M_{\mathrm{BH}}$ is the SMBH mass. In Ref.~\cite{MarascoMBHMH}, the authors use the above scaling relation to fit the $M_{\rm BH}-M_{\mathrm{vir}}$ correlation obtained from a sample of 55 galaxies with known dynamical masses of their SMBHs, and halo masses measured from either globular cluster or stellar dynamics. This relation has a scatter of 0.4 dex\,\cite{PowellMBHMH:, MarascoMBHMH}. Throughout our work we have used the lower error bar on blazar mass, as given in Table\,\ref{tab:Source-Coordinates}, to remain conservative.

The scaling relation in Eq.~\eqref{eq:MBH-Mhalo} is fitted for a sample of galaxies at low redshifts, and essentially needs to be modified for high redshift samples. However, we find that upto redshift $z \sim 2.2 $ the order of magnitude of the halo mass $M_{\mathrm{vir}}$ remains the same if estimated transitively from $M_{\rm BH}-M_{*}$ and $M_{*}-M_{\mathrm{vir}}$ scaling relations in the literature \cite{Kormendy:2013dxa, reines2015relations, 2018ApJ...863...42C,2010A&A...522L...3S, Shankar:2019yyr,suh2020no}. The redshifts of KM3-230213A candidate blazars in Table~\ref{tab:Source-Coordinates} fall within this range. Here, we assume that the virial mass is the same as the halo mass. From $M_{\mathrm{vir}}$, one can find out the values of $r_{s}$ and $\rho_{s}$ using two equations \cite{CirelliDM}.
\begin{eqnarray}
    \int_0^{R_{\rm vir}} d^3r \rho_{\mathrm{host}}(r) = M_{\mathrm{vir}},\;\; R_{\mathrm{vir}} = c_{200}\,r_{s}. \label{eq:ProfileParam}
\end{eqnarray}
We assume the virial radius, $R_{\mathrm{vir}} = 100\,\mathrm{kpc}$, of the same order as the MW virial radius. The parameter $c_{200}$ is called the concentration parameter, which for a halo of mass $M_{\mathrm{vir}}$ at a redshift $z \lesssim 4$ is given by an empirical formula given in Ref.\,\cite{Correa-ConcentrationParam} (Appendix B1),
\begin{eqnarray}
\log_{10} c_{200} &=& \alpha + \beta \log_{10} \left( \frac{M_{\rm vir}}{M_\odot} \right) \left[ 1 + \gamma' \left( \log_{10} \frac{M_{\rm vir}}{M_\odot} \right)^2 \right] \nonumber \,\,,\\
\alpha &=& 1.7543 - 0.2766(1 + z) + 0.02039(1 + z)^2 \nonumber\\
\beta  &=& 0.2753 + 0.00351(1 + z) - 0.3038(1 + z)^{0.0269} \nonumber \\
\gamma' &=& -0.01537 + 0.02102(1 + z)^{-0.1475}.\nonumber \\ \label{eq:Concentration-Param}
\end{eqnarray}
Using Eqs.~\eqref{eq:MBH-Mhalo} and \eqref{eq:ProfileParam} we estimate the parameters of the host halo DM profile. Our estimated parameter values are shown in Table~\ref{tab:HostProperties}. 

\begin{table}[h]
\caption{\label{tab:HostProperties}%
List of host DM halo parameters for the potential blazar candidates with known masses for the event KM3-230213A}
\begin{ruledtabular}
\begin{tabular}{lccc}
\textrm{Source Name} & $c_{200}$ & $r_{s}$ (kpc) & $\rho_{s}$ ($\mathrm{GeV/cm^{3}})$\\
\colrule
PKS 0605-085 & 5.60 & 17.84 & 3.02   \\
PMN J0609-0615 & 3.81 & 26.3 & 2.93  \\
\end{tabular}
\end{ruledtabular}
\end{table}

The black hole masses are estimated in various ways in the literature e.g., using dynamical measurement, reverberation mapping (RM) technique\,\cite{Cackett:2021gad}, and the $M_{\rm BH}-\sigma$ relation ($\sigma$ is the velocity dispersion of the bulge \cite{Xiong:2014hta}), among other methods. The mass of PKS 0605-085 blazar has been estimated in Ref.~\cite{PKSMBH}, using dynamical masses of the blazars, and in Ref.~\cite{PMNMBH}, using the $M_{\rm BH}-\sigma$ relation. In our work, we use the PKS 0605-085 mass from Ref.~\cite{PKSMBH}, which predicts a smaller blazar mass than estimated in Ref.~\cite{PMNMBH}. This choice makes our results conservative. For PMN J0609-0615, we use the mass estimate given in  Ref.~\cite{PMNMBH}. Due to lack of knowledge of the PMN
J0606-072 blazar mass in the literature, we include the host contribution only for blazars PKS 0605-085 and PMN J0609-0615, with and without a DM spike.

\begin{figure*}
	\begin{center}
		\includegraphics[width=\columnwidth]{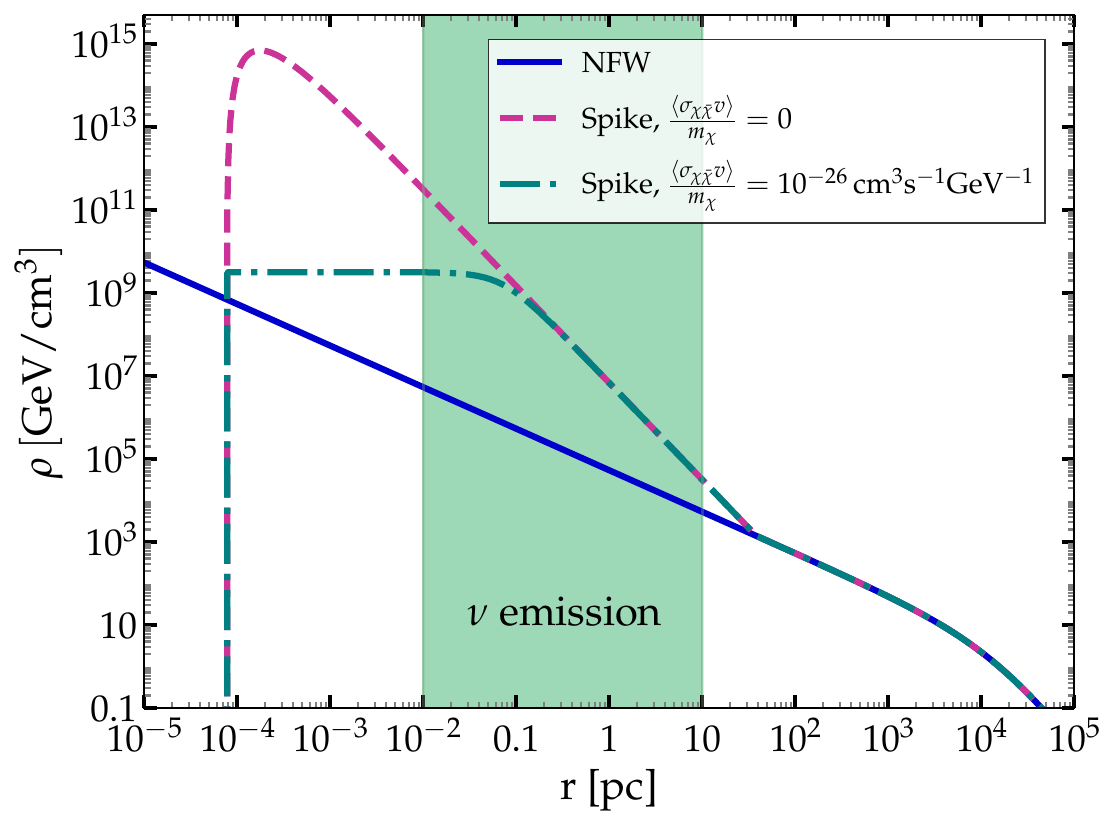}~~
		\includegraphics[width=\columnwidth]{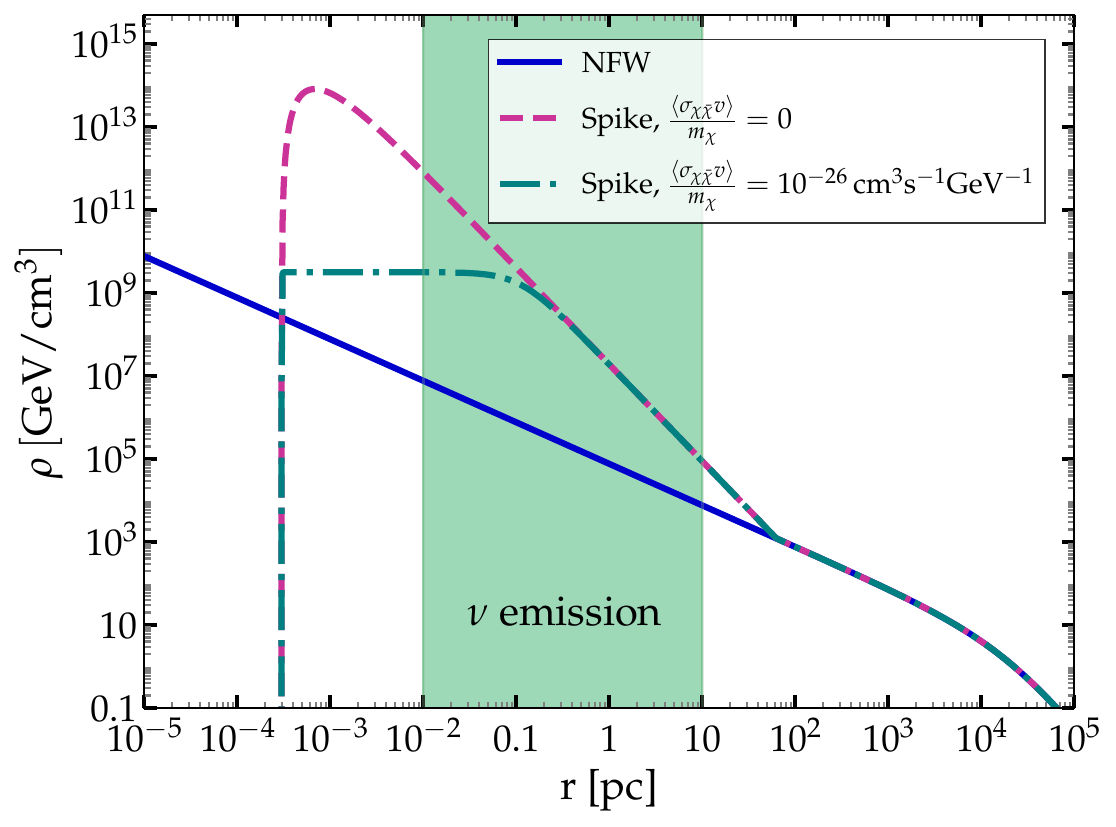}~~\\	
		\caption{DM density profiles of the blazars PKS 0605$-$085 (left panel) and PMN 0609$-$0615 (right panel), two possible candidate sources of KM3-230213A. In both the figures, the solid blue line represents the case where the host blazar DM follows NFW profile; the dashed pink line corresponds to the presence of a DM spike; and the teal dot-dashed line shows the scenario with DM spike along with self-annihilating DM, assuming a cross section of \( \langle\sigma v\rangle = 10^{-26}\,\mathrm{cm}^2 \). For both the blazars, neutrino is assumed to be originated from a region, shown in seagreen shade, located within  $10^{-2}-1$\,pc of the blazar.}
		\label{fig:Density-Profiles}
	\end{center}	
\end{figure*}

\subsection*{Host halo with a DM Spike}
 The presence of a central SMBH can significantly change the density profile of DM in its vicinity. As the black hole grows adiabatically, the surrounding DM can be gravitationally compressed into a dense region known as a `DM spike'~\cite{Gondolo-Spike,Merritt:2003qk,Gnedin:2003rj,Sadeghian:2013laa}. If the initial DM distribution follows a cuspy profile of the form $\rho(r) = \rho_s (r/r_s)^{-\gamma}$, the compression leads to a cuspier inner profile within a characteristic spike radius $ R_{\mathrm{sp}}$ around an SMBH of Schwarzschild radius $R_{s}$ as
\begin{equation}
    \rho_{\mathrm{sp}}(r) = \rho_{R}\,g_{\gamma}(r)\Big(\frac{R_{sp}}{r}\Big)^{\gamma_{\mathrm{sp}}},\label{SpikeWithoutAnn}
\end{equation}
where $\rho_{R} = \rho_{s}(R_{sp}/r_{s})^{-\gamma}$, spike radius $R_{\mathrm{sp}} = \alpha_{\gamma}r_{s}(M_{\mathrm{BH}}/\rho_{s}r_{s}^{3})^{1/(3-\gamma)}$, and $g_{\gamma}(r) = (1-4R_{s}/r)^{3}$. The parameter $\alpha_\gamma$ is a function of $\gamma$, with $\alpha_{\gamma} \approx 0.1$, for $\gamma = 0.7-1.4$\,\cite{Ferrera-TXS}. The other spike parameter, $\gamma_{sp} = (9\gamma-2)/(4-\gamma)$, controls the cuspiness of the spike profile. We assume that the initial inner halo profile is NFW, and therefore, we use $\gamma = 1$, and $\gamma_{\mathrm{sp}} = 7/3$\,\cite{Gondolo-Spike, Ferrera-TXS}. Recently, there have been some indirect evidences of DM spike\,\cite{Chan:2022gqd,Chan:2024yht}. We note that the survival of DM spike depends on the SMBH surroundings. Astrophysical effects like merger and kinetic heating by stars can destroy DM spike\,\cite{Ullio:2001fb,Merritt:2002vj,Bertone:2005hw}.

For typical WIMP DM scenario the current DM relic abundance is set by the DM self-annihilation\,\cite{Steigman:2012nb}. DM self-annihilation can make the density of the inner halo saturate at a value $\rho_{\mathrm{sat}} = m_{\chi}/(\langle\sigma v\rangle\, t_{\mathrm{BH}})$, where $\langle\sigma v\rangle$ is the velocity averaged self-annihilation cross section, and $t_{\mathrm{BH}}$ is the age of the black hole \cite{Gondolo-Spike, Ferrera-TXS}. The profile remains constant at $\rho_{\mathrm{sat}}$ approximately till a distance $r_{c}$, where $\rho_{\mathrm{sp}}(r_{c}) \sim \rho_{\mathrm{sat}}$. The DM spike density profile continues till $R_{\mathrm{sp}}$ beyond which it takes the form of the standard NFW profile. Thus, the full density profile inside a halo, with a self-annihilating DM can be written as\,\cite{Gondolo-Spike} 
\begin{equation}
\rho(r) =
\begin{cases}
0\,, & r \leq 4R_{s} \\
\frac{\rho_{\mathrm{sp}(r)}\,\rho_{\mathrm{sat}}}{\rho_{\mathrm{sp}(r)} +\rho_{\mathrm{sat}}}\,, & 4R_{s} < r \leq R_{\mathrm{sp}} \\
\rho_s \left( \dfrac{r}{r_s} \right)^{-\gamma} \left( 1 + \dfrac{r}{r_s} \right)^{-3 + \gamma}\,, & r > R_{\mathrm{sp}}\,\,.
\end{cases}
\label{eq:Spikeprofile}
\end{equation}

\begin{figure*}
	\begin{center}
		\includegraphics[width=\columnwidth]{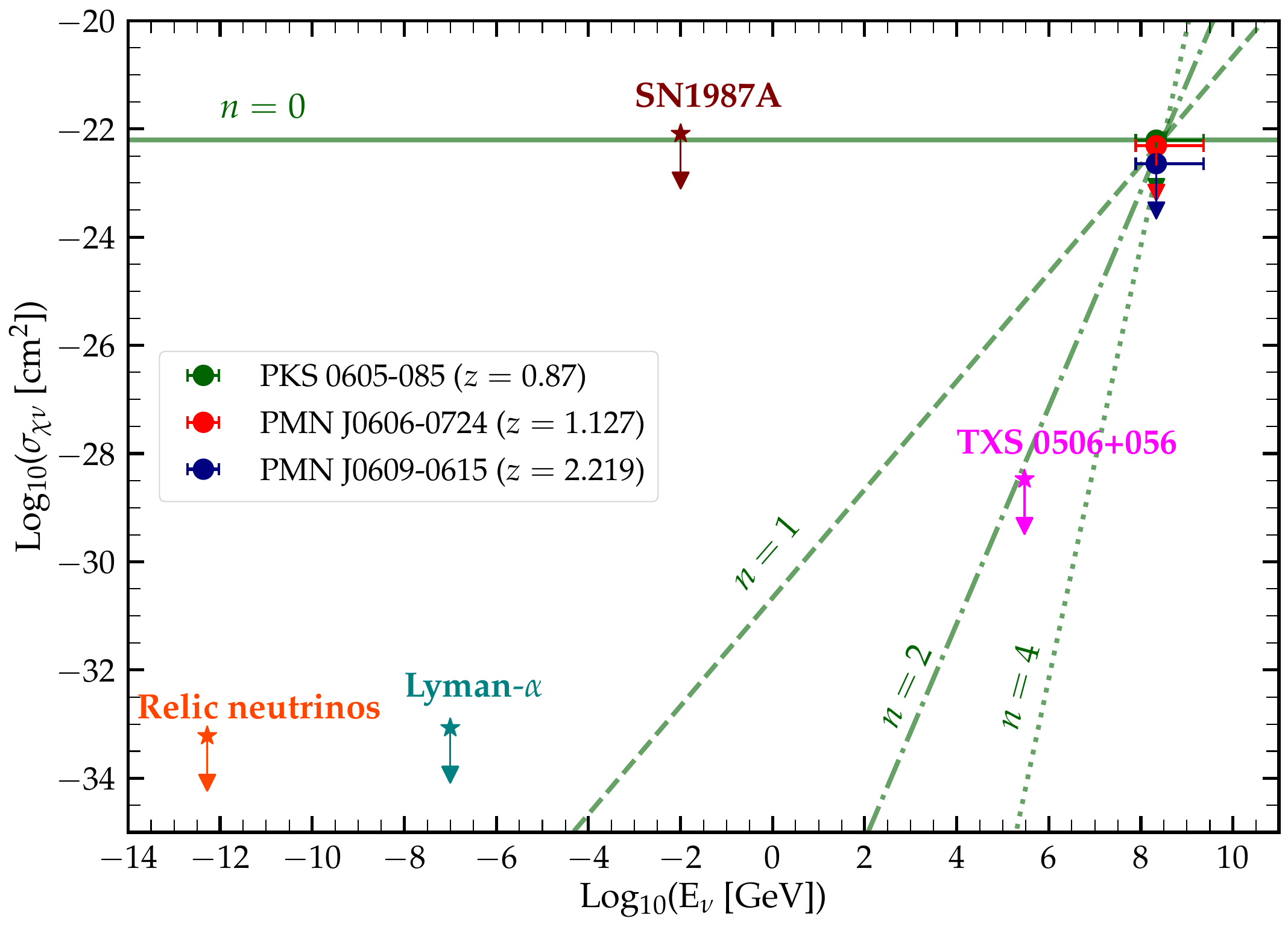}~~
		\includegraphics[width=\columnwidth]{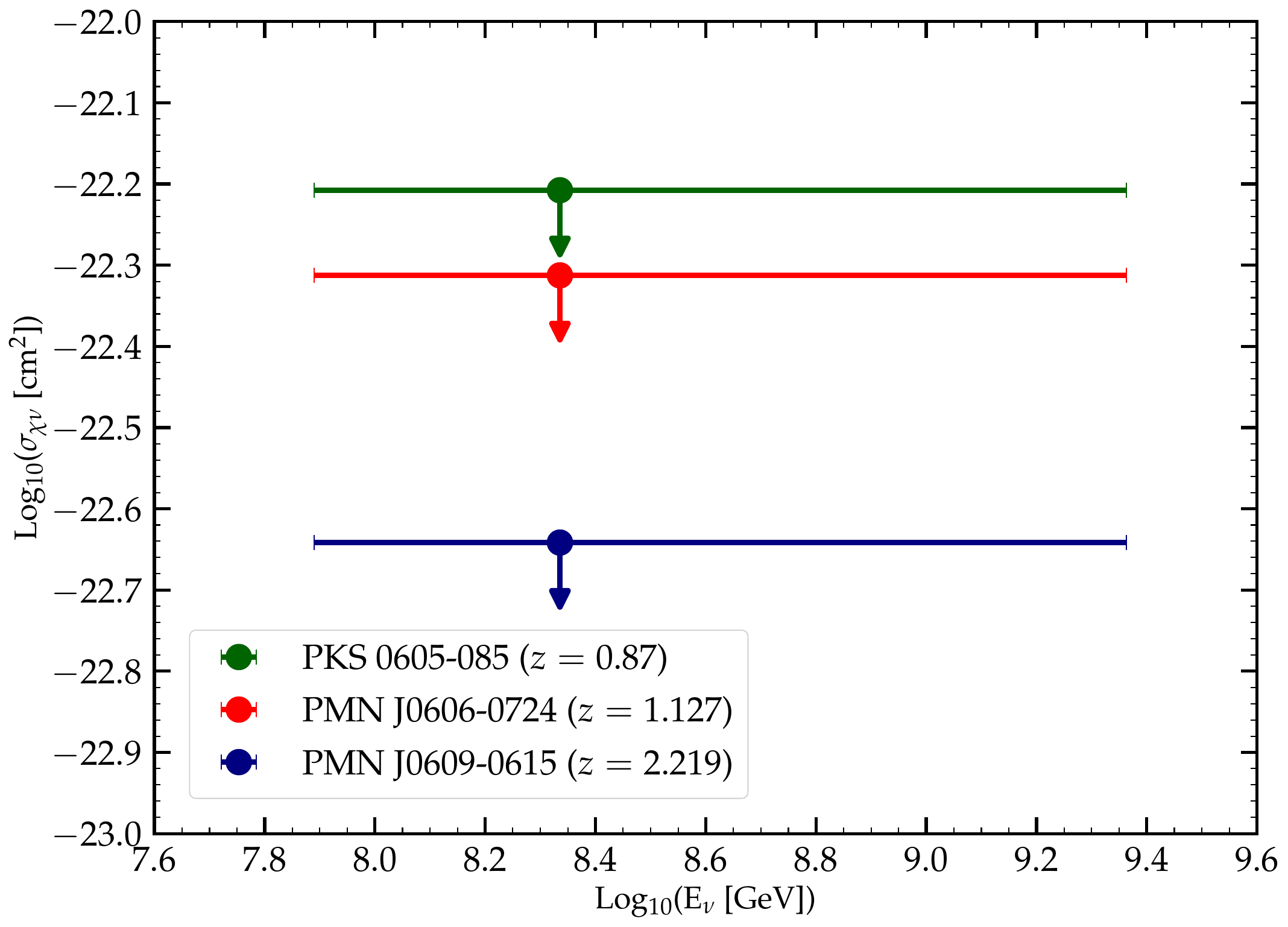}~~\\	
		\caption{(\textbf{Left panel}) Conservative constraints on energy-dependent DM-neutrino cross section, $\sigma_{\chi\nu}$, with contributions from the EG medium and MW halo. Solid circles denote bounds at KM3-230213A median energy $\sim 220$ PeV from blazars PKS 0605-085 (green circle), PMN J0606-0724 (red circle), and PMN J0609-0615 (blue circle). Dashed, dot-dashed, and dotted green lines show constraints for spectral indices $n= 0,1,2, 4$ for PKS 0605-085. For all the different spectral indices we fix the upper limits at the PKS 0605-085 median energy and extrapolate them to higher and lower neutrino energies. Here the DM mass is taken as, $m_\chi=1$ GeV. Previous limits include relic neutrinos\,\cite{Akita:2023yga} (orange star marker), Lyman-$\alpha$ forest\,\cite{Wilkinson:2014ksa} (teal star marker), SN1987A\,\cite{Mangano:2006mp} 
 (brown star marker), and TXS 0506+056\,\cite{Cline-TXS, Ferrera-TXS} (magenta star marker). (\textbf{Right panel}) Zoomed in version of the left panel showing the limits on DM-neutrino scattering obtained for three different candidate blazars. For visual clarity we do not show the different spectral index lines here.}
		\label{fig:E-dependent-Conservative}
	\end{center}	
\end{figure*}

DM particles reaching below a radius $4R_{s}$ get captured by the black hole and hence do not contribute to the energy density of the DM halo. We estimate the parameters $r_{s}$ and $\rho_{s}$ in the same way as described in the previous section by solving the equations in Eq.\,\eqref{eq:ProfileParam}. In several studies, the parameter $ \rho_{s} $ has been derived using the uncertainties in the black hole mass, often yielding stronger constraints\,\cite{Ferrera-TXS, Zapata:2025huq}. However, due to large uncertainties in both the blazar mass $ M_{\rm BH} $ and uncertainty $ \Delta M_{\rm BH} $, we adopt a more systematic approach, outlined in Ref.~\cite{Gondolo-Spike}. For simplicity, we do not take the spin of the black hole into account and consider spike formation in a Schwarschild geometry.

In Fig.~\ref{fig:Density-Profiles}, we show the host DM density profiles for the two blazars\,— PKS 0605-085 (left panel) and PMN J0609-0619 (right panel). In both panels, the solid blue lines represent the standard NFW profile in the absence of a DM spike, while the dashed pink and the dot-dashed teal lines correspond to density profiles with a DM spike, without and with DM self-annihilation, respectively. For DM self-annihilation the cross section is taken as, $\langle\sigma v\rangle = 10^{-26}\,\mathrm{cm}^3$/s. For both blazars, we assume that the neutrino is emitted from a region located between 0.01–1 pc from the blazar, highlighted by the green bands.

\subsection*{The Extragalactic Medium}
 After leaving the host halo, the neutrino propagates through the cosmological DM present in EG medium. The second term in Eqn.\,\eqref{eq:DiffCtbnLos} represents the EG contribution to $\tau$. The energy density for cosmological DM $\rho(z)$ evolves with the redshift as $\rho(z) = \rho_{0}(1+z)^{3}$, where $\rho_{0} = 1.3\times10^{-6}\,\mathrm{GeV/cm^{3}}$ \cite{ChoiIceCube}. The cosmic time $t$ varies with the redshift $z$ as $dt/dz = -1/(H(z)(1+z))$. With this, the EG component of $\tau$ can be rewritten as
\begin{equation}
    \int_{\mathrm{los}}n_{\chi}(z)\sigma_{\chi\nu}dl = \frac{\sigma_{\chi\nu}}{m_{\chi}}\int_{0}^{z_{b}}\frac{ \rho_{0}(1+z)^{2}dz}{H_{0}\sqrt{\Omega_{m}(1+z)^{3} + 1-\Omega_{m}}} \label{eq:ExtragalComponent},
\end{equation}
where $H_{0}$, $\Omega_{m}$, and $z_{b}$ are the Hubble constant, the matter density parameter, and the redshift of the blazar, respectively.\,\,In order to compute the integral in Eq.\,\eqref{eq:ExtragalComponent}, we take $H_{0} = 67.36\,\mathrm{km\, s^{-1}Mpc^{-1}}$ and $\Omega_{m} = 0.3153$\,\cite{Planck2018}. Despite the low energy density of cosmological DM, the EG contribution to the optical depth can become comparable to the MW contribution for high redshift blazars.

\begin{figure*}
	\begin{center}
		\includegraphics[width=\columnwidth]{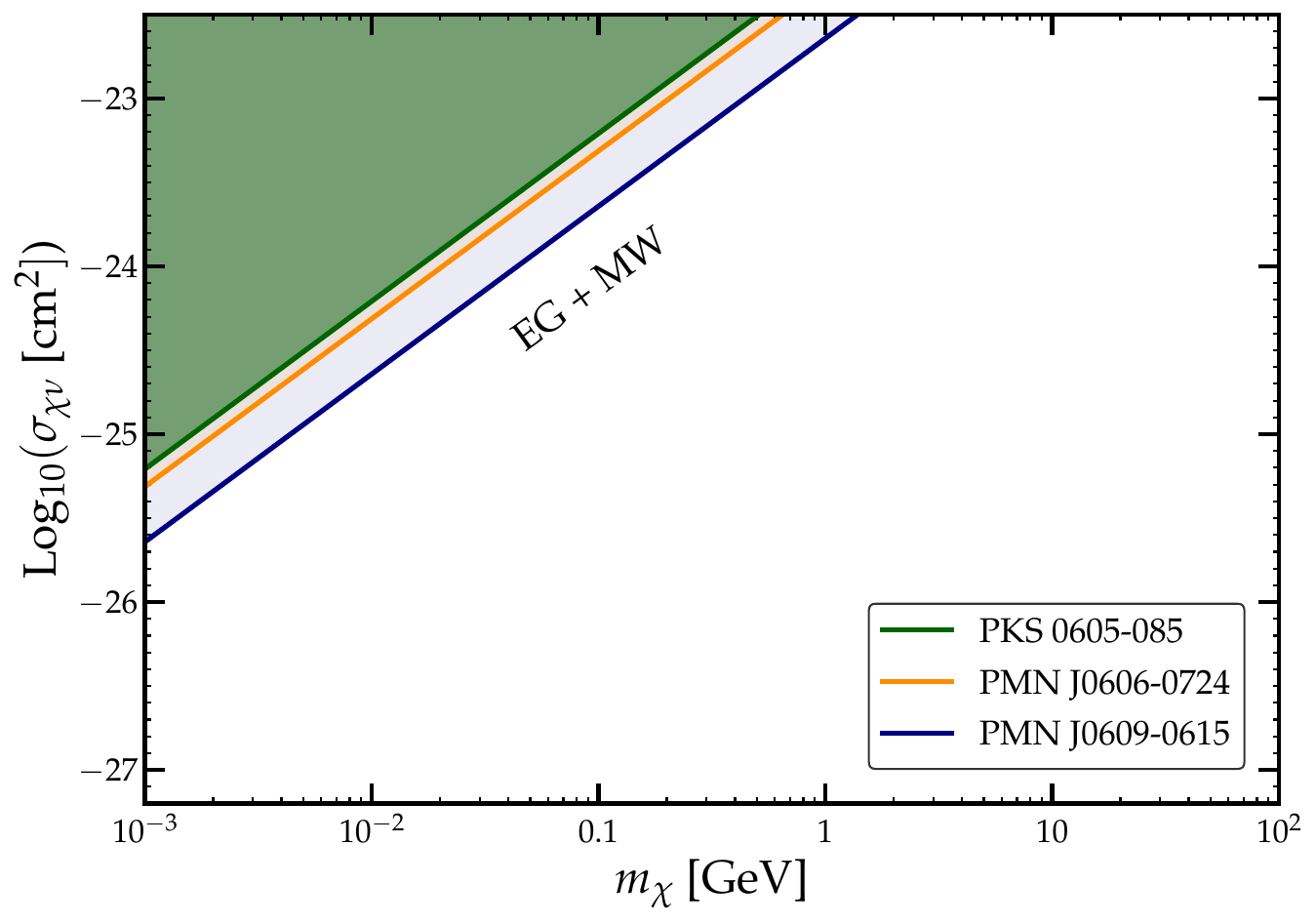}~~
		\includegraphics[width=\columnwidth]{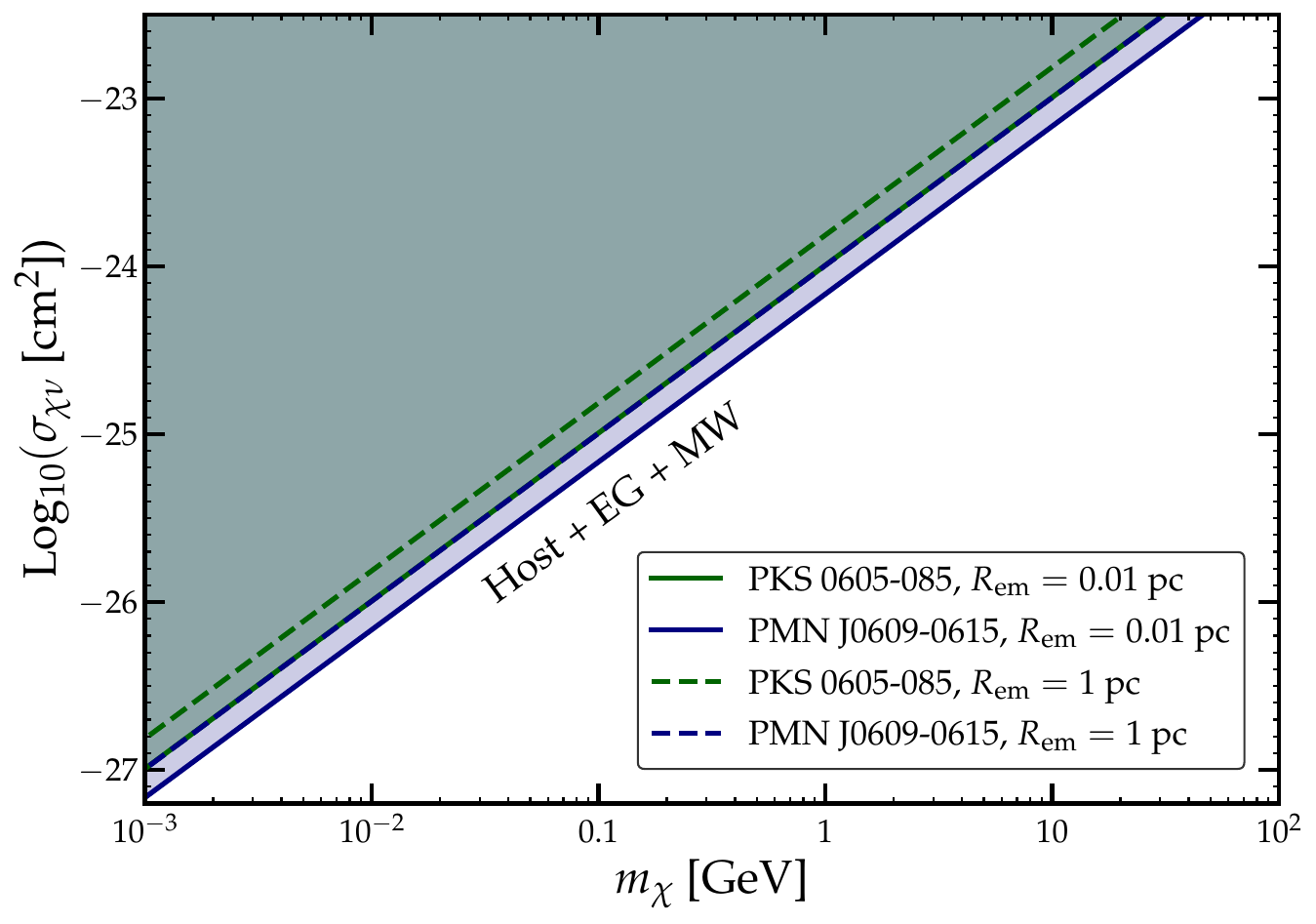}~~\\	
		\caption{(\textbf{Left panel}) Conservative constraints on the energy-independent DM-neutrino scattering cross section as a function of DM mass, assuming contribution from EG medium and MW DM halo for the three candidate blazars. The green, orange, and blue solid lines correspond to the constraints obtained for blazars PKS 0605-085, PMN J0606-0724, and PMN J0609-0615, respectively. ($\textbf{Right panel}$) Energy-independent constraints on DM-neutrino scattering cross section as a function of DM mass, considering host without DM spike, EG, and MW DM contribution. The green and blue solid lines show the constraints for blazars PKS 0605-085 and PMN J0609-0615, respectively, assuming neutrinos are emitted at $R_{\mathrm{em}} = 0.01\,\mathrm{pc}$
        near the blazar. The dashed lines show the corresponding constraints when the neutrino emission region is assumed to be $R_{\mathrm{em}} = 1\,\mathrm{pc}$. }
        \label{fig:E-Independent-Conservative}
	\end{center}	
\end{figure*}

\begin{figure*}
	\begin{center}
		\includegraphics[width=\columnwidth]{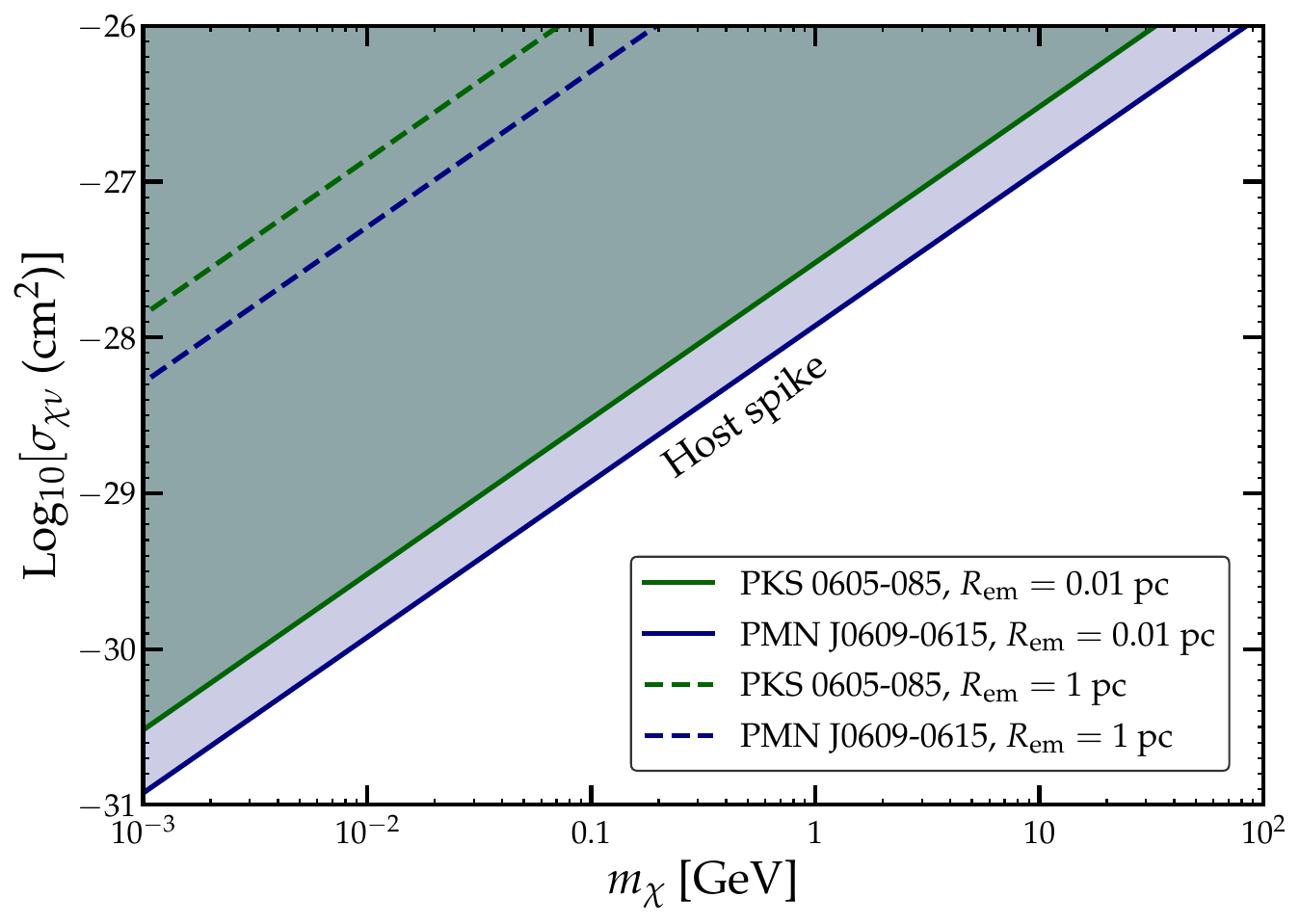}~~
		\includegraphics[width=\columnwidth]{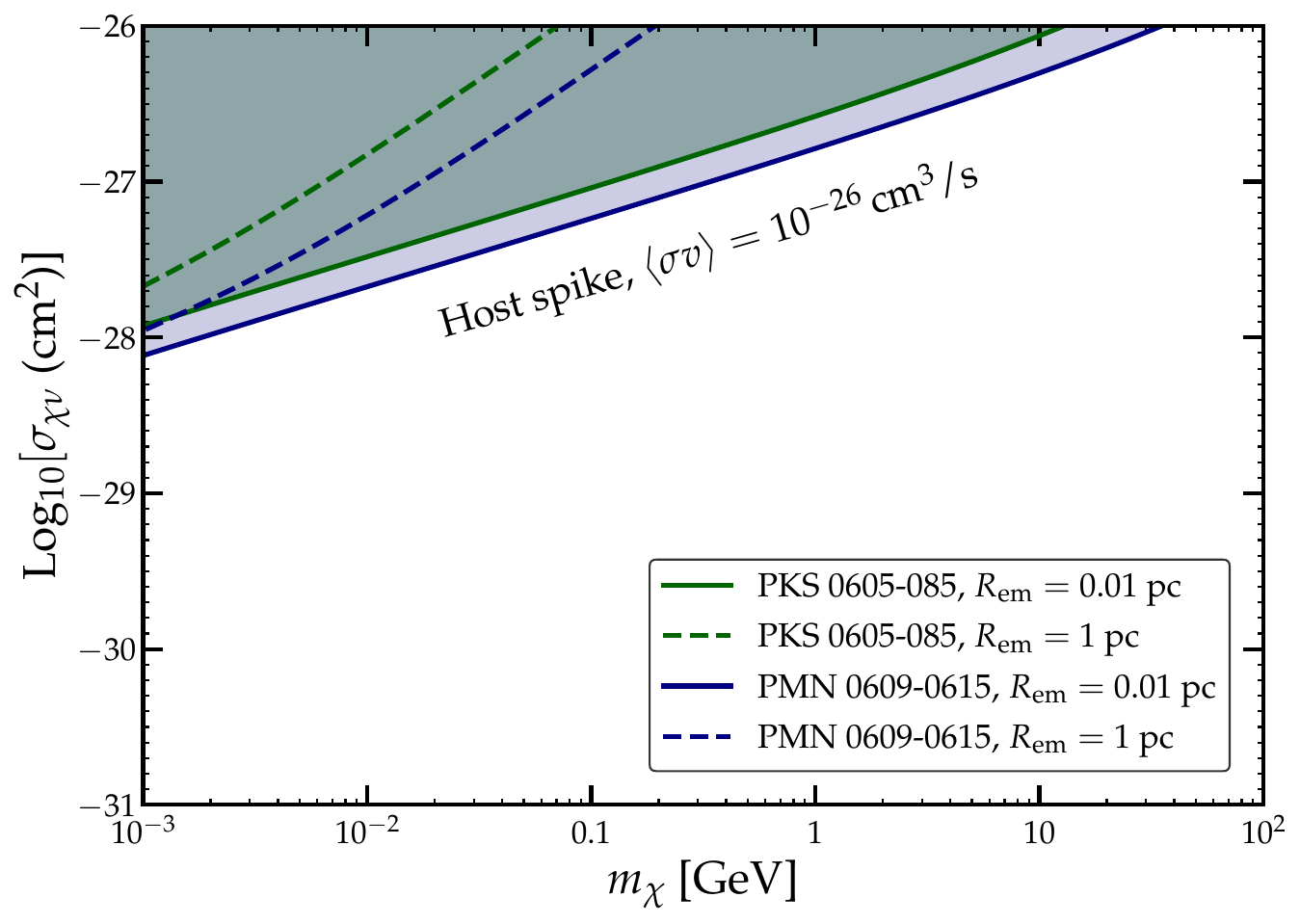}~~\\	
		\caption{(\textbf{Left panel})   Constraints on energy-independent DM-neutrino cross section assuming the presence of a DM spike in the host halo along with EG and MW halo contributions, for different emission radii $R_{\mathrm{em}}$. The green and blue lines represent the constraints derived from PKS 0605-085 and PMN J0609-0615, respectively. The solid and dashed lines represent the constraints assuming $R_{\mathrm{em}}=$ 0.01 pc, and 1 pc, respectively. (\textbf{Right panel}) Constraints on the energy-independent DM-neutrino scattering cross section, $\sigma_{\chi\nu}$, assuming self-annihilating DM with annihilation cross section, $\langle\sigma v\rangle = 10^{-26}\,\mathrm{cm}^{3}\,\mathrm{s}^{-1}$. Here host spike, EG medium, and MW halo DM contributions are taken into account. The solid blue and green lines represent the constraints derived from the blazars PKS 0605-085 and PMN 0609-0615, respectively, assuming an emission radius of $R_{\mathrm{em}} = 0.01\,\mathrm{pc}$. The corresponding dashed lines indicate the constraints at $R_{\mathrm{em}} = 1\,\mathrm{pc}$.  }
		\label{fig:E-Independent-Host-Spike}
	\end{center}	
\end{figure*}

\subsection*{The MW DM halo} 
 The third term in Eqn.\,\eqref{eq:DiffCtbnLos} represents the contribution to  $\tau$ from the MW DM halo. One of the most widely used parametric form for the DM density profile is \cite{Bringmann:2012ez, Ng:2013xha}
\begin{equation}
    \rho^{\alpha\beta\gamma}(r) = \rho_{\odot}\left(\frac{r}{r_{\odot}}\right)^{-\gamma}\left(\frac{1+\left(r_{\odot}/r_s\right)^{\alpha}}{1+\left(r/r_s\right)^{\alpha}}\right)^{(\beta-\gamma)/\alpha} \label{eq:MWParametric},
\end{equation}
where $\rho_{\odot} = 0.4$ GeV/cm$^3$, $r_{\odot} = 8$ kpc, and $r_s = 20$ kpc. Similar to the host halo, for MW DM we use the NFW profile, for which $\alpha = 1$, $\beta = 3$, $\gamma = 1$ in the above equation. 
The DM density profile in Eq.~\eqref{eq:MWParametric} is expressed as a function of the radial distance from the Galactic Center, which differs from our $l.o.s$ distance since we are located at an off-centered position within the galaxy. The radial distance is expressed in terms of the $l.o.s$ distance $s$, and galactic coordinates $(l,b)$ as
\begin{equation}
    r(s, b, \ell) = \sqrt{s^2 + r_{\odot}^2 - 2 s r_{\odot} \cos b \cos \ell}.
\end{equation}
With this, we calculate the contribution from the MW halo with NFW parameters:
\begin{equation}
    \int_{\mathrm{los}}\rho_{\mathrm{gal}}(r)ds = \int_{0}^{s_{\mathrm{max}}}\rho^{131}(r(s,b,\ell))ds,
\end{equation}
where $s_{\mathrm{max}}$ is the maximum $l.o.s$ distance obtained from the virial size of the MW halo, $\rho^{131}(r)$ is the density profile defined in Eq.\,(\ref{eq:MWParametric}), for $\alpha = 1$, $\beta = 3$, $\gamma = 1$ (NFW profile). In our work, we use $R_{\mathrm{vir} }^{\mathrm{MW}}= 200$ kpc \cite{Ng:2013xha}. We note that the value of $\Sigma_{\mathrm{MW}}$ depends on the directionality of $l.o.s$, and it increases if $l.o.s$ passes through or near the Galactic Center.

\section{Limits on DM-neutrino scattering}
\label{sec:scattering}
We assume that the neutrino flux originating from the blazar is at most attenuated by 90\% due to its interaction with the DM. From Eq.~\eqref{eq:OpticalDepth}, this assumption on the observed neutrino flux translates to an upper bound on the optical depth, $\tau \lesssim 2.3$. This sets a limit on $\sigma_{\chi\nu}/m_{\chi}$ as
\begin{equation}
    \frac{\sigma_{\chi\nu}}{m_{\chi}} \leq 2.3\, \Big(\int_{\mathrm{los}}\rho dl\Big)^{-1}\,\,.
    \label{eq:E-Independent}
\end{equation}
Energy-independent constraints on DM-neutrino cross sections can be derived for a wide range of DM masses using the above equation. However, more realistic models of DM-neutrino interaction suggest an energy dependence in the scattering cross section, e.g., DM-neutrino scattering mediated by $Z'$ mediator (as discussed below). Following realistic particle physics models, the cross section $\sigma_{\chi\nu}$ can be parameterized as \cite{Cline-TXS, ChoiIceCube, Ferrera-TXS}
\begin{eqnarray}
        \sigma_{\chi\nu}(E_{\nu}) = \sigma_{0}\Bigg(\frac{E_{\nu}}{1\,\mathrm{GeV}}\Bigg)^{n},\label{eq:Edependent}
\end{eqnarray}
In the above equation, the normalization parameter $\sigma_{0}$ is fixed such that the DM-neutrino cross section matches the energy-independent limits at the median energy of KM3-230213A, 
\begin{equation}
     \sigma(E_{\nu}) = 2.3\Big(\int_{\mathrm{los}}\rho dl\Big)^{-1}\,\, \mathrm{at}\,\, E_{\nu} = 220\,\mathrm{PeV}\,\,. \label{eq:E-Dependent-Normalization}
\end{equation}
The value of parameter $n$ depends on the particle physics model. For our model-independent limits, we evaluate the constraints for four different values of $n$; $n = 0, 1, 2 ,4$.

\subsection*{Particle Physics Model} Several particle physics models have been proposed for DM–neutrino interactions. One of the most widely studied scenarios involves interactions mediated by a new gauge boson $Z'$, which couples to a Dirac fermion DM particle with interaction strength $g_{\chi}$, and to neutrinos via a flavor-universal coupling strength $g_{\nu}$\,\cite{Fox:2008kb,Arguelles:2017atb,Bell:2014tta}. In addition to this, the literature also discusses alternative interaction models involving different types of DM particles and mediators\,\cite{Boehm:2003hm,Boehm:2006mi, Fox:2008kb,Boehm:2013jpa, Arhrib:2015dez, Belyaev:2022shr, Herms:2023cyy}.

For a Dirac fermion DM interacting with neutrino via $Z'$ boson\,\cite{Arguelles:2017atb},\\

\begin{widetext}
\begin{align}\label{eq:DiracDM}
\sigma_{\nu\chi} = \frac{g_\nu^2 g_\chi^2}{16 \pi E_\nu^2 m_\chi^2 m_{Z'}^2} \bigg[ 
\left( m_{Z'}^2 + m_\chi^2 + 2E_\nu m_\chi \right) 
\log\left(\frac{m_{Z'}^2 (2E_\nu + m_\chi)}{m_\chi(4E_\nu^2 + m_{Z'}^2) + 2E_\nu m_{Z'}^2}\right)
\nonumber
    \\
+ 4E_\nu^2 \bigg( 1 + \frac{m_\chi^2}{m_{Z'}^2} - 
\frac{2E_\nu (4E_\nu^2 m_\chi + E_\nu (m_\chi^2 + 2m_{Z'}^2) + m_\chi m_{Z'}^2)}
{(2E_\nu + m_\chi)(m_\chi(4E_\nu^2 + m_{Z'}^2) + 2E_\nu m_{Z'}^2)} \bigg) 
\bigg]\,\,,
\end{align}
\end{widetext}
where $g_\nu$, $g_\chi$ are the couplings of the $Z'$ mediator to the neutrino and DM, respectively, $m_{Z'}$ is the mass of the mediator, and $E_\nu$ = 220 PeV is the neutrino energy. With this cross section for the model, we can use Eq.\,(\ref{eq:E-Independent}) to put limits on the coupling $g_\nu g_\chi$ for a given $m_\chi$, $m_{Z'}$. We note that for EG and Host DM contributions one has to take into account the redshifting of neutrino energy in Eq.\,(\ref{eq:DiracDM}).

\section{Results}
\label{sec:Result}

In this section, we present both energy-dependent and energy-independent constraints on DM-neutrino cross section based on the high-energy neutrino event detected by KM3NeT.  In presence of DM-neutrino interactions, high-energy neutrinos get attenuated as it propagates through DM distributed across different astrophysical scales, as described in Sec.~\ref{sec:Distribution}. We derive our limits under the assumption that the neutrino originates from one of the three candidate blazars associated with the KM3-230213A event, listed in Table~\ref{tab:Source-Coordinates}. We derive conservative constraints on $\sigma_{\chi\nu}$ by accounting for flux attenuation due to neutrino propagation through cosmological DM in the EG medium and the MW DM halo. Additionally, we derive more optimistic constraints by including the attenuation effects from the DM halo of the host blazar. We examine both scenarios in which flux suppression arises from the DM halo of the host blazar—with and without a DM spike. For the DM spike case, we also derive constraints assuming that DM particles self-annihilate within the spike with a thermal velocity-averaged cross section, $\langle\sigma v\rangle = 10^{-26}\,\mathrm{cm}^3$/s. Finally, we explore a particle physics model and derive limits on the model parameters based on the event KM3-230213A. \par

In Fig.~\ref{fig:E-dependent-Conservative}, we present the conservative energy-dependent upper bounds on $\sigma_{\chi\nu}$, derived by considering flux attenuation only from the EG DM and the MW DM halo. The inset (right panel) highlights the constraints on $\sigma_{\chi\nu}$ at the median energy of the KM3-230213A event, $E_{\mathrm{KM3NeT}} \sim 220\,\mathrm{PeV}$, for the three blazar candidates. The green solid, dashed, dot-dashed, and dotted lines correspond to different energy-dependent models $\sigma_{\chi\nu}(E_\nu) \propto E_\nu^n$, for $n = 0, 1, 2,$ and $4$, respectively, shown for the blazar PKS 0605-085. All the spectral index curves are fixed such that the upper limit for each of them at the median KMKM3-230213A energy matches with the energy-independent DM-neutrino cross section. The same can be shown for blazars PMN J0606-0724 and PMN J0609-0615. The previous upper limits on DM-neutrino cross section are obtained from relic neutrinos\,\cite{Akita:2023yga} (orange star marker), Lyman-$\alpha$ forest\,\cite{Wilkinson:2014ksa} (teal star marker), SN1987A\,\cite{Mangano:2006mp} 
 (brown star marker), and TXS 0506+056\,\cite{Cline-TXS, Ferrera-TXS} (magenta star marker).  Given all three blazars lie within a small angular region,
the neutrinos originating from these sources traverse approximately the same path length through the MW DM
halo. Thus, in this conservative case, the upper limits primarily depend on the blazar redshift. With increasing
redshift, the neutrinos have higher probability of getting
attenuated by scattering with cosmological DM in the
EG medium, even though the EG DM density is low. 
\begin{figure*}[!htbp]
	\begin{center}
		\includegraphics[width=\columnwidth]{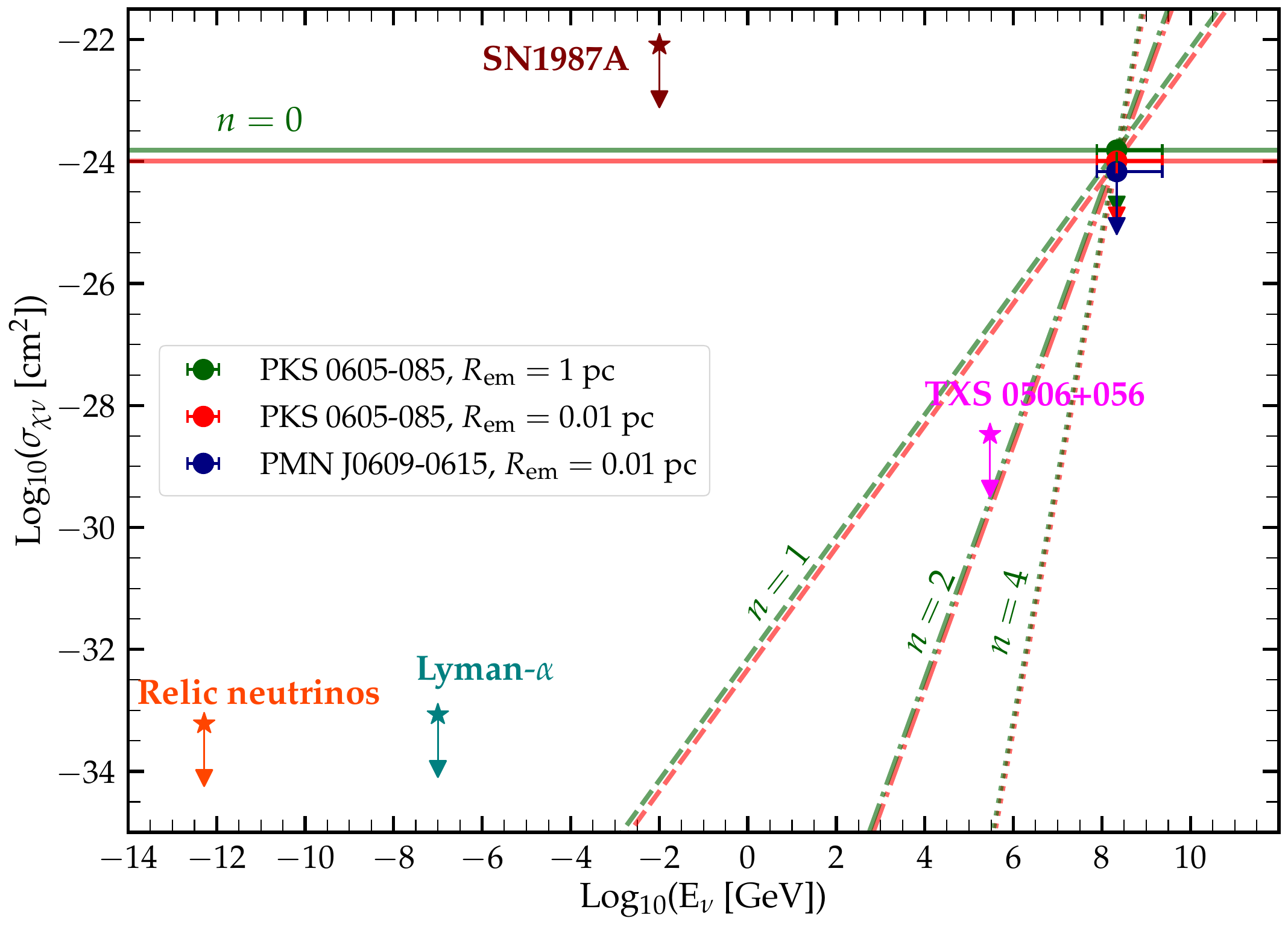}~~
		\includegraphics[width=\columnwidth]{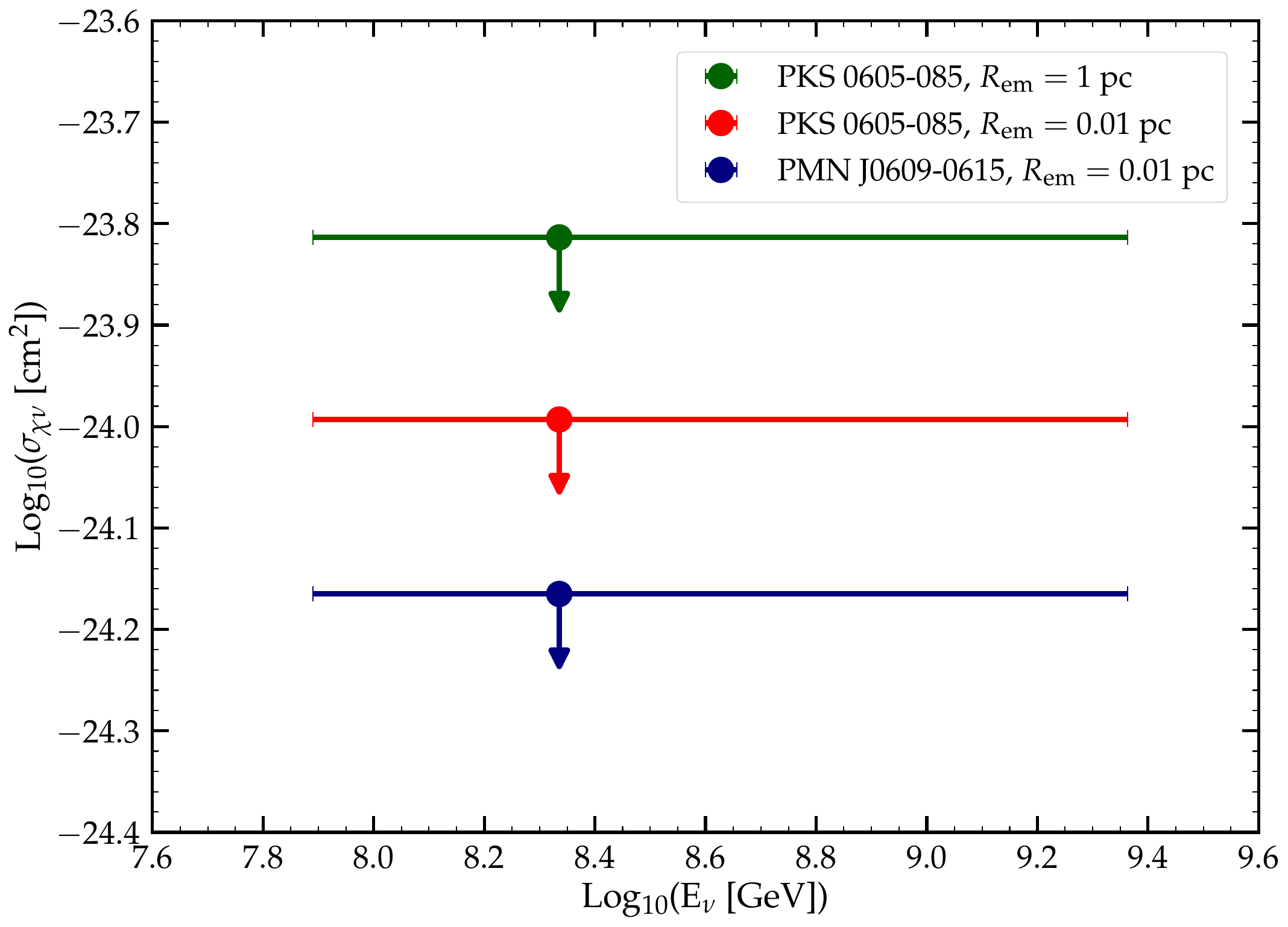}~~\\	
		\caption{(\textbf{Left panel}) Constraints on energy-dependent DM-neutrino cross-section are shown, accounting for contributions from the host halo, extragalactic (EG) medium, and MW DM for the blazars PKS 0605-085 and PMN J0609-0615. Here the DM mass is taken as $m_\chi=1$ GeV. The green and red solid circles represent the constraints on for blazars PKS 0605-085, assuming neutrino emission at $R_{\mathrm{em}} = 1\,\mathrm{pc}$ and $0.01\,\mathrm{pc}$, respectively, at the median energy of the event KM3-230213A. The corresponding solid, dashed, dot-dashed, and dotted lines (for the same color scheme) indicate the constraints for spectral indices $n = 0$, 1, 2, and 4, respectively. The blue solid circle represent the constraint for PMN J0609-0615, assuming $R_{\mathrm{em}} = $ 0.01 pc. The previous limits are same as Fig.\,\ref{fig:E-dependent-Conservative}. (\textbf{Right panel}) Zoomed in version of the left panel. For visual clarity we do not show the different spectral index lines here.  }
		\label{fig:Host-E-Dependent}
	\end{center}	
\end{figure*}

Our estimation of the conservative constraint for energy-independent case is shown in  Fig.~\ref{fig:E-Independent-Conservative} (left panel). Here the strongest upper limit on $\sigma_{\chi\nu}$ is obtained for the blazar PMN J0609-0615 shown in solid blue line, which is at the highest redshift ($z = 2.219$) among the three. In contrast, the weakest limit is obtained for PKS 0605-085, represented by the solid green line, consistent with its lowest redshift ($z = 0.87$) among the three. The limit for the blazar PMN J0606-0724 ($z = 1.277$), shown by the solid orange line, falls between these two limits. Similar to Fig.~\ref{fig:E-dependent-Conservative}, here the constraints are primarily sensitive to the redshift of the blazars. As shown in Table~\ref{tab:Upper-Bound-EIndependent-Conservative}, the conservative upper bounds on the DM-neutrino interaction cross section reflect the redshift dependence as discussed above. 
\begin{table}[h]
\caption{\label{tab:Upper-Bound-EIndependent-Conservative}%
We list the $l.o.s$ integral contributions to the optical depth from EG medium and MW halo for each blazar source, expressed in units of $\mathrm{cm^{2}/GeV}$. }
\begin{ruledtabular}
\begin{tabular}{lc}
\textrm{Source Name} & $\sigma_{\chi\nu}/m_{\chi}\,\mathrm{(\,cm^{2}/GeV)}$\\
\colrule
PKS 0605-085 &  $6.20\times 10^{-23}$    \\
PMN J0606-0724 & $4.87\times 10^{-23}$  \\
PMN J0609-0615 & $2.28\times 10^{-23}$   \\
\end{tabular}
\end{ruledtabular}
\end{table}

 In Fig.~\ref{fig:E-Independent-Conservative} (right panel) we show the constraints on $\sigma_{\chi\nu}$ in the case where flux attenuation is caused by neutrino interactions with the host DM halo as well as the EG medium and MW halo. For the host halo we assume the NFW DM density profile. The solid (dashed) green and blue lines show upper bounds for the two blazars PKS 0605-085 and PMN J0609-0615, when the neutrino is emitted at $R_{\mathrm{em}} = 0.01$ pc (1 pc), respectively, from the blazar. As discussed earlier, since the values of $M_{\mathrm{BH}}$ are available only for the blazars PKS 0605-085 and PMN J0609-0615, here we derive the upper limits on $\sigma_{\chi\nu}$ from flux suppression only for these two sources. \par

The limits presented in Fig.~\ref{fig:E-Independent-Conservative} (right panel) are stronger than those obtained in Fig.~\ref{fig:E-Independent-Conservative} (left panel), as for the former case, in addition to the MW halo and EG medium, we also take into account the host halo DM contribution. Thus, the host halo DM has a significant contribution to the total column density. For example, with host halo contribution, for $m_{\chi} = 1\,\mathrm{GeV}$, the column density reaches $\Sigma \sim 10^{24}\,\mathrm{GeV/cm^2}$, which is nearly two orders of magnitude higher than the conservative case where $\Sigma \sim 10^{22}\,\mathrm{GeV/cm^2}$.

In Fig.~\ref{fig:E-Independent-Host-Spike} we show the constraints on $\sigma_{\chi\nu}$ arising from neutrino flux suppression in the presence of a DM spike, without (left panel) and with (right panel) DM self-annihilation. Here the MW and EG DM contributions are also taken into account. The color coding and linestyles used in these figures follow the same convention as in Fig.~\ref{fig:E-Independent-Conservative}. 
For Fig.~\ref{fig:E-Independent-Host-Spike} (right panel), we adopt a thermal self-annihilation cross section of $\langle\sigma v\rangle  = 10^{-26}\,\mathrm{cm}^{3}/\rm s$\,\cite{Steigman:2012nb}.
\par

In the presence of a DM spike, the column density increases by several orders of magnitude due to the sharply enhanced DM density—particularly when the emission radius $R_{\mathrm{em}} \ll R_{\mathrm{sp}}$. When DM self-annihilation becomes significant, it flattens the central density to a saturation value $\rho_{\mathrm{sat}}$ within the region $r < r_{c}$. As a result, for a neutrino produced within the range $4R_{s} \leq R_{\mathrm{em}} \leq r_{c}$, the flux attenuation in presence of self-annihilating DM is reduced compared to non-annihilating DM, leading to a weaker upper bound on $\sigma_{\chi\nu}$, as can be seen when comparing the two panels of Fig.~\ref{fig:E-Independent-Host-Spike}. For example, the critical radius ($r_c$) for PKS 0605-085 is $\sim 0.1$ pc for $m_\chi$= 1 GeV and $\langle\sigma v\rangle = 10^{-26}\,\mathrm{cm}^3$/s. In this case $R_{\rm em}=0.01$ pc implies $R_{\rm em} < r_c$. As a result, for PKS 0605-085, the constraint on DM-neutrino scattering (green solid line) is almost an order of magnitude stronger when DM annihilation is absent. For $R_{\rm em}=1$ pc, we have $R_{\rm em} > r_c$. Thus, the constraint on DM-neutrino scattering (green dashed line) remains the same for both with and without self-annihilation. For each panel, the limits for $R_{\rm em} = 0.01$ pc is stronger as a result of the larger path length traversed by the neutrino as compared to the $R_{\rm em} = 1$ pc case. For $m_\chi \gtrsim 2\times10^{-3}$ GeV, the critical radius is smaller than 1 pc. Thus, the limits for $R_{\rm em} = 1$ pc become significantly weaker than $R_{\rm em} = 0.01$ pc. The above discussion is valid for PMN 0609-0615 also. The non-linear nature of the limits in Fig.~\ref{fig:E-Independent-Host-Spike} (right panel) arises due to DM mass dependence in the critical radius.

\begin{figure*}
	\begin{center}
		\includegraphics[width=\columnwidth]{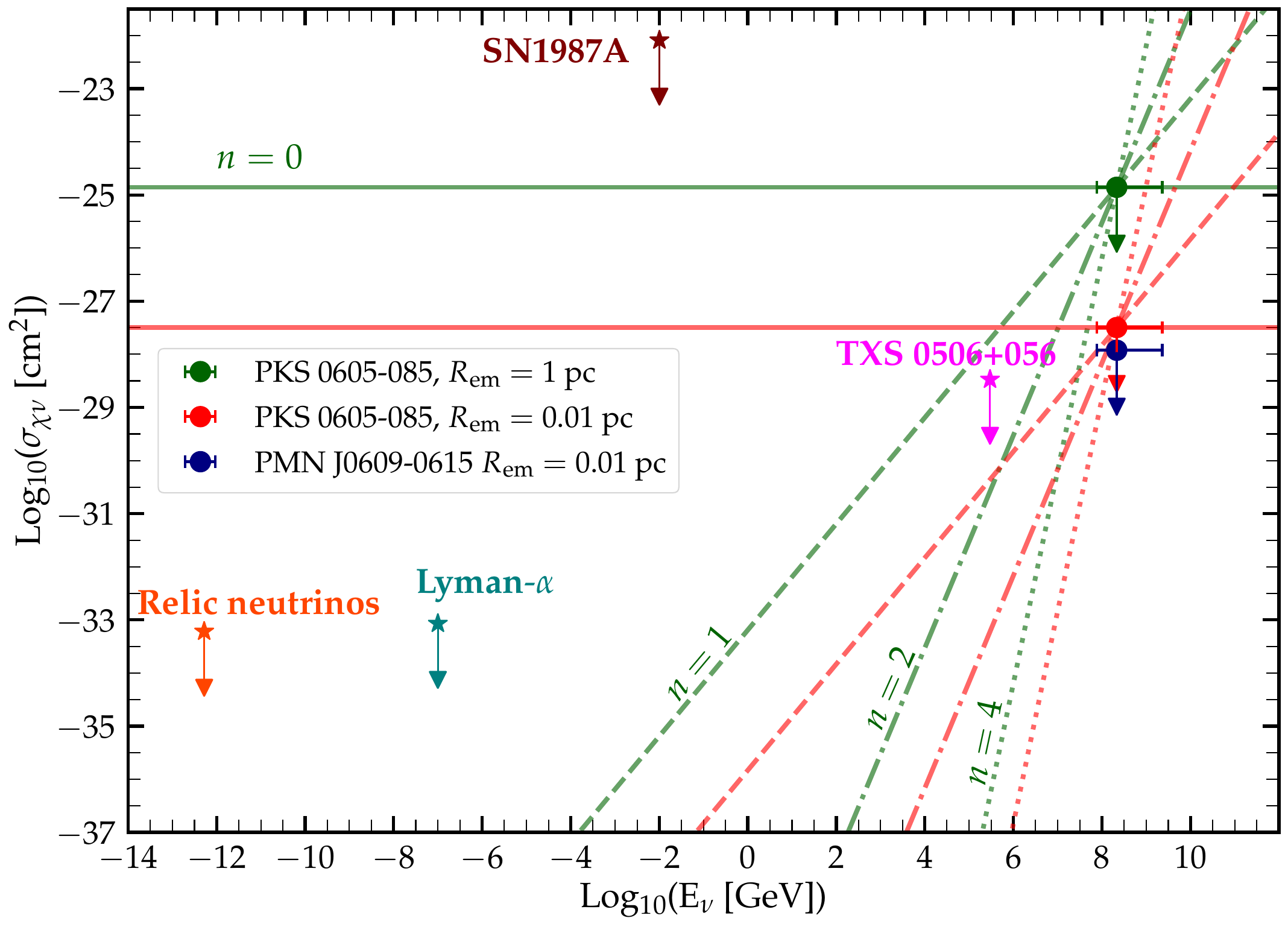}~~
		\includegraphics[width=\columnwidth]{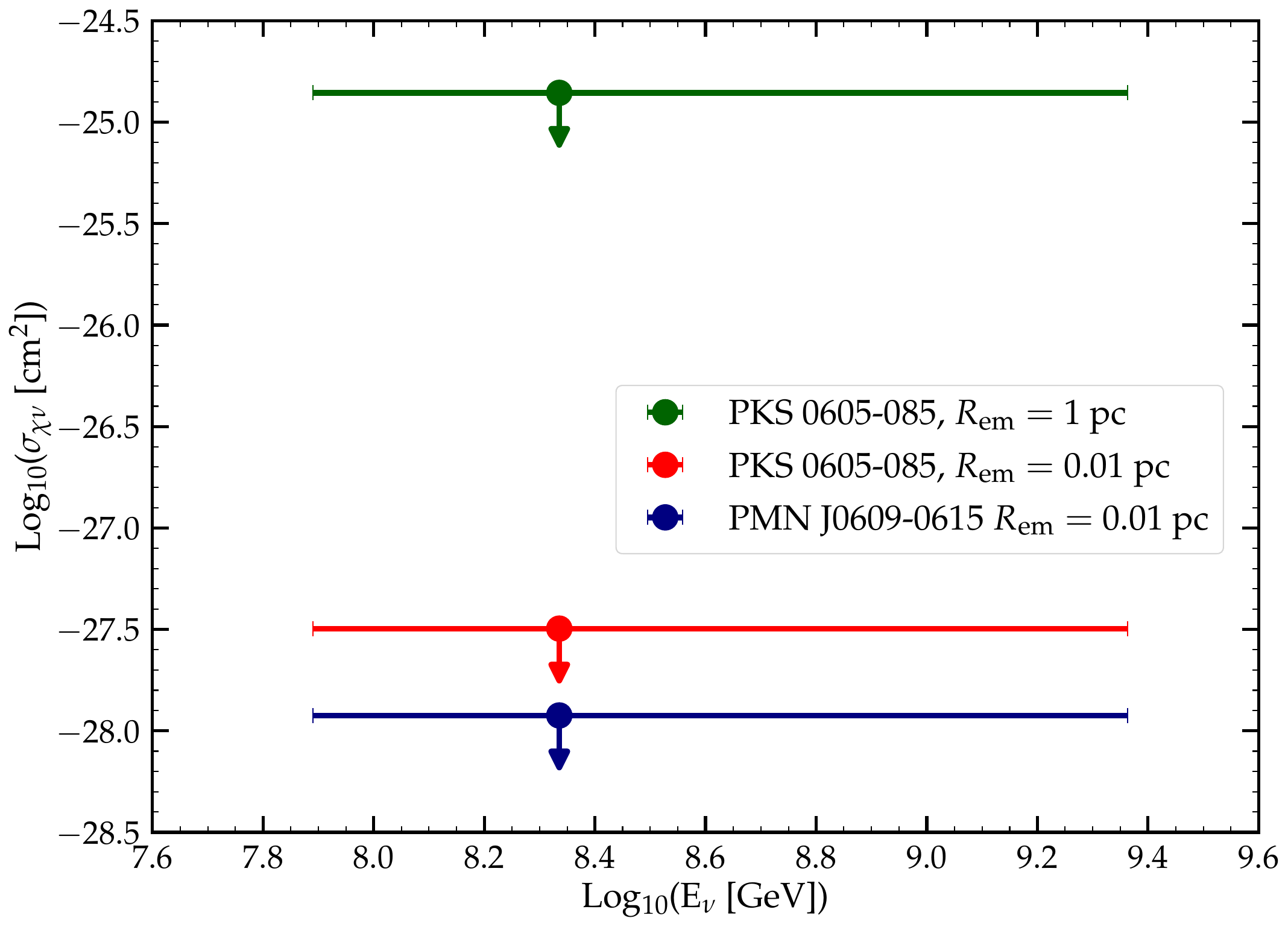}~~\\	
		\caption{(\textbf{Left panel}) Constraints on energy-dependent DM-neutrino cross-section, taking into account contributions from the host halo spike, extragalactic (EG) medium, and MW DM for the blazars PKS 0605-085 and PMN J0609-0615. Here the DM mass is taken as, $m_\chi=1$ GeV. The color scheme and  previous limits are same as Fig.\,\ref{fig:E-dependent-Conservative}. (\textbf{Right panel}) Zoomed in version of the left panel. For visual clarity we do not show the different spectral index lines here. }
		\label{fig:Host-Spike-E-Dependent}
	\end{center}	
\end{figure*}

\begin{figure*}[!htbp]
	\begin{center}
		\includegraphics[width=\columnwidth]{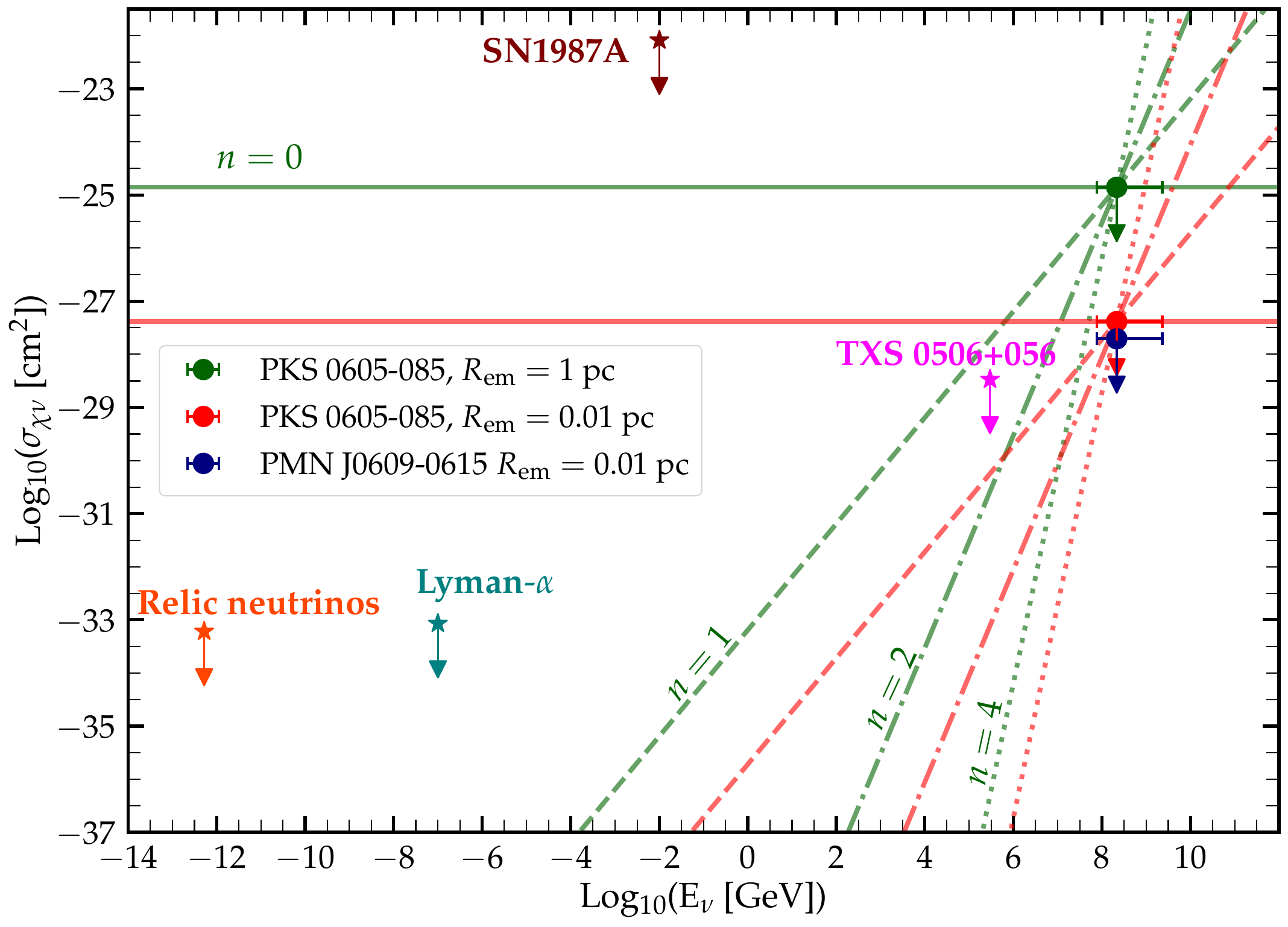}~~
		\includegraphics[width=\columnwidth]{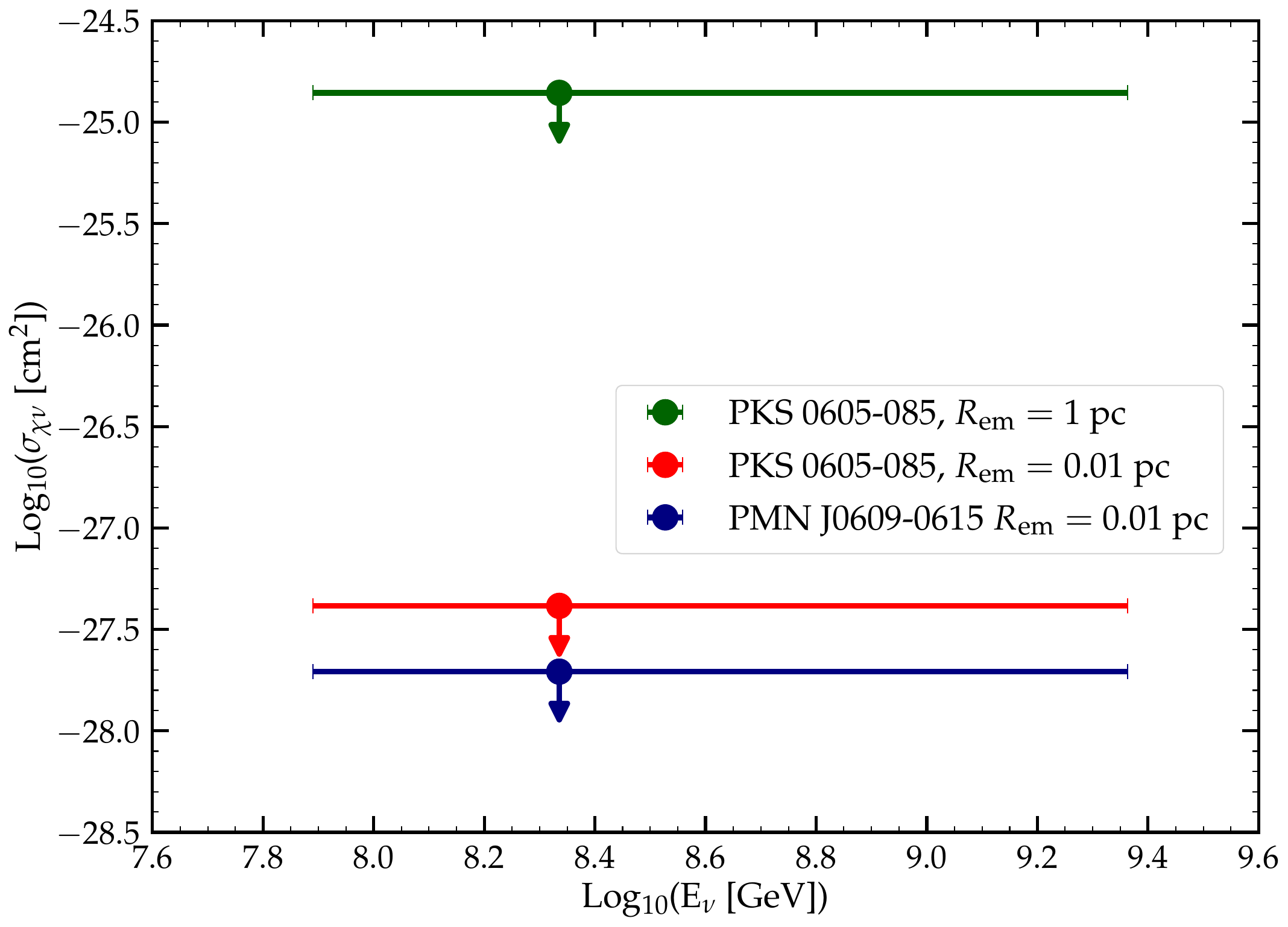}~~\\	
		\caption{(\textbf{Left panel}) Constraints on energy-dependent DM-neutrino cross-section, taking into account contributions from the host halo DM spike, extragalactic (EG) medium, and MW DM for the blazars PKS 0605-085 and PMN J0609-0615. Here the DM mass is taken as, $m_\chi=1$ GeV and DM is assumed to be have thermal self-annihilating cross section, $\langle\sigma v\rangle = 10^{-26}\,\mathrm{cm}^{3}\,\mathrm{s}^{-1}$. The color scheme and  previous limits are same as Fig.\,\ref{fig:E-dependent-Conservative}. (\textbf{Right panel}) Zoomed in version of the left panel. For visual clarity we do not show the different spectral index lines here. 
    }
		\label{fig:Host-SpikeAnn-E-Dependent}
	\end{center}	
\end{figure*}

In \,\,Figs.~\ref{fig:Host-E-Dependent},\,\ref{fig:Host-Spike-E-Dependent}, and \ref{fig:Host-SpikeAnn-E-Dependent} we show the energy-dependent constraints for neutrino flux suppression through \textit{i)} a host DM halo without a spike, \textit{ii)} a host DM halo with a central spike and no DM self-annihilation ($\langle\sigma v\rangle  = 0$), and \textit{iii)} a host DM halo with a spike and self-annihilation cross section $\langle\sigma v\rangle  =  10^{-26}\,\mathrm{cm}^{3}/$s, respectively, for the two blazar sources. In these plots we fix the DM mass, $m_\chi = 1$ GeV. We show the constraints for different spectral indices $n$, as defined in Eq.~(\ref{eq:Edependent}). In each of the three figures, the inset box (right panel) highlights the upper bounds on $\sigma_{\chi\nu}$ obtained at the median energy of the KM3-230213A event, showing the variation across different blazars and the impact of the neutrino emission radius $R_{\mathrm{em}}$. The green and red solid circles indicate the upper bounds on the cross section for neutrino emission at $R_{\mathrm{em}} = 1\,\mathrm{pc}$ and $0.01\,\mathrm{pc}$, respectively, for the blazar PKS 0605-085. The associated green and red lines illustrate the upper bounds for the same cases across different spectral indices: $n = 0$ (solid line), $n = 1$ (dashed line), $n = 2$ (dot-dashed line), and $n = 4$ (dotted line). Similarly, the blue solid circle shows the corresponding bound for the blazar PMN J0609-0615, assuming $R_{\mathrm{em}} = 0.01\,\mathrm{pc}$. The energy dependence of the cross-section exhibits similar features for both the blazars. We do not show the different spectral index curves for PMN J0609-0615 to avoid visual clutter. The previous limits in the parameter space are same as Fig.~\ref{fig:E-dependent-Conservative}. From Eqs.~\eqref{eq:Edependent} and \eqref{eq:E-Dependent-Normalization}, the energy-dependent cross section scales as, $\sigma_{\chi\nu}(E_{\nu}) \propto (E_{\nu}/E_{\mathrm{KM3NeT}})^{n}$. This implies that for neutrino energies $E_{\nu} < E_{\mathrm{KM3NeT}}$, larger values of $n$ lead to more stringent upper bounds on $\sigma_{\chi\nu}$. In contrast, for $E_{\nu} > E_{\mathrm{KM3NeT}}$, the bounds become weaker as $n$ increases.
Although the energy-independent constraints can be weaker compared to existing limits, energy-dependent interaction can yield bounds that are stronger than the most stringent existing constraints from other observables like relic neutrinos\,\cite{Akita:2023yga} and Lyman $\alpha$ forest measurements\,\cite{Akita:2023yga}.

In Fig.~\ref{fig:Host-Spike-E-Dependent},  the upper bounds on $\sigma_{\chi\nu}$ (shown by the green and red solid circles) change significantly when $R_{\mathrm{em}}$ is varied from 1 pc to 0.01 pc, respectively, unlike in Fig.~\ref{fig:Host-E-Dependent}, where the change is only marginal. This sharp variation in the DM spike case arises due to the many orders of magnitude increase in the DM density within this radial range, as illustrated in Fig.~\ref{fig:Density-Profiles}
(dashed pink line), compared to the NFW profile depicted in the same figure (solid blue line). 

In Fig.\,\ref{fig:Host-SpikeAnn-E-Dependent}, the constraints on the DM–neutrino scattering cross section for blazars PKS 0605-085 and PMN J 0609–0615 are observed to be weaker when the emission radius satisfies $R_{\mathrm{em}} \leq 0.1$ pc, compared to the scenario without dark matter annihilation. For $R_{\rm em} \gtrsim 1$ pc, the constraints remain unchanged as the density profile beyond $r_{c}$ remains the same. The relative strength of the limits depending on $R_{\rm em}$ follows the same trend as shown in Fig.\,\ref{fig:E-Independent-Host-Spike} (right panel).

\begin{figure*}[!htbp]
	\begin{center}
		\includegraphics[width=\columnwidth]{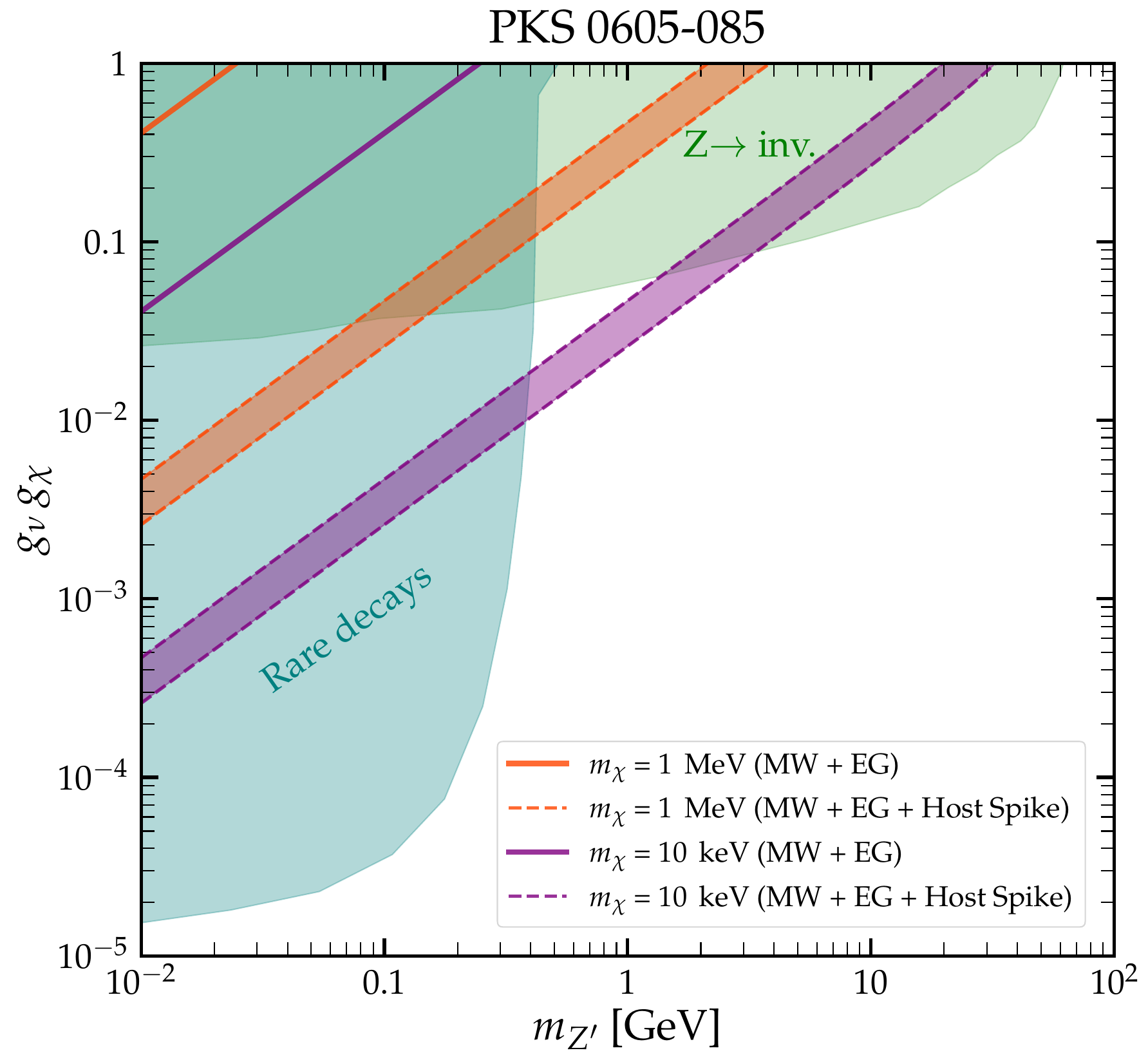}~~
		\includegraphics[width=\columnwidth]{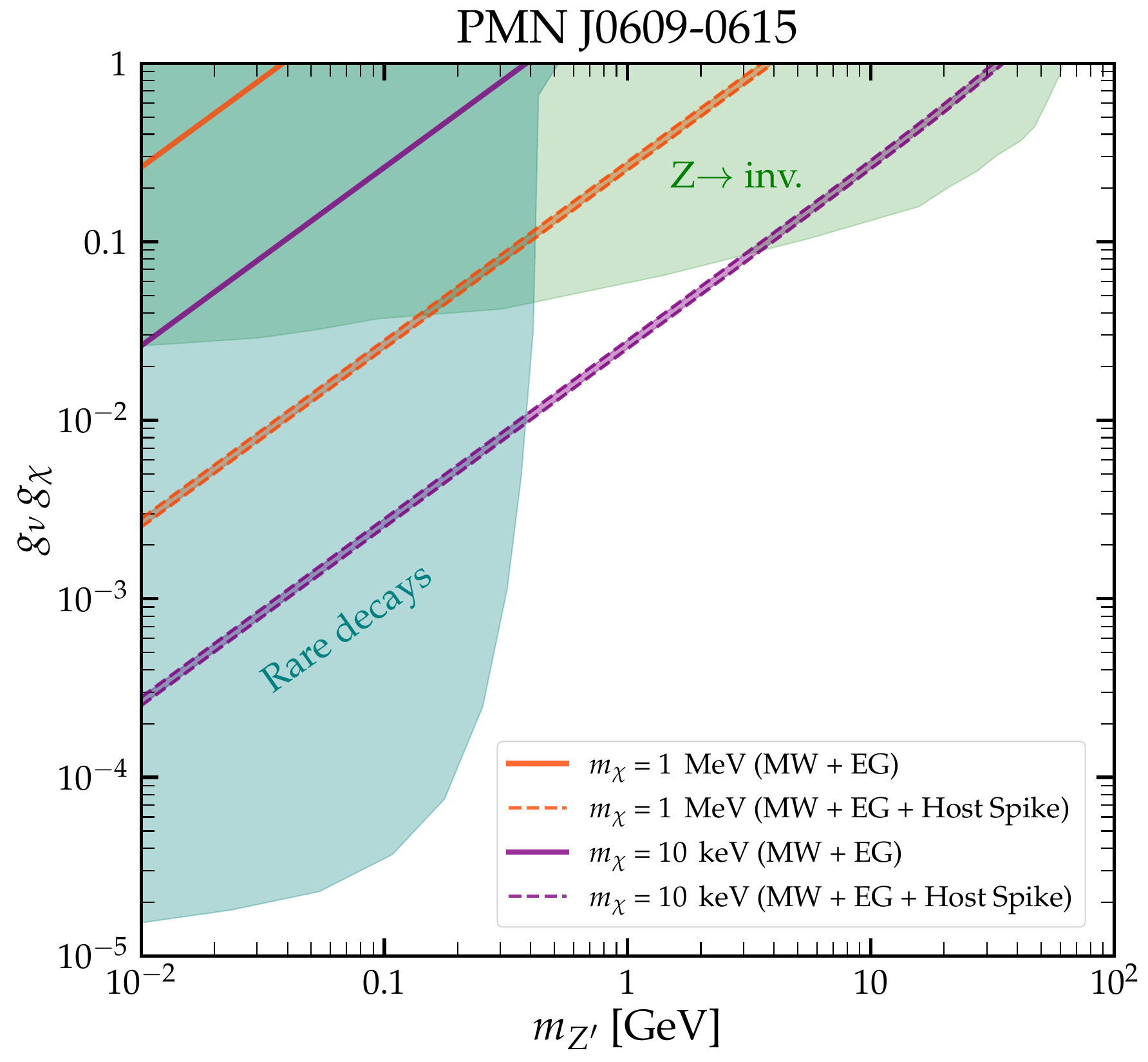}~~\\	
		\caption{Constraints on the coupling product $g_\nu g_\chi$ as a function of $Z'$ mediator mass for blazars PKS 0605-085 (\textbf{left panel}) and PMN J0609-0615 (\textbf{right panel}). Limits for DM masses 1 MeV and 10 keV are shown by orange and purple lines, respectively. The solid lines assume contributions from MW halo and EG medium DM, whereas the dashed lines take into account contributions from halo spike along with EG medium and MW halo DM. For the host spike contribution, the neutrino emission region is taken as, $R_{\rm em}=0.1$ pc. In both the panels, the shaded orange and purple regions show the spread in our limits depending on the possible mass ranges of the corresponding blazars. Previous limits include constraints from invisible $Z$ decays (green shaded region) and rare meson decays (teal shaded region)\,\cite{Berryman:2022hds}.}.
		\label{fig:model}
	\end{center}	
\end{figure*}

In Table~\ref{tab:Constraint-Host}, we present the upper limits on $\sigma_{\chi\nu} / m_{\chi}$, taking into account the contribution of the host DM halo to neutrino flux attenuation. The results are shown for different density profiles and emission radii $R_{\mathrm{em}}$ for the two blazars.

We show our model-dependent limits on DM-neutrino scattering for blazars PKS 0605-085 and PMN J0609-0615 in Fig.\,\ref{fig:model}, left and right panels, respectively.  Here the bounds on the coupling product $g_\nu g_\chi$ is plotted as a function of the mediator mass $m_{Z'}$, for fixed DM masses. In both the panels, the orange and purple solid lines show the limits for DM masses 1 MeV and 1 keV, respectively, when MW halo and EG DM contributions are taken. The orange and purple dashed lines show the limits when the host halo spike is also taken into account along with the MW and EG contributions. Existing limits in the parameter space include limits from invisible $Z$ decays and rare meson decays\,\cite{Berryman:2022hds}. For 10 keV DM mass, for both the blazars, we are able to constrain new regions of the parameter space, provided we take into account the MW, EG, and host DM spike contributions. The orange and purple shaded regions show the spread in our limits depending on the mass of PKS 0605-08 (left panel) and PMN J0609-0615 (right panel), provided in Table\,\ref{tab:Source-Coordinates}. Given the stringent lower limit on fermionic DM mass at $\sim$ keV\,\cite{PhysRevLett.42.407,Dalcanton:2000hn, Boyarsky:2008ju,Alvey:2020xsk}, we choose not to go below $m_\chi =10$ keV. Among the two blazars, PMN J0609-0615 provides stronger constrain which is consistent with the blazar being at higher redshift and its host having a larger virial mass (see Table\,\ref{tab:Source-Coordinates}). We note that in our work we remain agnostic of the DM production mechanism in the early universe. Depending on the exact production mechanism, DM particles can have a wide range of masses\,\cite{Dvorkin:2019zdi,Chang:2019xva,DEramo:2020gpr,Berbig:2022nre}. We leave this for future work.

\begin{table*}[t]
\caption{\label{tab:Constraint-Host}%
Constraints on $\sigma_{\chi\nu}$ for different DM propagation scenarios. Results are shown for various values of $R_{\mathrm{em}}$ for PKS 0605-085 and PMN J0609-0615. Here DM mass is taken as, $m_\chi= 1$ GeV. All limits are in units of cm$^2$.}
\begin{ruledtabular}
\begin{tabular}{lcccc}
\textrm{Source Name}&  $R_{\mathrm{em\,}}(\rm pc)$ & $\sigma_{\chi\nu}^{\mathrm{Host}}$ & $\sigma_{\chi\nu}^{\mathrm{Spike}}$, ($\langle\sigma v\rangle = 0$) & $\sigma_{\chi\nu}^{\mathrm{Spike}}$, ($\langle\sigma v\rangle = 10^{-26}\,\,\mathrm{cm^{3}}/s$) \\
\colrule
\multirow{3}{*}{PKS 0605-085} 
  & 0.01 & $1.03  \times10^{-24}$ & $3.19\times10^{-28}$ &$2.62\times10^{-27}$\\
  & 0.1  & $1.25\times10^{-24}$ & $6.78\times10^{-27}$ &$7.80\times10^{-27}$\\
  & 1.0  & $1.57\times10^{-24}$ & $1.39\times10^{-25}$ &$1.39\times10^{-25}$\\
\multirow{3}{*}{PMN J0609-0615} 
  & 0.01 & $7.05\times10^{-25}$ & $1.19\times10^{-28}$ &$1.63\times10^{-27}$\\
  & 0.1  & $8.47\times10^{-25}$ & $2.45\times10^{-27}$ &$3.35\times10^{-27}$\\
  & 1.0  & $1.06 \times10^{-24}$ & $5.14\times10^{-26}$ &$5.14\times10^{-26}$\\
\end{tabular}
\end{ruledtabular}
\end{table*}

\section{Discussion and Conclusion}
\label{sec:conclusion}

In this work, we place upper bounds on the DM-neutrino cross section from the detection of the high-energy neutrino event KM3-230213A, assuming its blazar origin. In the presence of DM-neutrino interaction, the high-energy neutrino can scatter with the DM present in its path and the neutrino flux can be attenuated/\,deflected. Assuming that the observed event flux at KM3NeT is not significantly modified due to DM-neutrino scattering, we obtain constraints on the DM-neutrino cross section. For our limits, we show the effects of three different DM distributions\,--\,MW, EG medium, and host halo. We also take into account the possibility of the host halo having a DM spike, with and without DM self-annihilation. Finally, we constrain a well-motivated fermionic DM model with KM3-230213A. KM3-230213A is the highest energy neutrino event to date. Thus, we find that for energy-dependent DM-neutrino scattering, KM3-230213A can provide the most stringent limits on DM-neutrino scattering for specific energy dependence on the cross section.  For Dirac fermion DM interacting with neutrino via $Z'$ mediator, we are able to probe new regions of parameter space for light DM masses ($\sim$ 10 keV).  Our work shows the potential of high-energy neutrino detectors like KM3NeT in probing DM-neutrino interactions.

Cosmogenic neutrinos remain a viable candidate for KM3-230213A\,\cite{KM3NeT:2025vut,Zhang:2025abk}. Ultra-high-energy cosmic rays (UHECRs) can interact with the background Cosmic Microwave Background (CMB)/ extragalactic background light (EBL) photon and produce high-energy neutrinos and contribute to the diffuse neutrino flux known as cosmogenic neutrinos\,\cite{Berezinsky:1969erk,1979ApJ...228..919S}. According to Ref.\,\cite{KM3NeT:2025vut}, the cosmogenic origin can explain KM3-230213A, if there is a strong source evolution and a non-negligible fraction of protons in high-energy cosmic rays. Assuming a cosmogenic origin of KM3-230213A, one can again derive novel bounds on DM-neutrino scattering cross section using the same technique outlined here. We leave this for future exploration.    

For halo DM contributions, the limits derived in this work depend on the production site of the high-energy neutrinos around the blazar. The existing tension between IceCube and KM3NeT detection may hold the key to understanding the production site of such high-energy neutrinos\,\cite{Li:2025tqf, Neronov:2025jfj, Wang:2025lgn}. Besides, even with the blazar origin, the blazar mass and the corresponding virial mass of the host halo have significant uncertainty. Future multi-messenger observations as well as more events in KM3NeT and IceCube will be essential in pinpointing the exact source and nature of such an event. This can result in more robust limits on DM-neutrino scattering cross sections.

Besides attenuation due to scattering, DM-neutrino interactions can have important effects on both cosmological and astrophysical observables\,\cite{DiValentino:2017oaw,Olivares-DelCampo:2017feq,Lattanzi:2017ubx,Boehm:2014vja}. The particle physics models allowing DM-neutrino interactions can also induce alternate scenarios like DM self-annihilation to neutrinos. These BSM scenarios can yield characteristic observational signatures for indirect searches of DM\,\cite{Yuksel:2007ac, Palomares-Ruiz:2007trf, Palomares-Ruiz:2007egs, Kile:2009nn, Covi:2009xn, Primulando:2017kxf, Arguelles:2019ouk,Blennow:2019fhy,Arguelles:2022nbl}, as well as account for the DM relic abundance\,\cite{Aoki:2012ub,Berlin:2018bsc}. The diffuse supernova neutrino background (DSNB) can also act as an important probe for DM-neutrino searches, as they would form dips in the background and result in creating anisotropic signatures of high-energy neutrino distributions\,\cite{Farzan:2014gza,Chauhan:2025hoz}. In the context of cosmology, these interactions can change both the temperature and matter power spectrum through collisional damping and free streaming. These effects will be reflected in cosmological observations, e.g., CMB\,\cite{Wilkinson:2014ksa}, Large Scale Structure (LSS)\,\cite{Boehm:2003xr,Escudero:2018thh}, $\mathrm{Lyman\,\alpha}$\,\cite{Hooper:2021rjc}, 21 cm observations\,\cite{Mosbech:2022uud,Escudero:2018thh}, etc. Future astrophysical and cosmological complementary probes will be essential in discovering exotic interactions between neutrinos and DM. 

{\bf \em Note added.}
While this work was being finalized, Ref.\,\cite{Bertolez-Martinez:2025trs} appeared on arXiv exploring a scenario similar to ours.

\section*{Acknowledgements} 
 We thank Bhavesh Chauhan, Debtosh Chowdhury, Amol Dighe, and Maria Cristina Volpe for discussions and insightful comments. We also thank Debajit Bose for detailed discussions regarding the particle physics model. We thank Deep Jyoti Das and Rinchen Sherpa for comments on the manuscript. S.B. acknowledges the Council of Scientific and Industrial Research (CSIR), Government of India, for supporting his research under the CSIR Junior/Senior Research Fellowship program through grant no. 09/0079(15488)/2022-EMR-I. A.K.S. acknowledges the Ministry of Human Resource Development, Government of India, for financial support via the Prime Ministers’ Research Fellowship (PMRF). R.L.\,\,acknowledges financial support from
the institute start-up funds and ISRO-IISc STC for the
grant no. ISTC/PHY/RL/499.

\newpage
\bibliographystyle{JHEP}
\bibliography{ref.bib}

\providecommand{\href}[2]{#2}\begingroup\raggedright\begin{thebibliography}{100}

\bibitem{Super-Kamiokande:1998kpq}
{\scshape Super-Kamiokande} collaboration, Y.~Fukuda et~al., \emph{{Evidence
  for oscillation of atmospheric neutrinos}},
  \href{http://dx.doi.org/10.1103/PhysRevLett.81.1562}{\emph{Phys. Rev. Lett.}
  {\bf 81} (1998) 1562--1567},
  [\href{https://arxiv.org/abs/hep-ex/9807003}{{\tt hep-ex/9807003}}].

\bibitem{Super-Kamiokande:2001ljr}
{\scshape Super-Kamiokande} collaboration, S.~Fukuda et~al., \emph{{Solar B-8
  and hep neutrino measurements from 1258 days of Super-Kamiokande data}},
  \href{http://dx.doi.org/10.1103/PhysRevLett.86.5651}{\emph{Phys. Rev. Lett.}
  {\bf 86} (2001) 5651--5655},
  [\href{https://arxiv.org/abs/hep-ex/0103032}{{\tt hep-ex/0103032}}].

\bibitem{Super-Kamiokande:2005mbp}
{\scshape Super-Kamiokande} collaboration, Y.~Ashie et~al., \emph{{A
  Measurement of atmospheric neutrino oscillation parameters by
  SUPER-KAMIOKANDE I}},
  \href{http://dx.doi.org/10.1103/PhysRevD.71.112005}{\emph{Phys. Rev. D} {\bf
  71} (2005) 112005}, [\href{https://arxiv.org/abs/hep-ex/0501064}{{\tt
  hep-ex/0501064}}].

\bibitem{Borexino:2008dzn}
{\scshape Borexino} collaboration, C.~Arpesella et~al., \emph{{Direct
  Measurement of the Be-7 Solar Neutrino Flux with 192 Days of Borexino Data}},
  \href{http://dx.doi.org/10.1103/PhysRevLett.101.091302}{\emph{Phys. Rev.
  Lett.} {\bf 101} (2008) 091302}, [\href{https://arxiv.org/abs/0805.3843}{{\tt
  0805.3843}}].

\bibitem{SNO:2002tuh}
{\scshape SNO} collaboration, Q.~R. Ahmad et~al., \emph{{Direct evidence for
  neutrino flavor transformation from neutral current interactions in the
  Sudbury Neutrino Observatory}},
  \href{http://dx.doi.org/10.1103/PhysRevLett.89.011301}{\emph{Phys. Rev.
  Lett.} {\bf 89} (2002) 011301},
  [\href{https://arxiv.org/abs/nucl-ex/0204008}{{\tt nucl-ex/0204008}}].

\bibitem{SNO:2001kpb}
{\scshape SNO} collaboration, Q.~R. Ahmad et~al., \emph{{Measurement of the
  rate of $\nu_e+d \to p+p+e^-$ interactions produced by $^8$B solar neutrinos
  at the Sudbury Neutrino Observatory}},
  \href{http://dx.doi.org/10.1103/PhysRevLett.87.071301}{\emph{Phys. Rev.
  Lett.} {\bf 87} (2001) 071301},
  [\href{https://arxiv.org/abs/nucl-ex/0106015}{{\tt nucl-ex/0106015}}].

\bibitem{IceCube:2013low}
{\scshape IceCube} collaboration, M.~G. Aartsen et~al., \emph{{Evidence for
  High-Energy Extraterrestrial Neutrinos at the IceCube Detector}},
  \href{http://dx.doi.org/10.1126/science.1242856}{\emph{Science} {\bf 342}
  (2013) 1242856}, [\href{https://arxiv.org/abs/1311.5238}{{\tt 1311.5238}}].

\bibitem{IceCube:2013cdw}
{\scshape IceCube} collaboration, M.~G. Aartsen et~al., \emph{{First
  observation of PeV-energy neutrinos with IceCube}},
  \href{http://dx.doi.org/10.1103/PhysRevLett.111.021103}{\emph{Phys. Rev.
  Lett.} {\bf 111} (2013) 021103}, [\href{https://arxiv.org/abs/1304.5356}{{\tt
  1304.5356}}].

\bibitem{IceCube:2014stg}
{\scshape IceCube} collaboration, M.~G. Aartsen et~al., \emph{{Observation of
  High-Energy Astrophysical Neutrinos in Three Years of IceCube Data}},
  \href{http://dx.doi.org/10.1103/PhysRevLett.113.101101}{\emph{Phys. Rev.
  Lett.} {\bf 113} (2014) 101101}, [\href{https://arxiv.org/abs/1405.5303}{{\tt
  1405.5303}}].

\bibitem{IceCube:2018cha}
{\scshape IceCube} collaboration, M.~G. Aartsen et~al., \emph{{Neutrino
  emission from the direction of the blazar TXS 0506+056 prior to the
  IceCube-170922A alert}},
  \href{http://dx.doi.org/10.1126/science.aat2890}{\emph{Science} {\bf 361}
  (2018) 147--151}, [\href{https://arxiv.org/abs/1807.08794}{{\tt
  1807.08794}}].

\bibitem{IceCube:2018dnn}
{\scshape IceCube, Fermi-LAT, MAGIC, AGILE, ASAS-SN, HAWC, H.E.S.S., INTEGRAL,
  Kanata, Kiso, Kapteyn, Liverpool Telescope, Subaru, Swift NuSTAR, VERITAS,
  VLA/17B-403} collaboration, M.~G. Aartsen et~al., \emph{{Multimessenger
  observations of a flaring blazar coincident with high-energy neutrino
  IceCube-170922A}},
  \href{http://dx.doi.org/10.1126/science.aat1378}{\emph{Science} {\bf 361}
  (2018) eaat1378}, [\href{https://arxiv.org/abs/1807.08816}{{\tt
  1807.08816}}].

\bibitem{LIGOScientific:2017ync}
{\scshape LIGO} collaboration, B.~P. Abbott et~al., \emph{{Multi-messenger
  Observations of a Binary Neutron Star Merger}},
  \href{http://dx.doi.org/10.3847/2041-8213/aa91c9}{\emph{Astrophys. J. Lett.}
  {\bf 848} (2017) L12}, [\href{https://arxiv.org/abs/1710.05833}{{\tt
  1710.05833}}].

\bibitem{IceCube:2021rpz}
{\scshape IceCube} collaboration, M.~G. Aartsen et~al., \emph{{Detection of a
  particle shower at the Glashow resonance with IceCube}},
  \href{http://dx.doi.org/10.1038/s41586-021-03256-1}{\emph{Nature} {\bf 591}
  (2021) 220--224}, [\href{https://arxiv.org/abs/2110.15051}{{\tt
  2110.15051}}].

\bibitem{IceCube:2023ame}
{\scshape IceCube} collaboration, R.~Abbasi et~al., \emph{{Observation of
  high-energy neutrinos from the Galactic plane}},
  \href{http://dx.doi.org/10.1126/science.adc9818}{\emph{Science} {\bf 380}
  (2023) adc9818}, [\href{https://arxiv.org/abs/2307.04427}{{\tt 2307.04427}}].

\bibitem{IceCube:2015gsk}
{\scshape IceCube} collaboration, M.~G. Aartsen et~al., \emph{{A combined
  maximum-likelihood analysis of the high-energy astrophysical neutrino flux
  measured with IceCube}},
  \href{http://dx.doi.org/10.1088/0004-637X/809/1/98}{\emph{Astrophys. J.} {\bf
  809} (2015) 98}, [\href{https://arxiv.org/abs/1507.03991}{{\tt 1507.03991}}].

\bibitem{ANTARES:2012xie}
{\scshape ANTARES} collaboration, S.~Adrian-Martinez et~al., \emph{{Search for
  Cosmic Neutrino Point Sources with Four Year Data of the ANTARES Telescope}},
  \href{http://dx.doi.org/10.1088/0004-637X/760/1/53}{\emph{Astrophys. J.} {\bf
  760} (2012) 53}, [\href{https://arxiv.org/abs/1207.3105}{{\tt 1207.3105}}].

\bibitem{Baikal-GVD:2022fmn}
{\scshape Baikal-GVD} collaboration, V.~A. Allakhverdyan et~al.,
  \emph{{High-energy neutrino-induced cascade from the direction of the flaring
  radio blazar TXS 0506+056 observed by Baikal-GVD in 2021}},
  \href{http://dx.doi.org/10.1093/mnras/stad3653}{\emph{Mon. Not. Roy. Astron.
  Soc.} {\bf 527} (2023) 8784--8792},
  [\href{https://arxiv.org/abs/2210.01650}{{\tt 2210.01650}}].

\bibitem{Baikal-GVD:2022fis}
{\scshape Baikal-GVD} collaboration, V.~A. Allakhverdyan et~al., \emph{{Diffuse
  neutrino flux measurements with the Baikal-GVD neutrino telescope}},
  \href{http://dx.doi.org/10.1103/PhysRevD.107.042005}{\emph{Phys. Rev. D} {\bf
  107} (2023) 042005}, [\href{https://arxiv.org/abs/2211.09447}{{\tt
  2211.09447}}].

\bibitem{KM3-Blazar}
{\scshape KM3NeT, MessMapp Group, Fermi-LAT, Owens Valley Radio Observatory
  40-m Telescope Group, SVOM} collaboration, O.~Adriani et~al.,
  \emph{{Characterising Candidate Blazar Counterparts of the Ultra-High-Energy
  Event KM3-230213A}},  \href{https://arxiv.org/abs/2502.08484}{{\tt
  2502.08484}}.

\bibitem{KM3NeT-Nature}
{\scshape KM3NeT} collaboration, S.~Aiello et~al., \emph{{Observation of an
  ultra-high-energy cosmic neutrino with KM3NeT}},
  \href{http://dx.doi.org/10.1038/s41586-024-08543-1}{\emph{Nature} {\bf 638}
  (2025) 376--382}.

\bibitem{KM3NeT:2025vut}
{\scshape KM3NeT} collaboration, O.~Adriani et~al., \emph{{On the Potential
  Cosmogenic Origin of the Ultra-high-energy Event KM3-230213A}},
  \href{http://dx.doi.org/10.3847/2041-8213/adcc29}{\emph{Astrophys. J. Lett.}
  {\bf 984} (2025) L41}, [\href{https://arxiv.org/abs/2502.08508}{{\tt
  2502.08508}}].

\bibitem{KM3NeT:2025bxl}
{\scshape KM3NeT, MessMapp Group, Fermi-LAT, Owens Valley Radio Observatory
  40-m Telescope Group, SVOM} collaboration, O.~Adriani et~al.,
  \emph{{Characterising Candidate Blazar Counterparts of the Ultra-High-Energy
  Event KM3-230213A}},  \href{https://arxiv.org/abs/2502.08484}{{\tt
  2502.08484}}.

\bibitem{KM3NeT:2025aps}
{\scshape KM3NeT} collaboration, O.~Adriani et~al., \emph{{On the Potential
  Galactic Origin of the Ultra-High-Energy Event KM3-230213A}},
  \href{https://arxiv.org/abs/2502.08387}{{\tt 2502.08387}}.

\bibitem{KM3NeT-Oscillation}
V.~Brdar and D.~S. Chattopadhyay, \emph{{Does the 220 PeV Event at KM3NeT Point
  to New Physics?}},  \href{https://arxiv.org/abs/2502.21299}{{\tt
  2502.21299}}.

\bibitem{Li:2025tqf}
S.~W. Li, P.~Machado, D.~Naredo-Tuero and T.~Schwemberger, \emph{{Clash of the
  Titans: ultra-high energy KM3NeT event versus IceCube data}},
  \href{https://arxiv.org/abs/2502.04508}{{\tt 2502.04508}}.

\bibitem{Muzio:2025gbr}
M.~S. Muzio, T.~Yuan and L.~Lu, \emph{{Emergence of a neutrino flux above 5 PeV
  and implications for ultrahigh energy cosmic rays}},
  \href{https://arxiv.org/abs/2502.06944}{{\tt 2502.06944}}.

\bibitem{Fang:2025nzg}
K.~Fang, F.~Halzen and D.~Hooper, \emph{{Cascaded Gamma-Ray Emission Associated
  with the KM3NeT Ultrahigh-energy Event KM3-230213A}},
  \href{http://dx.doi.org/10.3847/2041-8213/adbbec}{\emph{Astrophys. J. Lett.}
  {\bf 982} (2025) L16}, [\href{https://arxiv.org/abs/2502.09545}{{\tt
  2502.09545}}].

\bibitem{Filipovic:2025ulm}
M.~D. Filipovi\'c et~al., \emph{{ASKAP and VLASS Search for a Radio-continuum
  Counterpart of Ultra-high-energy Neutrino Event KM3\textendash{}230213A}},
  \href{http://dx.doi.org/10.3847/2041-8213/adca37}{\emph{Astrophys. J. Lett.}
  {\bf 984} (2025) L52}, [\href{https://arxiv.org/abs/2503.09108}{{\tt
  2503.09108}}].

\bibitem{Neronov:2025jfj}
A.~Neronov, F.~Oikonomou and D.~Semikoz, \emph{{KM3-230213A: An Ultra-High
  Energy Neutrino from a Year-Long Astrophysical Transient}},
  \href{https://arxiv.org/abs/2502.12986}{{\tt 2502.12986}}.

\bibitem{Wang:2025lgn}
R.~Wang, J.~Zhu, H.~Li and B.-Q. Ma, \emph{{Association of 220\,PeV Neutrino
  KM3-230213A with Gamma-Ray Bursts}},
  \href{http://dx.doi.org/10.3847/2515-5172/adc452}{\emph{Res. Notes AAS} {\bf
  9} (2025) 65}, [\href{https://arxiv.org/abs/2503.14471}{{\tt 2503.14471}}].

\bibitem{Boccia:2025hpm}
A.~Boccia and F.~Iocco, \emph{{A strike of luck: could the KM3-230213A event be
  caused by an evaporating primordial black hole?}},
  \href{https://arxiv.org/abs/2502.19245}{{\tt 2502.19245}}.

\bibitem{Borah:2025igh}
D.~Borah, N.~Das, N.~Okada and P.~Sarmah, \emph{{Possible origin of the
  KM3-230213A neutrino event from dark matter decay}},
  \href{http://dx.doi.org/10.1103/shsw-mct6}{\emph{Phys. Rev. D} {\bf 111}
  (2025) 123022}, [\href{https://arxiv.org/abs/2503.00097}{{\tt 2503.00097}}].

\bibitem{Brdar:2025azm}
V.~Brdar and D.~S. Chattopadhyay, \emph{{Does the 220 PeV Event at KM3NeT Point
  to New Physics?}},  \href{https://arxiv.org/abs/2502.21299}{{\tt
  2502.21299}}.

\bibitem{Airoldi:2025opo}
L.~F.~T. Airoldi, G.~F.~S. Alves, Y.~F. Perez-Gonzalez, G.~M. Salla and R.~Z.
  Funchal, \emph{{Could a Primordial Black Hole Explosion Explain the KM3NeT
  Event?}},  \href{https://arxiv.org/abs/2505.24666}{{\tt 2505.24666}}.

\bibitem{Dev:2025czz}
P.~S.~B. Dev, B.~Dutta, A.~Karthikeyan, W.~Maitra, L.~E. Strigari and A.~Verma,
  \emph{{`Dark' Matter Effect as a Novel Solution to the KM3-230213A Puzzle}},
  \href{https://arxiv.org/abs/2505.22754}{{\tt 2505.22754}}.

\bibitem{Farzan:2025ydi}
Y.~Farzan and M.~Hostert, \emph{{Astrophysical sources of dark particles as a
  solution to the KM3NeT and IceCube tension over KM3-230213A}},
  \href{https://arxiv.org/abs/2505.22711}{{\tt 2505.22711}}.

\bibitem{Bertone:2016nfn}
G.~Bertone and D.~Hooper, \emph{{History of dark matter}},
  \href{http://dx.doi.org/10.1103/RevModPhys.90.045002}{\emph{Rev. Mod. Phys.}
  {\bf 90} (2018) 045002}, [\href{https://arxiv.org/abs/1605.04909}{{\tt
  1605.04909}}].

\bibitem{Cirelli:2024ssz}
M.~Cirelli, A.~Strumia and J.~Zupan, \emph{{Dark Matter}},
  \href{https://arxiv.org/abs/2406.01705}{{\tt 2406.01705}}.

\bibitem{Strigari:2012acq}
L.~E. Strigari, \emph{{Galactic Searches for Dark Matter}},
  \href{http://dx.doi.org/10.1016/j.physrep.2013.05.004}{\emph{Phys. Rept.}
  {\bf 531} (2013) 1--88}, [\href{https://arxiv.org/abs/1211.7090}{{\tt
  1211.7090}}].

\bibitem{Slatyer:2017sev}
T.~R. Slatyer, \emph{{Indirect detection of dark matter.}},  in
  \emph{{Theoretical Advanced Study Institute in Elementary Particle Physics}:
  {Anticipating the Next Discoveries in Particle Physics}}, pp.~297--353, 2018.
\newblock \href{https://arxiv.org/abs/1710.05137}{{\tt 1710.05137}}.
\newblock \href{http://dx.doi.org/10.1142/9789813233348_0005}{DOI}.

\bibitem{Lin:2019uvt}
T.~Lin, \emph{{Dark matter models and direct detection}},
  \href{http://dx.doi.org/10.22323/1.333.0009}{\emph{PoS} {\bf 333} (2019)
  009}, [\href{https://arxiv.org/abs/1904.07915}{{\tt 1904.07915}}].

\bibitem{Berryman:2022hds}
J.~M. Berryman et~al., \emph{{Neutrino self-interactions: A white paper}},
  \href{http://dx.doi.org/10.1016/j.dark.2023.101267}{\emph{Phys. Dark Univ.}
  {\bf 42} (2023) 101267}, [\href{https://arxiv.org/abs/2203.01955}{{\tt
  2203.01955}}].

\bibitem{Dev:2024twk}
P.~S.~B. Dev, D.~Kim, D.~Sathyan, K.~Sinha and Y.~Zhang, \emph{{New Laboratory
  Constraints on Neutrinophilic Mediators}},
  \href{https://arxiv.org/abs/2407.12738}{{\tt 2407.12738}}.

\bibitem{Boehm:2000gq}
C.~Boehm, P.~Fayet and R.~Schaeffer, \emph{{Constraining dark matter candidates
  from structure formation}},
  \href{http://dx.doi.org/10.1016/S0370-2693(01)01060-7}{\emph{Phys. Lett. B}
  {\bf 518} (2001) 8--14}, [\href{https://arxiv.org/abs/astro-ph/0012504}{{\tt
  astro-ph/0012504}}].

\bibitem{Boehm:2003xr}
C.~Boehm, H.~Mathis, J.~Devriendt and J.~Silk, \emph{{Non-linear evolution of
  suppressed dark matter primordial power spectra}},
  \href{http://dx.doi.org/10.1111/j.1365-2966.2005.09032.x}{\emph{Mon. Not.
  Roy. Astron. Soc.} {\bf 360} (2005) 282--287},
  [\href{https://arxiv.org/abs/astro-ph/0309652}{{\tt astro-ph/0309652}}].

\bibitem{Mangano:2006mp}
G.~Mangano, A.~Melchiorri, P.~Serra, A.~Cooray and M.~Kamionkowski,
  \emph{{Cosmological bounds on dark matter-neutrino interactions}},
  \href{http://dx.doi.org/10.1103/PhysRevD.74.043517}{\emph{Phys. Rev. D} {\bf
  74} (2006) 043517}, [\href{https://arxiv.org/abs/astro-ph/0606190}{{\tt
  astro-ph/0606190}}].

\bibitem{Wilkinson:2014ksa}
R.~J. Wilkinson, C.~Boehm and J.~Lesgourgues, \emph{{Constraining Dark
  Matter-Neutrino Interactions using the CMB and Large-Scale Structure}},
  \href{http://dx.doi.org/10.1088/1475-7516/2014/05/011}{\emph{JCAP} {\bf 05}
  (2014) 011}, [\href{https://arxiv.org/abs/1401.7597}{{\tt 1401.7597}}].

\bibitem{Escudero:2018thh}
M.~Escudero, L.~Lopez-Honorez, O.~Mena, S.~Palomares-Ruiz and
  P.~Villanueva-Domingo, \emph{{A fresh look into the interacting dark matter
  scenario}},
  \href{http://dx.doi.org/10.1088/1475-7516/2018/06/007}{\emph{JCAP} {\bf 06}
  (2018) 007}, [\href{https://arxiv.org/abs/1803.08427}{{\tt 1803.08427}}].

\bibitem{Boehm:2014vja}
C.~Boehm, J.~A. Schewtschenko, R.~J. Wilkinson, C.~M. Baugh and S.~Pascoli,
  \emph{{Using the Milky Way satellites to study interactions between cold dark
  matter and radiation}},
  \href{http://dx.doi.org/10.1093/mnrasl/slu115}{\emph{Mon. Not. Roy. Astron.
  Soc.} {\bf 445} (2014) L31--L35},
  [\href{https://arxiv.org/abs/1404.7012}{{\tt 1404.7012}}].

\bibitem{Brax:2023tvn}
P.~Brax, C.~van~de Bruck, E.~Di~Valentino, W.~Giar\`e and S.~Trojanowski,
  \emph{{Extended analysis of neutrino-dark matter interactions with
  small-scale CMB experiments}},
  \href{http://dx.doi.org/10.1016/j.dark.2023.101321}{\emph{Phys. Dark Univ.}
  {\bf 42} (2023) 101321}, [\href{https://arxiv.org/abs/2305.01383}{{\tt
  2305.01383}}].

\bibitem{Farzan:2014gza}
Y.~Farzan and S.~Palomares-Ruiz, \emph{{Dips in the Diffuse Supernova Neutrino
  Background}},
  \href{http://dx.doi.org/10.1088/1475-7516/2014/06/014}{\emph{JCAP} {\bf 06}
  (2014) 014}, [\href{https://arxiv.org/abs/1401.7019}{{\tt 1401.7019}}].

\bibitem{Arguelles:2017atb}
C.~A. Arg\"uelles, A.~Kheirandish and A.~C. Vincent, \emph{{Imaging Galactic
  Dark Matter with High-Energy Cosmic Neutrinos}},
  \href{http://dx.doi.org/10.1103/PhysRevLett.119.201801}{\emph{Phys. Rev.
  Lett.} {\bf 119} (2017) 201801},
  [\href{https://arxiv.org/abs/1703.00451}{{\tt 1703.00451}}].

\bibitem{Kelly:2018tyg}
K.~J. Kelly and P.~A.~N. Machado, \emph{{Multimessenger Astronomy and New
  Neutrino Physics}},
  \href{http://dx.doi.org/10.1088/1475-7516/2018/10/048}{\emph{JCAP} {\bf 10}
  (2018) 048}, [\href{https://arxiv.org/abs/1808.02889}{{\tt 1808.02889}}].

\bibitem{ChoiIceCube}
K.-Y. Choi, J.~Kim and C.~Rott, \emph{{Constraining dark matter-neutrino
  interactions with IceCube-170922A}},
  \href{http://dx.doi.org/10.1103/PhysRevD.99.083018}{\emph{Phys. Rev. D} {\bf
  99} (2019) 083018}, [\href{https://arxiv.org/abs/1903.03302}{{\tt
  1903.03302}}].

\bibitem{Murase-Neutrino-AGN}
K.~Murase and F.~W. Stecker, \emph{{Chapter 10: High-Energy Neutrinos from
  Active Galactic Nuclei}},  \href{https://arxiv.org/abs/2202.03381}{{\tt
  2202.03381}}.

\bibitem{Murase:2019xqi}
K.~Murase and I.~M. Shoemaker, \emph{{Neutrino Echoes from Multimessenger
  Transient Sources}},
  \href{http://dx.doi.org/10.1103/PhysRevLett.123.241102}{\emph{Phys. Rev.
  Lett.} {\bf 123} (2019) 241102},
  [\href{https://arxiv.org/abs/1903.08607}{{\tt 1903.08607}}].

\bibitem{Ferrera-TXS}
F.~Ferrer, G.~Herrera and A.~Ibarra, \emph{{New constraints on the dark
  matter-neutrino and dark matter-photon scattering cross sections from TXS
  0506+056}},
  \href{http://dx.doi.org/10.1088/1475-7516/2023/05/057}{\emph{JCAP} {\bf 05}
  (2023) 057}, [\href{https://arxiv.org/abs/2209.06339}{{\tt 2209.06339}}].

\bibitem{Carpio:2022sml}
J.~A. Carpio, A.~Kheirandish and K.~Murase, \emph{{Time-delayed neutrino
  emission from supernovae as a probe of dark matter-neutrino interactions}},
  \href{http://dx.doi.org/10.1088/1475-7516/2023/04/019}{\emph{JCAP} {\bf 04}
  (2023) 019}, [\href{https://arxiv.org/abs/2204.09650}{{\tt 2204.09650}}].

\bibitem{Cline-TXS}
J.~M. Cline, S.~Gao, F.~Guo, Z.~Lin, S.~Liu, M.~Puel et~al., \emph{{Blazar
  Constraints on Neutrino-Dark Matter Scattering}},
  \href{http://dx.doi.org/10.1103/PhysRevLett.130.091402}{\emph{Phys. Rev.
  Lett.} {\bf 130} (2023) 091402},
  [\href{https://arxiv.org/abs/2209.02713}{{\tt 2209.02713}}].

\bibitem{Eskenasy:2022aup}
R.~Eskenasy, A.~Kheirandish and K.~Murase, \emph{{Light curves of BSM-induced
  neutrino echoes in the optically thin limit}},
  \href{http://dx.doi.org/10.1103/PhysRevD.107.103038}{\emph{Phys. Rev. D} {\bf
  107} (2023) 103038}, [\href{https://arxiv.org/abs/2204.08924}{{\tt
  2204.08924}}].

\bibitem{Fujiwara:2023lsv}
M.~Fujiwara and G.~Herrera, \emph{{Tidal disruption events and dark matter
  scatterings with neutrinos and photons}},
  \href{http://dx.doi.org/10.1016/j.physletb.2024.138573}{\emph{Phys. Lett. B}
  {\bf 851} (2024) 138573}, [\href{https://arxiv.org/abs/2312.11670}{{\tt
  2312.11670}}].

\bibitem{KA:2023dyz}
S.~K.~A., A.~Das, G.~Lambiase, T.~Nomura and Y.~Orikasa, \emph{{Probing chiral
  and flavored $Z^\prime $ from cosmic bursts through neutrino interactions}},
  \href{http://dx.doi.org/10.1140/epjc/s10052-024-13530-x}{\emph{Eur. Phys. J.
  C} {\bf 84} (2024) 1224}, [\href{https://arxiv.org/abs/2308.14483}{{\tt
  2308.14483}}].

\bibitem{Choi:2019ixb}
K.-Y. Choi, J.~Kim and C.~Rott, \emph{{Constraining dark matter-neutrino
  interactions with IceCube-170922A}},
  \href{http://dx.doi.org/10.1103/PhysRevD.99.083018}{\emph{Phys. Rev. D} {\bf
  99} (2019) 083018}, [\href{https://arxiv.org/abs/1903.03302}{{\tt
  1903.03302}}].

\bibitem{Cline:NGC-1068}
J.~M. Cline and M.~Puel, \emph{{NGC 1068 constraints on neutrino-dark matter
  scattering}},
  \href{http://dx.doi.org/10.1088/1475-7516/2023/06/004}{\emph{JCAP} {\bf 06}
  (2023) 004}, [\href{https://arxiv.org/abs/2301.08756}{{\tt 2301.08756}}].

\bibitem{Heston:2024ljf}
S.~Heston, S.~Horiuchi and S.~Shirai, \emph{{Constraining neutrino-DM
  interactions with Milky~Way dwarf spheroidals and supernova neutrinos}},
  \href{http://dx.doi.org/10.1103/PhysRevD.110.023004}{\emph{Phys. Rev. D} {\bf
  110} (2024) 023004}, [\href{https://arxiv.org/abs/2402.08718}{{\tt
  2402.08718}}].

\bibitem{Fujiwara:2024qos}
M.~Fujiwara, G.~Herrera and S.~Horiuchi, \emph{{Neutrino Diffusion within Dark
  Matter Spikes}},  \href{https://arxiv.org/abs/2412.00805}{{\tt 2412.00805}}.

\bibitem{Chauhan:2025hoz}
G.~Chauhan, R.~A. Gustafson, G.~Herrera, T.~Johnson and I.~Shoemaker,
  \emph{{The Dark Matter Diffused Supernova Neutrino Background}},
  \href{https://arxiv.org/abs/2505.03882}{{\tt 2505.03882}}.

\bibitem{Zapata:2025huq}
G.~D. Zapata, J.~Jones-P\'erez and A.~M. Gago, \emph{{Bounds on neutrino-DM
  interactions from TXS 0506+056 neutrino outburst}},
  \href{https://arxiv.org/abs/2503.03823}{{\tt 2503.03823}}.

\bibitem{Trojanowski:2025oro}
S.~Trojanowski and L.~Zu, \emph{{Cosmological impact of $\nu$DM interactions
  enhanced in narrow redshift ranges}},
  \href{https://arxiv.org/abs/2505.20396}{{\tt 2505.20396}}.

\bibitem{Ng:2014pca}
K.~C.~Y. Ng and J.~F. Beacom, \emph{{Cosmic neutrino cascades from secret
  neutrino interactions}},
  \href{http://dx.doi.org/10.1103/PhysRevD.90.065035}{\emph{Phys. Rev. D} {\bf
  90} (2014) 065035}, [\href{https://arxiv.org/abs/1404.2288}{{\tt
  1404.2288}}].

\bibitem{Mazumdar:2020ibx}
A.~Mazumdar, S.~Mohanty and P.~Parashari, \emph{{Flavour specific neutrino
  self-interaction: H $_{0}$ tension and IceCube}},
  \href{http://dx.doi.org/10.1088/1475-7516/2022/10/011}{\emph{JCAP} {\bf 10}
  (2022) 011}, [\href{https://arxiv.org/abs/2011.13685}{{\tt 2011.13685}}].

\bibitem{Chauhan:2024fas}
B.~Chauhan and P.~Parashari, \emph{{Probing the cosmic sterile-neutrino
  background with IceCube}},  \href{https://arxiv.org/abs/2409.12145}{{\tt
  2409.12145}}.

\bibitem{Das:2021lcr}
A.~Das and M.~Sen, \emph{{Boosted dark matter from diffuse supernova
  neutrinos}}, \href{http://dx.doi.org/10.1103/PhysRevD.104.075029}{\emph{Phys.
  Rev. D} {\bf 104} (2021) 075029},
  [\href{https://arxiv.org/abs/2104.00027}{{\tt 2104.00027}}].

\bibitem{Esteban:2021tub}
I.~Esteban, S.~Pandey, V.~Brdar and J.~F. Beacom, \emph{{Probing secret
  interactions of astrophysical neutrinos in the high-statistics era}},
  \href{http://dx.doi.org/10.1103/PhysRevD.104.123014}{\emph{Phys. Rev. D} {\bf
  104} (2021) 123014}, [\href{https://arxiv.org/abs/2107.13568}{{\tt
  2107.13568}}].

\bibitem{Das:2024ghw}
A.~Das, T.~Herbermann, M.~Sen and V.~Takhistov, \emph{{Energy-dependent boosted
  dark matter from diffuse supernova neutrino background}},
  \href{http://dx.doi.org/10.1088/1475-7516/2024/07/045}{\emph{JCAP} {\bf 07}
  (2024) 045}, [\href{https://arxiv.org/abs/2403.15367}{{\tt 2403.15367}}].

\bibitem{KM3NeT:2025mfl}
{\scshape KM3NeT} collaboration, O.~Adriani et~al., \emph{{KM3NeT Constraint on
  Lorentz-Violating Superluminal Neutrino Velocity}},
  \href{https://arxiv.org/abs/2502.12070}{{\tt 2502.12070}}.

\bibitem{Satunin:2025uui}
P.~Satunin, \emph{{Ultra-high-energy event KM3-230213A constraints on Lorentz
  Invariance Violation in neutrino sector}},
  \href{http://dx.doi.org/10.1140/epjc/s10052-025-14240-8}{\emph{Eur. Phys. J.
  C} {\bf 85} (2025) 545}, [\href{https://arxiv.org/abs/2502.09548}{{\tt
  2502.09548}}].

\bibitem{Cattaneo:2025uxk}
P.~W. Cattaneo, \emph{{Constraints on Lorentz invariance from the event
  KM3-230213A}},
  \href{http://dx.doi.org/10.1140/epjc/s10052-025-14264-0}{\emph{Eur. Phys. J.
  C} {\bf 85} (2025) 529}, [\href{https://arxiv.org/abs/2502.13201}{{\tt
  2502.13201}}].

\bibitem{Kohri:2025bsn}
K.~Kohri, P.~K. Paul and N.~Sahu, \emph{{Super heavy dark matter origin of the
  PeV neutrino event: KM3-230213A}},
  \href{https://arxiv.org/abs/2503.04464}{{\tt 2503.04464}}.

\bibitem{Narita:2025udw}
Y.~Narita and W.~Yin, \emph{{Explaining the KM3-230213A Detection without
  Gamma-Ray Emission: Cosmic-Ray Dark Radiation}},
  \href{https://arxiv.org/abs/2503.07776}{{\tt 2503.07776}}.

\bibitem{He:2025bex}
Y.~He, J.~Liu, X.-P. Wang and Y.-M. Zhong, \emph{{Implications of the KM3NeT
  Ultrahigh-energy Event on Neutrino Self-interactions}},
  \href{https://arxiv.org/abs/2504.20163}{{\tt 2504.20163}}.

\bibitem{Murase:2019vdl}
K.~Murase, S.~S. Kimura and P.~Meszaros, \emph{{Hidden Cores of Active Galactic
  Nuclei as the Origin of Medium-Energy Neutrinos: Critical Tests with the MeV
  Gamma-Ray Connection}},
  \href{http://dx.doi.org/10.1103/PhysRevLett.125.011101}{\emph{Phys. Rev.
  Lett.} {\bf 125} (2020) 011101},
  [\href{https://arxiv.org/abs/1904.04226}{{\tt 1904.04226}}].

\bibitem{Smith:2020oac}
D.~Smith, D.~Hooper and A.~Vieregg, \emph{{Revisiting AGN as the source of
  IceCube\textquoteright{}s diffuse neutrino flux}},
  \href{http://dx.doi.org/10.1088/1475-7516/2021/03/031}{\emph{JCAP} {\bf 03}
  (2021) 031}, [\href{https://arxiv.org/abs/2007.12706}{{\tt 2007.12706}}].

\bibitem{IceCube:2021pgw}
{\scshape IceCube} collaboration, R.~Abbasi et~al., \emph{{Search for neutrino
  emission from cores of active galactic nuclei}},
  \href{http://dx.doi.org/10.1103/PhysRevD.106.022005}{\emph{Phys. Rev. D} {\bf
  106} (2022) 022005}, [\href{https://arxiv.org/abs/2111.10169}{{\tt
  2111.10169}}].

\bibitem{Kheirandish:2021wkm}
A.~Kheirandish, K.~Murase and S.~S. Kimura, \emph{{High-energy Neutrinos from
  Magnetized Coronae of Active Galactic Nuclei and Prospects for Identification
  of Seyfert Galaxies and Quasars in Neutrino Telescopes}},
  \href{http://dx.doi.org/10.3847/1538-4357/ac1c77}{\emph{Astrophys. J.} {\bf
  922} (2021) 45}, [\href{https://arxiv.org/abs/2102.04475}{{\tt 2102.04475}}].

\bibitem{Murase:2022feu}
K.~Murase and F.~W. Stecker, \emph{{Chapter 10: High-Energy Neutrinos from
  Active Galactic Nuclei}},  \href{https://arxiv.org/abs/2202.03381}{{\tt
  2202.03381}}.

\bibitem{Oikonomou-Neutrino-Blazar}
F.~Oikonomou, \emph{{High-energy neutrino emission from blazars}},
  \href{http://dx.doi.org/10.22323/1.395.0030}{\emph{PoS} {\bf ICRC2021} (2022)
  030}, [\href{https://arxiv.org/abs/2201.05623}{{\tt 2201.05623}}].

\bibitem{Cavaliere:2001gb}
A.~Cavaliere and V.~D'Elia, \emph{{The blazar main sequence}},
  \href{http://dx.doi.org/10.1086/339778}{\emph{Astrophys. J.} {\bf 571} (2002)
  226--233}, [\href{https://arxiv.org/abs/astro-ph/0106512}{{\tt
  astro-ph/0106512}}].

\bibitem{Hovatta:2019ulp}
T.~Hovatta and E.~Lindfors, \emph{{Relativistic Jets of Blazars}},
  \href{http://dx.doi.org/10.1016/j.newar.2020.101541}{\emph{New Astron. Rev.}
  {\bf 87} (2019) 101541}, [\href{https://arxiv.org/abs/2003.06322}{{\tt
  2003.06322}}].

\bibitem{Dzhatdoev:2025sdi}
T.~A. Dzhatdoev, \emph{{The blazar PKS 0605-085 as the origin of the
  KM3-230213A ultra high energy neutrino event}},
  \href{https://arxiv.org/abs/2502.11434}{{\tt 2502.11434}}.

\bibitem{2025ApJS..276...38P}
L.~Y. {Petrov} and Y.~Y. {Kovalev}, \emph{{The Radio Fundamental Catalog. I.
  Astrometry}}, \href{http://dx.doi.org/10.3847/1538-4365/ad8c36}{\emph{\apjs}
  {\bf 276} (Feb., 2025) 38}, [\href{https://arxiv.org/abs/2410.11794}{{\tt
  2410.11794}}].

\bibitem{2018ApJS..234...12L}
M.~L. {Lister}, M.~F. {Aller}, H.~D. {Aller}, M.~A. {Hodge}, D.~C. {Homan},
  Y.~Y. {Kovalev} et~al., \emph{{MOJAVE. XV. VLBA 15 GHz Total Intensity and
  Polarization Maps of 437 Parsec-scale AGN Jets from 1996 to 2017}},
  \href{http://dx.doi.org/10.3847/1538-4365/aa9c44}{\emph{\apjs} {\bf 234}
  (Jan., 2018) 12}, [\href{https://arxiv.org/abs/1711.07802}{{\tt
  1711.07802}}].

\bibitem{Aller:2014nwa}
M.~F. Aller, P.~A. Hughes, H.~D. Aller, G.~E. Latimer and T.~Hovatta,
  \emph{{Constraining the Physical Conditions in the Jets of $\Gamma$-Ray
  Flaring Blazars using Centimeter-Band Polarimetry and Radiative Transfer
  Simulations. I. Data and Models for 0420-014, OJ 287, and 1156+295}},
  \href{http://dx.doi.org/10.1088/0004-637X/791/1/53}{\emph{Astrophys. J.} {\bf
  791} (2014) 53}, [\href{https://arxiv.org/abs/1407.2194}{{\tt 1407.2194}}].

\bibitem{vallenari2023gaia}
A.~Vallenari, A.~G. Brown, T.~Prusti, J.~H. De~Bruijne, F.~Arenou, C.~Babusiaux
  et~al., \emph{Gaia data release 3-summary of the content and survey
  properties}, {\emph{Astronomy \& Astrophysics} {\bf 674} (2023) A1}.

\bibitem{2010AJ....140.1868W}
E.~L. {Wright}, P.~R.~M. {Eisenhardt}, A.~K. {Mainzer}, M.~E. {Ressler}, R.~M.
  {Cutri}, T.~{Jarrett} et~al., \emph{{The Wide-field Infrared Survey Explorer
  (WISE): Mission Description and Initial On-orbit Performance}},
  \href{http://dx.doi.org/10.1088/0004-6256/140/6/1868}{\emph{\aj} {\bf 140}
  (Dec., 2010) 1868--1881}, [\href{https://arxiv.org/abs/1008.0031}{{\tt
  1008.0031}}].

\bibitem{2011ApJ...731...53M}
A.~{Mainzer}, J.~{Bauer}, T.~{Grav}, J.~{Masiero}, R.~M. {Cutri}, J.~{Dailey}
  et~al., \emph{{Preliminary Results from NEOWISE: An Enhancement to the
  Wide-field Infrared Survey Explorer for Solar System Science}},
  \href{http://dx.doi.org/10.1088/0004-637X/731/1/53}{\emph{\apj} {\bf 731}
  (Apr., 2011) 53}, [\href{https://arxiv.org/abs/1102.1996}{{\tt 1102.1996}}].

\bibitem{2000yCat.9031....0W}
N.~E. {White}, P.~{Giommi} and L.~{Angelini}, ``{VizieR Online Data Catalog:
  The WGACAT version of ROSAT sources (White+ 2000)}.'' VizieR On-line Data
  Catalog: IX/31. Originally published in: Laboratory for High Energy
  Astrophysics (LHEA/NASA), Greenbelt (2000), June, 2000.

\bibitem{Voges:1999ju}
W.~Voges et~al., \emph{{The ROSAT all - sky survey bright source catalogue}},
  {\emph{Astron. Astrophys.} {\bf 349} (1999) 389},
  [\href{https://arxiv.org/abs/astro-ph/9909315}{{\tt astro-ph/9909315}}].

\bibitem{2004ApJ...611.1005G}
N.~{Gehrels}, G.~{Chincarini}, P.~{Giommi}, K.~O. {Mason}, J.~A. {Nousek},
  A.~A. {Wells} et~al., \emph{{The Swift Gamma-Ray Burst Mission}},
  \href{http://dx.doi.org/10.1086/422091}{\emph{\apj} {\bf 611} (Aug., 2004)
  1005--1020}, [\href{https://arxiv.org/abs/astro-ph/0405233}{{\tt
  astro-ph/0405233}}].

\bibitem{2005SSRv..120..165B}
D.~N. {Burrows}, J.~E. {Hill}, J.~A. {Nousek}, J.~A. {Kennea}, A.~{Wells},
  J.~P. {Osborne} et~al., \emph{{The Swift X-Ray Telescope}},
  \href{http://dx.doi.org/10.1007/s11214-005-5097-2}{\emph{\ssr} {\bf 120}
  (Oct., 2005) 165--195}, [\href{https://arxiv.org/abs/astro-ph/0508071}{{\tt
  astro-ph/0508071}}].

\bibitem{2021A&A...656A.132S}
R.~{Sunyaev}, V.~{Arefiev}, V.~{Babyshkin}, A.~{Bogomolov}, K.~{Borisov},
  M.~{Buntov} et~al., \emph{{SRG X-ray orbital observatory. Its telescopes and
  first scientific results}},
  \href{http://dx.doi.org/10.1051/0004-6361/202141179}{\emph{\aap} {\bf 656}
  (Dec., 2021) A132}, [\href{https://arxiv.org/abs/2104.13267}{{\tt
  2104.13267}}].

\bibitem{2012ApJ...748...49S}
M.~S. {Shaw}, R.~W. {Romani}, G.~{Cotter}, S.~E. {Healey}, P.~F. {Michelson},
  A.~C.~S. {Readhead} et~al., \emph{{Spectroscopy of Broad-line Blazars from
  1LAC}}, \href{http://dx.doi.org/10.1088/0004-637X/748/1/49}{\emph{\apj} {\bf
  748} (Mar., 2012) 49}, [\href{https://arxiv.org/abs/1201.0999}{{\tt
  1201.0999}}].

\bibitem{2015Ap&SS.357...75M}
E.~{Massaro}, A.~{Maselli}, C.~{Leto}, P.~{Marchegiani}, M.~{Perri},
  P.~{Giommi} et~al., \emph{{The 5th edition of the Roma-BZCAT. A short
  presentation}},
  \href{http://dx.doi.org/10.1007/s10509-015-2254-2}{\emph{\apss} {\bf 357}
  (May, 2015) 75}, [\href{https://arxiv.org/abs/1502.07755}{{\tt 1502.07755}}].

\bibitem{Ballet:2023qzs}
{\scshape Fermi-LAT} collaboration, J.~Ballet, P.~Bruel, T.~H. Burnett and
  B.~Lott, \emph{{Fermi Large Area Telescope Fourth Source Catalog Data Release
  4 (4FGL-DR4)}},  \href{https://arxiv.org/abs/2307.12546}{{\tt 2307.12546}}.

\bibitem{PKSMBH}
X.~Zhang, D.-r. Xiong, Q.-g. Gao, G.-q. Yang, F.-w. Lu, W.-w. Na et~al.,
  \emph{{The fundamental plane of blazars based on the black hole spin-mass
  energy}}, \href{http://dx.doi.org/10.1093/mnras/stae765}{\emph{Mon. Not. Roy.
  Astron. Soc.} {\bf 529} (2024) 3699--3711},
  [\href{https://arxiv.org/abs/2403.09088}{{\tt 2403.09088}}].

\bibitem{PMNMBH}
G.~Ghisellini and F.~Tavecchio, \emph{{Fermi/LAT broad emission line blazars}},
  \href{http://dx.doi.org/10.1093/mnras/stv055}{\emph{Mon. Not. Roy. Astron.
  Soc.} {\bf 448} (2015) 1060--1077},
  [\href{https://arxiv.org/abs/1501.03504}{{\tt 1501.03504}}].

\bibitem{Padovani:2018acg}
P.~Padovani, P.~Giommi, E.~Resconi, T.~Glauch, B.~Arsioli, N.~Sahakyan et~al.,
  \emph{{Dissecting the region around IceCube-170922A: the blazar TXS 0506+056
  as the first cosmic neutrino source}},
  \href{http://dx.doi.org/10.1093/mnras/sty1852}{\emph{Mon. Not. Roy. Astron.
  Soc.} {\bf 480} (2018) 192--203},
  [\href{https://arxiv.org/abs/1807.04461}{{\tt 1807.04461}}].

\bibitem{Dermer:2014vaa}
C.~D. Dermer, K.~Murase and Y.~Inoue, \emph{{Photopion Production in Black-Hole
  Jets and Flat-Spectrum Radio Quasars as PeV Neutrino Sources}},
  \href{http://dx.doi.org/10.1016/j.jheap.2014.09.001}{\emph{JHEAp} {\bf 3-4}
  (2014) 29--40}, [\href{https://arxiv.org/abs/1406.2633}{{\tt 1406.2633}}].

\bibitem{Padovani-BLR-Size-TXS}
P.~Padovani, F.~Oikonomou, M.~Petropoulou, P.~Giommi and E.~Resconi, \emph{{TXS
  0506+056, the first cosmic neutrino source, is not a BL Lac}},
  \href{http://dx.doi.org/10.1093/mnrasl/slz011}{\emph{Mon. Not. Roy. Astron.
  Soc.} {\bf 484} (2019) L104--L108},
  [\href{https://arxiv.org/abs/1901.06998}{{\tt 1901.06998}}].

\bibitem{Ghisellini-BLR-Size}
G.~Ghisellini and F.~Tavecchio, \emph{{The blazar sequence: a new
  perspective}},
  \href{http://dx.doi.org/10.1111/j.1365-2966.2008.13360.x}{\emph{Mon. Not.
  Roy. Astron. Soc.} {\bf 387} (2008) 1669},
  [\href{https://arxiv.org/abs/0802.1918}{{\tt 0802.1918}}].

\bibitem{Dzhatdoev-PKS}
T.~A. Dzhatdoev, \emph{{The blazar PKS 0605-085 as the origin of the
  KM3-230213A ultra high energy neutrino event}},
  \href{https://arxiv.org/abs/2502.11434}{{\tt 2502.11434}}.

\bibitem{Tavecchio:2016dcj}
F.~Tavecchio, \emph{{Gamma Rays From Blazars}},
  \href{http://dx.doi.org/10.1063/1.4968892}{\emph{AIP Conf. Proc.} {\bf 1792}
  (2017) 020007}, [\href{https://arxiv.org/abs/1609.04260}{{\tt 1609.04260}}].

\bibitem{Zheng-FSRQ_BLR}
Y.~G. Zheng, C.~Y. Yang, L.~Zhang and J.~C. Wang, \emph{{Discerning the
  $\gamma$-ray emitting region in the flat spectrum radio quasars}},
  \href{http://dx.doi.org/10.3847/1538-4365/228/1/1}{\emph{Astrophys. J.
  Suppl.} {\bf 228} (2017) 1}, [\href{https://arxiv.org/abs/1612.02394}{{\tt
  1612.02394}}].

\bibitem{Fan:2023fzj}
J.~Fan, H.~Xiao, W.~Yang, L.~Zhang, A.~A. Strigachev, R.~S. Bachev et~al.,
  \emph{{Characterizing the Emission Region Properties of Blazars}},
  \href{http://dx.doi.org/10.3847/1538-4365/ace7c8}{\emph{Astrophys. J. Suppl.}
  {\bf 268} (2023) 23}, [\href{https://arxiv.org/abs/2307.07163}{{\tt
  2307.07163}}].

\bibitem{Viale-Neutrino-Emission}
{\scshape Fermi/LAT} collaboration, I.~Viale, C.~Righi, G.~Principe,
  M.~Cerruti, F.~Tavecchio and E.~Bernardini, \emph{{Candidate
  neutrino-emitting blazars sharing physical properties with TXS 0506+056}},
  \href{http://dx.doi.org/10.22323/1.444.1526}{\emph{PoS} {\bf ICRC2023} (2023)
  1526}, [\href{https://arxiv.org/abs/2410.07905}{{\tt 2410.07905}}].

\bibitem{Xue-Emission-BLR}
R.~Xue, R.-Y. Liu, M.~Petropoulou, F.~Oikonomou, Z.-R. Wang, K.~Wang et~al.,
  \emph{{A two-zone model for blazar emission: implications for TXS 0506+056
  and the neutrino event IceCube-170922A}},
  \href{https://arxiv.org/abs/1908.10190}{{\tt 1908.10190}}.

\bibitem{Zhu-BLR-Jet-Production}
J.~Zhu, H.~Cao, H.~Xiao, Z.~Pei, J.~Fan and D.~Bastieri, \emph{{Chasing the
  Neutrino Blazar Candidates}},
  \href{http://dx.doi.org/10.3847/1538-4365/ad7730}{\emph{Astrophys. J. Suppl.}
  {\bf 275} (2024) 11}.

\bibitem{Murase:2018iyl}
K.~Murase, F.~Oikonomou and M.~Petropoulou, \emph{{Blazar Flares as an Origin
  of High-Energy Cosmic Neutrinos?}},
  \href{http://dx.doi.org/10.3847/1538-4357/aada00}{\emph{Astrophys. J.} {\bf
  865} (2018) 124}, [\href{https://arxiv.org/abs/1807.04748}{{\tt
  1807.04748}}].

\bibitem{Xue:2019txw}
R.~Xue, R.-Y. Liu, M.~Petropoulou, F.~Oikonomou, Z.-R. Wang, K.~Wang et~al.,
  \emph{{A two-zone model for blazar emission: implications for TXS 0506+056
  and the neutrino event IceCube-170922A}},
  \href{https://arxiv.org/abs/1908.10190}{{\tt 1908.10190}}.

\bibitem{Zhang:2019dob}
H.~Zhang, K.~Fang, H.~Li, D.~Giannios, M.~B\"ottcher and S.~Buson,
  \emph{{Probing the Emission Mechanism and Magnetic Field of Neutrino Blazars
  with Multiwavelength Polarization Signatures}},
  \href{http://dx.doi.org/10.3847/1538-4357/ab158d}{\emph{Astrophys. J.} {\bf
  876} (2019) 109}, [\href{https://arxiv.org/abs/1903.01956}{{\tt
  1903.01956}}].

\bibitem{Gasparyan-Neutrino-TXS}
S.~Gasparyan, D.~B\'egu\'e and N.~Sahakyan, \emph{{Time-dependent
  lepto-hadronic modelling of the emission from blazar jets with SOPRANO: the
  case of TXS 0506~+~056, 3HSP J095507.9~+~355101, and 3C 279}},
  \href{http://dx.doi.org/10.1093/mnras/stab2688}{\emph{Mon. Not. Roy. Astron.
  Soc.} {\bf 509} (2021) 2102--2121},
  [\href{https://arxiv.org/abs/2110.01549}{{\tt 2110.01549}}].

\bibitem{Navarro:1996gj}
J.~F. Navarro, C.~S. Frenk and S.~D.~M. White, \emph{{A Universal density
  profile from hierarchical clustering}},
  \href{http://dx.doi.org/10.1086/304888}{\emph{Astrophys. J.} {\bf 490} (1997)
  493--508}, [\href{https://arxiv.org/abs/astro-ph/9611107}{{\tt
  astro-ph/9611107}}].

\bibitem{2010arXiv1010.2539D}
N.~{Dalal}, Y.~{Lithwick} and M.~{Kuhlen}, \emph{{The Origin of Dark Matter
  Halo Profiles}},
  \href{http://dx.doi.org/10.48550/arXiv.1010.2539}{\emph{arXiv e-prints}
  (Oct., 2010) arXiv:1010.2539}, [\href{https://arxiv.org/abs/1010.2539}{{\tt
  1010.2539}}].

\bibitem{Enomoto:2024twc}
Y.~Enomoto, A.~Taruya, S.~Tanaka and T.~Nishimichi, \emph{{Inner structure of
  cold and warm dark matter halos from particle dynamics}},
  \href{http://dx.doi.org/10.1093/pasj/psaf001}{\emph{Publ. Astron. Soc. Jap.}
  {\bf 77} (2025) 337--355}, [\href{https://arxiv.org/abs/2410.00195}{{\tt
  2410.00195}}].

\bibitem{Angulo:2021kes}
R.~E. Angulo and O.~Hahn, \emph{{Large-scale dark matter simulations}},
  \href{http://dx.doi.org/10.1007/s41115-021-00013-z}{\emph{Liv. Rev. Comput.
  Astrophys.} {\bf 8} (2022) 1}, [\href{https://arxiv.org/abs/2112.05165}{{\tt
  2112.05165}}].

\bibitem{PowellMBHMH:}
M.~C. Powell et~al., \emph{{BASS. XXXVI. Constraining the Local Supermassive
  Black Hole\textendash{}Halo Connection with BASS DR2 AGNs}},
  \href{http://dx.doi.org/10.3847/1538-4357/ac8f8e}{\emph{Astrophys. J.} {\bf
  938} (2022) 77}, [\href{https://arxiv.org/abs/2209.02728}{{\tt 2209.02728}}].

\bibitem{MarascoMBHMH}
A.~{Marasco}, G.~{Cresci}, L.~{Posti}, F.~{Fraternali}, F.~{Mannucci},
  A.~{Marconi} et~al., \emph{{A universal relation between the properties of
  supermassive black holes, galaxies, and dark matter haloes}},
  \href{http://dx.doi.org/10.1093/mnras/stab2317}{\emph{\mnras} {\bf 507}
  (Nov., 2021) 4274--4293}, [\href{https://arxiv.org/abs/2105.10508}{{\tt
  2105.10508}}].

\bibitem{Kormendy:2013dxa}
J.~Kormendy and L.~C. Ho, \emph{{Coevolution (Or Not) of Supermassive Black
  Holes and Host Galaxies}},
  \href{http://dx.doi.org/10.1146/annurev-astro-082708-101811}{\emph{Ann. Rev.
  Astron. Astrophys.} {\bf 51} (2013) 511--653},
  [\href{https://arxiv.org/abs/1304.7762}{{\tt 1304.7762}}].

\bibitem{reines2015relations}
A.~E. Reines and M.~Volonteri, \emph{Relations between central black hole mass
  and total galaxy stellar mass in the local universe}, {\emph{The
  Astrophysical Journal} {\bf 813} (2015) 82}.

\bibitem{2018ApJ...863...42C}
C.~J. {Conselice}, J.~W. {Twite}, D.~P. {Palamara} and W.~{Hartley}, \emph{{The
  Halo Masses of Galaxies to z {\ensuremath{\sim}} 3: A Hybrid Observational
  and Theoretical Approach}},
  \href{http://dx.doi.org/10.3847/1538-4357/aacda8}{\emph{\apj} {\bf 863}
  (Aug., 2018) 42}, [\href{https://arxiv.org/abs/1806.07752}{{\tt
  1806.07752}}].

\bibitem{2010A&A...522L...3S}
J.~E. {Sarria}, R.~{Maiolino}, F.~{La Franca}, F.~{Pozzi}, F.~{Fiore},
  A.~{Marconi} et~al., \emph{{The M$_{BH}$ - M$_{star}$ relation of obscured
  AGNs at high redshift}},
  \href{http://dx.doi.org/10.1051/0004-6361/201015696}{\emph{\aap} {\bf 522}
  (Nov., 2010) L3}, [\href{https://arxiv.org/abs/1010.0768}{{\tt 1010.0768}}].

\bibitem{Shankar:2019yyr}
F.~Shankar et~al., \emph{{Black hole scaling relations of active and quiescent
  galaxies: Addressing selection effects and constraining virial factors}},
  \href{http://dx.doi.org/10.1093/mnras/stz376}{\emph{Mon. Not. Roy. Astron.
  Soc.} {\bf 485} (2019) 1278--1292},
  [\href{https://arxiv.org/abs/1901.11036}{{\tt 1901.11036}}].

\bibitem{suh2020no}
H.~Suh, F.~Civano, B.~Trakhtenbrot, F.~Shankar, G.~Hasinger, D.~B. Sanders
  et~al., \emph{No significant evolution of relations between black hole mass
  and galaxy total stellar mass up to z~ 2.5}, {\emph{The Astrophysical
  Journal} {\bf 889} (2020) 32}.

\bibitem{CirelliDM}
M.~Cirelli, A.~Strumia and J.~Zupan, \emph{{Dark Matter}},
  \href{https://arxiv.org/abs/2406.01705}{{\tt 2406.01705}}.

\bibitem{Correa-ConcentrationParam}
C.~A. Correa, J.~S.~B. Wyithe, J.~Schaye and A.~R. Duffy, \emph{{The accretion
  history of dark matter haloes \textendash{} III. A physical model for the
  concentration\textendash{}mass relation}},
  \href{http://dx.doi.org/10.1093/mnras/stv1363}{\emph{Mon. Not. Roy. Astron.
  Soc.} {\bf 452} (2015) 1217--1232},
  [\href{https://arxiv.org/abs/1502.00391}{{\tt 1502.00391}}].

\bibitem{Cackett:2021gad}
E.~M. Cackett, M.~C. Bentz and E.~Kara, \emph{{Reverberation mapping of active
  galactic nuclei: From X-ray corona to dusty torus}},
  \href{http://dx.doi.org/10.1016/j.isci.2021.102557}{\emph{iScience} {\bf 24}
  (2021) 102557}, [\href{https://arxiv.org/abs/2105.06926}{{\tt 2105.06926}}].

\bibitem{Xiong:2014hta}
D.~R. Xiong and X.~Zhang, \emph{{Intrinsic \ensuremath{\gamma}-ray luminosity,
  black hole mass, jet and accretion in Fermi blazars}},
  \href{http://dx.doi.org/10.1093/mnras/stu755}{\emph{Mon. Not. Roy. Astron.
  Soc.} {\bf 441} (2014) 3375--3395},
  [\href{https://arxiv.org/abs/1404.3556}{{\tt 1404.3556}}].

\bibitem{Gondolo-Spike}
P.~Gondolo and J.~Silk, \emph{{Dark matter annihilation at the galactic
  center}}, \href{http://dx.doi.org/10.1103/PhysRevLett.83.1719}{\emph{Phys.
  Rev. Lett.} {\bf 83} (1999) 1719--1722},
  [\href{https://arxiv.org/abs/astro-ph/9906391}{{\tt astro-ph/9906391}}].

\bibitem{Merritt:2003qk}
D.~Merritt, \emph{{Evolution of the dark matter distribution at the galactic
  center}}, \href{http://dx.doi.org/10.1103/PhysRevLett.92.201304}{\emph{Phys.
  Rev. Lett.} {\bf 92} (2004) 201304},
  [\href{https://arxiv.org/abs/astro-ph/0311594}{{\tt astro-ph/0311594}}].

\bibitem{Gnedin:2003rj}
O.~Y. Gnedin and J.~R. Primack, \emph{{Dark Matter Profile in the Galactic
  Center}}, \href{http://dx.doi.org/10.1103/PhysRevLett.93.061302}{\emph{Phys.
  Rev. Lett.} {\bf 93} (2004) 061302},
  [\href{https://arxiv.org/abs/astro-ph/0308385}{{\tt astro-ph/0308385}}].

\bibitem{Sadeghian:2013laa}
L.~Sadeghian, F.~Ferrer and C.~M. Will, \emph{{Dark matter distributions around
  massive black holes: A general relativistic analysis}},
  \href{http://dx.doi.org/10.1103/PhysRevD.88.063522}{\emph{Phys. Rev. D} {\bf
  88} (2013) 063522}, [\href{https://arxiv.org/abs/1305.2619}{{\tt
  1305.2619}}].

\bibitem{Chan:2022gqd}
M.~H. Chan and C.~M. Lee, \emph{{Indirect Evidence for Dark Matter Density
  Spikes around Stellar-mass Black Holes}},
  \href{http://dx.doi.org/10.3847/2041-8213/acaafa}{\emph{Astrophys. J. Lett.}
  {\bf 943} (2023) L11}, [\href{https://arxiv.org/abs/2212.05664}{{\tt
  2212.05664}}].

\bibitem{Chan:2024yht}
M.~H. Chan and C.~M. Lee, \emph{{The First Robust Evidence Showing a Dark
  Matter Density Spike Around the Supermassive Black Hole in OJ 287}},
  \href{http://dx.doi.org/10.3847/2041-8213/ad2465}{\emph{Astrophys. J. Lett.}
  {\bf 962} (2024) L40}, [\href{https://arxiv.org/abs/2402.03751}{{\tt
  2402.03751}}].

\bibitem{Ullio:2001fb}
P.~Ullio, H.~Zhao and M.~Kamionkowski, \emph{{A Dark matter spike at the
  galactic center?}},
  \href{http://dx.doi.org/10.1103/PhysRevD.64.043504}{\emph{Phys. Rev. D} {\bf
  64} (2001) 043504}, [\href{https://arxiv.org/abs/astro-ph/0101481}{{\tt
  astro-ph/0101481}}].

\bibitem{Merritt:2002vj}
D.~Merritt, M.~Milosavljevic, L.~Verde and R.~Jimenez, \emph{{Dark matter
  spikes and annihilation radiation from the galactic center}},
  \href{http://dx.doi.org/10.1103/PhysRevLett.88.191301}{\emph{Phys. Rev.
  Lett.} {\bf 88} (2002) 191301},
  [\href{https://arxiv.org/abs/astro-ph/0201376}{{\tt astro-ph/0201376}}].

\bibitem{Bertone:2005hw}
G.~Bertone and D.~Merritt, \emph{{Time-dependent models for dark matter at the
  Galactic Center}},
  \href{http://dx.doi.org/10.1103/PhysRevD.72.103502}{\emph{Phys. Rev. D} {\bf
  72} (2005) 103502}, [\href{https://arxiv.org/abs/astro-ph/0501555}{{\tt
  astro-ph/0501555}}].

\bibitem{Steigman:2012nb}
G.~Steigman, B.~Dasgupta and J.~F. Beacom, \emph{{Precise Relic WIMP Abundance
  and its Impact on Searches for Dark Matter Annihilation}},
  \href{http://dx.doi.org/10.1103/PhysRevD.86.023506}{\emph{Phys. Rev. D} {\bf
  86} (2012) 023506}, [\href{https://arxiv.org/abs/1204.3622}{{\tt
  1204.3622}}].

\bibitem{Akita:2023yga}
K.~Akita and S.~Ando, \emph{{Constraints on dark matter-neutrino scattering
  from the Milky-Way satellites and subhalo modeling for dark acoustic
  oscillations}},
  \href{http://dx.doi.org/10.1088/1475-7516/2023/11/037}{\emph{JCAP} {\bf 11}
  (2023) 037}, [\href{https://arxiv.org/abs/2305.01913}{{\tt 2305.01913}}].

\bibitem{Planck2018}
{\scshape Planck} collaboration, N.~Aghanim et~al., \emph{{Planck 2018 results.
  VI. Cosmological parameters}},
  \href{http://dx.doi.org/10.1051/0004-6361/201833910}{\emph{Astron.
  Astrophys.} {\bf 641} (2020) A6},
  [\href{https://arxiv.org/abs/1807.06209}{{\tt 1807.06209}}].

\bibitem{Bringmann:2012ez}
T.~Bringmann and C.~Weniger, \emph{{Gamma Ray Signals from Dark Matter:
  Concepts, Status and Prospects}},
  \href{http://dx.doi.org/10.1016/j.dark.2012.10.005}{\emph{Phys. Dark Univ.}
  {\bf 1} (2012) 194--217}, [\href{https://arxiv.org/abs/1208.5481}{{\tt
  1208.5481}}].

\bibitem{Ng:2013xha}
K.~C.~Y. Ng, R.~Laha, S.~Campbell, S.~Horiuchi, B.~Dasgupta, K.~Murase et~al.,
  \emph{{Resolving small-scale dark matter structures using multisource
  indirect detection}},
  \href{http://dx.doi.org/10.1103/PhysRevD.89.083001}{\emph{Phys. Rev. D} {\bf
  89} (2014) 083001}, [\href{https://arxiv.org/abs/1310.1915}{{\tt
  1310.1915}}].

\bibitem{Fox:2008kb}
P.~J. Fox and E.~Poppitz, \emph{{Leptophilic Dark Matter}},
  \href{http://dx.doi.org/10.1103/PhysRevD.79.083528}{\emph{Phys. Rev. D} {\bf
  79} (2009) 083528}, [\href{https://arxiv.org/abs/0811.0399}{{\tt
  0811.0399}}].

\bibitem{Bell:2014tta}
N.~F. Bell, Y.~Cai, R.~K. Leane and A.~D. Medina, \emph{{Leptophilic dark
  matter with $Z'$ interactions}},
  \href{http://dx.doi.org/10.1103/PhysRevD.90.035027}{\emph{Phys. Rev. D} {\bf
  90} (2014) 035027}, [\href{https://arxiv.org/abs/1407.3001}{{\tt
  1407.3001}}].

\bibitem{Boehm:2003hm}
C.~Boehm and P.~Fayet, \emph{{Scalar dark matter candidates}},
  \href{http://dx.doi.org/10.1016/j.nuclphysb.2004.01.015}{\emph{Nucl. Phys. B}
  {\bf 683} (2004) 219--263}, [\href{https://arxiv.org/abs/hep-ph/0305261}{{\tt
  hep-ph/0305261}}].

\bibitem{Boehm:2006mi}
C.~Boehm, Y.~Farzan, T.~Hambye, S.~Palomares-Ruiz and S.~Pascoli, \emph{{Is it
  possible to explain neutrino masses with scalar dark matter?}},
  \href{http://dx.doi.org/10.1103/PhysRevD.77.043516}{\emph{Phys. Rev. D} {\bf
  77} (2008) 043516}, [\href{https://arxiv.org/abs/hep-ph/0612228}{{\tt
  hep-ph/0612228}}].

\bibitem{Boehm:2013jpa}
C.~Boehm, M.~J. Dolan and C.~McCabe, \emph{{A Lower Bound on the Mass of Cold
  Thermal Dark Matter from Planck}},
  \href{http://dx.doi.org/10.1088/1475-7516/2013/08/041}{\emph{JCAP} {\bf 08}
  (2013) 041}, [\href{https://arxiv.org/abs/1303.6270}{{\tt 1303.6270}}].

\bibitem{Arhrib:2015dez}
A.~Arhrib, C.~B\oe{}hm, E.~Ma and T.-C. Yuan, \emph{{Radiative Model of
  Neutrino Mass with Neutrino Interacting MeV Dark Matter}},
  \href{http://dx.doi.org/10.1088/1475-7516/2016/04/049}{\emph{JCAP} {\bf 04}
  (2016) 049}, [\href{https://arxiv.org/abs/1512.08796}{{\tt 1512.08796}}].

\bibitem{Belyaev:2022shr}
A.~Belyaev, A.~Deandrea, S.~Moretti, L.~Panizzi, D.~A. Ross and N.~Thongyoi,
  \emph{{Fermionic portal to vector dark matter from a new gauge sector}},
  \href{http://dx.doi.org/10.1103/PhysRevD.108.095001}{\emph{Phys. Rev. D} {\bf
  108} (2023) 095001}, [\href{https://arxiv.org/abs/2204.03510}{{\tt
  2204.03510}}].

\bibitem{Herms:2023cyy}
J.~Herms, S.~Jana, V.~P. K. and S.~Saad, \emph{{Light neutrinophilic dark
  matter from a scotogenic model}},
  \href{http://dx.doi.org/10.1016/j.physletb.2023.138167}{\emph{Phys. Lett. B}
  {\bf 845} (2023) 138167}, [\href{https://arxiv.org/abs/2307.15760}{{\tt
  2307.15760}}].

\bibitem{PhysRevLett.42.407}
S.~Tremaine and J.~E. Gunn, \emph{Dynamical role of light neutral leptons in
  cosmology}, \href{http://dx.doi.org/10.1103/PhysRevLett.42.407}{\emph{Phys.
  Rev. Lett.} {\bf 42} (Feb, 1979) 407--410}.

\bibitem{Dalcanton:2000hn}
J.~J. Dalcanton and C.~J. Hogan, \emph{{Halo cores and phase space densities:
  Observational constraints on dark matter physics and structure formation}},
  \href{http://dx.doi.org/10.1086/323207}{\emph{Astrophys. J.} {\bf 561} (2001)
  35--45}, [\href{https://arxiv.org/abs/astro-ph/0004381}{{\tt
  astro-ph/0004381}}].

\bibitem{Boyarsky:2008ju}
A.~Boyarsky, O.~Ruchayskiy and D.~Iakubovskyi, \emph{{A Lower bound on the mass
  of Dark Matter particles}},
  \href{http://dx.doi.org/10.1088/1475-7516/2009/03/005}{\emph{JCAP} {\bf 03}
  (2009) 005}, [\href{https://arxiv.org/abs/0808.3902}{{\tt 0808.3902}}].

\bibitem{Alvey:2020xsk}
J.~Alvey, N.~Sabti, V.~Tiki, D.~Blas, K.~Bondarenko, A.~Boyarsky et~al.,
  \emph{{New constraints on the mass of fermionic dark matter from dwarf
  spheroidal galaxies}},
  \href{http://dx.doi.org/10.1093/mnras/staa3640}{\emph{Mon. Not. Roy. Astron.
  Soc.} {\bf 501} (2021) 1188--1201},
  [\href{https://arxiv.org/abs/2010.03572}{{\tt 2010.03572}}].

\bibitem{Dvorkin:2019zdi}
C.~Dvorkin, T.~Lin and K.~Schutz, \emph{{Making dark matter out of light:
  freeze-in from plasma effects}},
  \href{http://dx.doi.org/10.1103/PhysRevD.99.115009}{\emph{Phys. Rev. D} {\bf
  99} (2019) 115009}, [\href{https://arxiv.org/abs/1902.08623}{{\tt
  1902.08623}}].

\bibitem{Chang:2019xva}
J.~H. Chang, R.~Essig and A.~Reinert, \emph{{Light(ly)-coupled Dark Matter in
  the keV Range: Freeze-In and Constraints}},
  \href{http://dx.doi.org/10.1007/JHEP03(2021)141}{\emph{JHEP} {\bf 03} (2021)
  141}, [\href{https://arxiv.org/abs/1911.03389}{{\tt 1911.03389}}].

\bibitem{DEramo:2020gpr}
F.~D'Eramo and A.~Lenoci, \emph{{Lower mass bounds on FIMP dark matter produced
  via freeze-in}},
  \href{http://dx.doi.org/10.1088/1475-7516/2021/10/045}{\emph{JCAP} {\bf 10}
  (2021) 045}, [\href{https://arxiv.org/abs/2012.01446}{{\tt 2012.01446}}].

\bibitem{Berbig:2022nre}
M.~Berbig, \emph{{Freeze-In of radiative keV-scale neutrino dark matter from a
  new U(1)$_{B-L}$}},
  \href{http://dx.doi.org/10.1007/JHEP09(2022)101}{\emph{JHEP} {\bf 09} (2022)
  101}, [\href{https://arxiv.org/abs/2203.04276}{{\tt 2203.04276}}].

\bibitem{Zhang:2025abk}
Q.~Zhang, T.-Q. Huang and Z.~Li, \emph{{Cosmogenic Neutrino Point Source and
  KM3-230213A}},  \href{https://arxiv.org/abs/2504.10378}{{\tt 2504.10378}}.

\bibitem{Berezinsky:1969erk}
V.~S. Berezinsky and G.~T. Zatsepin, \emph{{Cosmic rays at ultrahigh-energies
  (neutrino?)}},
  \href{http://dx.doi.org/10.1016/0370-2693(69)90341-4}{\emph{Phys. Lett. B}
  {\bf 28} (1969) 423--424}.

\bibitem{1979ApJ...228..919S}
F.~W. {Stecker}, \emph{{Diffuse fluxes of cosmic high-energy neutrinos.}},
  \href{http://dx.doi.org/10.1086/156919}{\emph{\apj} {\bf 228} (Mar., 1979)
  919--927}.

\bibitem{DiValentino:2017oaw}
E.~Di~Valentino, C.~B\o{}ehm, E.~Hivon and F.~R. Bouchet, \emph{{Reducing the
  $H_0$ and $\sigma_8$ tensions with Dark Matter-neutrino interactions}},
  \href{http://dx.doi.org/10.1103/PhysRevD.97.043513}{\emph{Phys. Rev. D} {\bf
  97} (2018) 043513}, [\href{https://arxiv.org/abs/1710.02559}{{\tt
  1710.02559}}].

\bibitem{Olivares-DelCampo:2017feq}
A.~Olivares-Del~Campo, C.~B\oe{}hm, S.~Palomares-Ruiz and S.~Pascoli,
  \emph{{Dark matter-neutrino interactions through the lens of their
  cosmological implications}},
  \href{http://dx.doi.org/10.1103/PhysRevD.97.075039}{\emph{Phys. Rev. D} {\bf
  97} (2018) 075039}, [\href{https://arxiv.org/abs/1711.05283}{{\tt
  1711.05283}}].

\bibitem{Lattanzi:2017ubx}
M.~Lattanzi and M.~Gerbino, \emph{{Status of neutrino properties and future
  prospects - Cosmological and astrophysical constraints}},
  \href{http://dx.doi.org/10.3389/fphy.2017.00070}{\emph{Front. in Phys.} {\bf
  5} (2018) 70}, [\href{https://arxiv.org/abs/1712.07109}{{\tt 1712.07109}}].

\bibitem{Yuksel:2007ac}
H.~Yuksel, S.~Horiuchi, J.~F. Beacom and S.~Ando, \emph{{Neutrino Constraints
  on the Dark Matter Total Annihilation Cross Section}},
  \href{http://dx.doi.org/10.1103/PhysRevD.76.123506}{\emph{Phys. Rev. D} {\bf
  76} (2007) 123506}, [\href{https://arxiv.org/abs/0707.0196}{{\tt
  0707.0196}}].

\bibitem{Palomares-Ruiz:2007trf}
S.~Palomares-Ruiz and S.~Pascoli, \emph{{Testing MeV dark matter with neutrino
  detectors}}, \href{http://dx.doi.org/10.1103/PhysRevD.77.025025}{\emph{Phys.
  Rev. D} {\bf 77} (2008) 025025}, [\href{https://arxiv.org/abs/0710.5420}{{\tt
  0710.5420}}].

\bibitem{Palomares-Ruiz:2007egs}
S.~Palomares-Ruiz, \emph{{Model-independent bound on the dark matter
  lifetime}},
  \href{http://dx.doi.org/10.1016/j.physletb.2008.05.040}{\emph{Phys. Lett. B}
  {\bf 665} (2008) 50--53}, [\href{https://arxiv.org/abs/0712.1937}{{\tt
  0712.1937}}].

\bibitem{Kile:2009nn}
J.~Kile and A.~Soni, \emph{{Hidden MeV-Scale Dark Matter in Neutrino
  Detectors}}, \href{http://dx.doi.org/10.1103/PhysRevD.80.115017}{\emph{Phys.
  Rev. D} {\bf 80} (2009) 115017}, [\href{https://arxiv.org/abs/0908.3892}{{\tt
  0908.3892}}].

\bibitem{Covi:2009xn}
L.~Covi, M.~Grefe, A.~Ibarra and D.~Tran, \emph{{Neutrino Signals from Dark
  Matter Decay}},
  \href{http://dx.doi.org/10.1088/1475-7516/2010/04/017}{\emph{JCAP} {\bf 04}
  (2010) 017}, [\href{https://arxiv.org/abs/0912.3521}{{\tt 0912.3521}}].

\bibitem{Primulando:2017kxf}
R.~Primulando and P.~Uttayarat, \emph{{Dark Matter-Neutrino Interaction in
  Light of Collider and Neutrino Telescope Data}},
  \href{http://dx.doi.org/10.1007/JHEP06(2018)026}{\emph{JHEP} {\bf 06} (2018)
  026}, [\href{https://arxiv.org/abs/1710.08567}{{\tt 1710.08567}}].

\bibitem{Arguelles:2019ouk}
C.~A. Arg\"uelles, A.~Diaz, A.~Kheirandish, A.~Olivares-Del-Campo, I.~Safa and
  A.~C. Vincent, \emph{{Dark matter annihilation to neutrinos}},
  \href{http://dx.doi.org/10.1103/RevModPhys.93.035007}{\emph{Rev. Mod. Phys.}
  {\bf 93} (2021) 035007}, [\href{https://arxiv.org/abs/1912.09486}{{\tt
  1912.09486}}].

\bibitem{Blennow:2019fhy}
M.~Blennow, E.~Fernandez-Martinez, A.~Olivares-Del~Campo, S.~Pascoli,
  S.~Rosauro-Alcaraz and A.~V. Titov, \emph{{Neutrino Portals to Dark Matter}},
  \href{http://dx.doi.org/10.1140/epjc/s10052-019-7060-5}{\emph{Eur. Phys. J.
  C} {\bf 79} (2019) 555}, [\href{https://arxiv.org/abs/1903.00006}{{\tt
  1903.00006}}].

\bibitem{Arguelles:2022nbl}
C.~A. Arg\"uelles, D.~Delgado, A.~Friedlander, A.~Kheirandish, I.~Safa, A.~C.
  Vincent et~al., \emph{{Dark matter decay to neutrinos}},
  \href{http://dx.doi.org/10.1103/PhysRevD.108.123021}{\emph{Phys. Rev. D} {\bf
  108} (2023) 123021}, [\href{https://arxiv.org/abs/2210.01303}{{\tt
  2210.01303}}].

\bibitem{Aoki:2012ub}
M.~Aoki, M.~Duerr, J.~Kubo and H.~Takano, \emph{{Multi-Component Dark Matter
  Systems and Their Observation Prospects}},
  \href{http://dx.doi.org/10.1103/PhysRevD.86.076015}{\emph{Phys. Rev. D} {\bf
  86} (2012) 076015}, [\href{https://arxiv.org/abs/1207.3318}{{\tt
  1207.3318}}].

\bibitem{Berlin:2018bsc}
A.~Berlin, N.~Blinov, G.~Krnjaic, P.~Schuster and N.~Toro, \emph{{Dark Matter,
  Millicharges, Axion and Scalar Particles, Gauge Bosons, and Other New Physics
  with LDMX}}, \href{http://dx.doi.org/10.1103/PhysRevD.99.075001}{\emph{Phys.
  Rev. D} {\bf 99} (2019) 075001},
  [\href{https://arxiv.org/abs/1807.01730}{{\tt 1807.01730}}].

\bibitem{Hooper:2021rjc}
D.~C. Hooper and M.~Lucca, \emph{{Hints of dark matter-neutrino interactions in
  Lyman-\ensuremath{\alpha} data}},
  \href{http://dx.doi.org/10.1103/PhysRevD.105.103504}{\emph{Phys. Rev. D} {\bf
  105} (2022) 103504}, [\href{https://arxiv.org/abs/2110.04024}{{\tt
  2110.04024}}].

\bibitem{Mosbech:2022uud}
M.~R. Mosbech, C.~Boehm and Y.~Y.~Y. Wong, \emph{{Probing dark matter
  interactions with 21cm observations}},
  \href{http://dx.doi.org/10.1088/1475-7516/2023/03/047}{\emph{JCAP} {\bf 03}
  (2023) 047}, [\href{https://arxiv.org/abs/2207.03107}{{\tt 2207.03107}}].

\bibitem{Bertolez-Martinez:2025trs}
T.~Bert\'olez-Mart\'\i{}nez, G.~Herrera, P.~Mart\'\i{}nez-Mirav\'e and J.~T.
  Calvo, \emph{{The Highest-Energy Neutrino Event Constrains Dark
  Matter-Neutrino Interactions}},  \href{https://arxiv.org/abs/2506.08993}{{\tt
  2506.08993}}.

\end{thebibliography}\endgroup
\end{document}